\documentclass[nofootinbib,aps,prd,superscriptaddress,preprintnumbers,%
 showpacs,showkeys,floatfix,amssymb,amsfonts]{revtex4-1}  

\usepackage{amssymb} 
\newenvironment{frcseries}{\fontfamily{frc}\selectfont}{}

\usepackage{graphicx}
\usepackage{bm}
\usepackage{amsmath}
\usepackage{amssymb}
\usepackage{amsfonts}
\usepackage{float}
\usepackage{dsfont}  
\usepackage{slashed}  
\usepackage{booktabs}
\usepackage{multirow}
\usepackage{subfigure}
\usepackage[sort&compress]{natbib}
\usepackage{ulem}
\usepackage{empheq}
\usepackage{color}

\newcommand{\zmax}{z_{\rm max}}

\newcommand{\be}{\begin{equation}}  
\newcommand{\ee}{\end{equation}}  
\newcommand{\beq}{\begin{eqnarray}} 
\newcommand{\eeq}{\end{eqnarray}}

\newcommand{\bea}{\begin{eqnarray}}
\newcommand{\eea}{\end{eqnarray}}

\newcommand{\MSb}{{\overline{\rm MS}}}

\usepackage{xr-hyper}

\usepackage[svgnames,x11names,table]{xcolor}
\usepackage[colorlinks=true,linkcolor=MediumBlue,citecolor=MediumBlue,urlcolor=MediumBlue]{hyperref}
\usepackage{array}
\usepackage{hyperref}

\begin{document}

\title{Pion and Kaon PDFs from Lattice QCD  \\via Large Momentum Effective Theory \\ and Short-Distance Factorization}
\author{Joshua Miller}
\email{joshua.miller0007@temple.edu}
\affiliation{Department of Physics,  Temple University,  Philadelphia,  PA 19122 - 1801,  USA}

\author{Joseph Torsiello}
\email{joseph.torsiello@temple.edu}
\affiliation{Department of Physics,  Temple University,  Philadelphia,  PA 19122 - 1801,  USA}

\author{Isaac Anderson}
\affiliation{Department of Physics and Astronomy,  The University of Alabama in Huntsville,  Huntsville,  AL 35899,  USA}
\affiliation{Department of Physics,  Temple University,  Philadelphia,  PA 19122 - 1801,  USA}

\author{Krzysztof Cichy}
\affiliation{Faculty of Physics and Astronomy, Adam Mickiewicz University, ul.\ Uniwersytetu Pozna\'nskiego 2, 61-614 Pozna\'{n}, Poland\\[5ex]}

\author{Martha Constantinou}
\email{marthac@temple.edu}
\affiliation{Department of Physics,  Temple University,  Philadelphia,  PA 19122 - 1801,  USA} 

\author{Joseph Delmar}
\affiliation{Department of Physics,  Temple University,  Philadelphia,  PA 19122 - 1801,  USA}

\author{Sarah Lampreich}
\affiliation{Department of Physics,  Temple University,  Philadelphia,  PA 19122 - 1801,  USA}

\vspace*{0.25cm}
\begin{abstract}
In this work, we present a first-principles lattice-QCD calculation of the unpolarized quark PDF for the pion and the kaon. The lattice data rely on matrix elements calculated for boosted mesons coupled to non-local operators containing a Wilson line. The calculations on this lattice ensemble correspond to two degenerate light, a strange, and a charm quark ($N_f=2+1+1$), using maximally twisted mass fermions with a clover term. The lattice volume is $32^3\times 64$, with a lattice spacing of 0.0934 fm, and a pion mass of 260 MeV. Matrix elements are calculated for hadron boosts of $|P_3| = 0,~0.41,~0.83,~1.25,~1.66,$ and 2.07 GeV. To match lattice QCD results to their light-cone counterparts, we employ two complementary frameworks: the large-momentum effective theory (LaMET) and the short-distance factorization (SDF). Using these approaches in parallel, we also test the lattice data to identify methodology-driven systematics.
Results are presented for the standard quark PDFs, as well as the valence sector.
Beyond obtaining the PDFs, we also explore the possibility of extracting information on SU(3) flavor-symmetry-breaking effects. For LaMET, we also parametrize the momentum dependence to obtain the infinite-momentum PDFs. Since the present calculation is performed on a single ensemble at a pion mass of 260 MeV and fixed lattice spacing, the uncertainties reported are statistical only, and systematic uncertainties remain to be addressed in future multi-ensemble studies.
\end{abstract}


\maketitle

\section{Introduction}

Understanding the internal structure of hadrons is an important component of nuclear and particle physics. Among these are the pions and kaons, which are relevant to important questions, such as spontaneous and explicit chiral-symmetry breaking, as well as the emergence of mass.
As pseudo-Goldstone bosons of spontaneously broken chiral symmetry (pions) and their SU(3) partner (kaons), these mesons provide a unique platform for understanding how quarks of different masses generate the particle's properties. A comparison of the structure of these mesons with the nucleon is crucial to understanding the Standard Model mechanisms that produce hadron masses. In the chiral limit, the masses of pions and kaons vanish, while nucleons retain a mass on the order of 1 GeV. Consequently, the trace anomaly must vanish in the pion/kaon in the chiral limit but remains non-vanishing in the nucleon. Also, the pions, as the lightest hadronic states in the QCD spectrum, play a significant role in chiral symmetry breaking, particularly in nucleon-nucleon interactions. The presence of a pion cloud, for example, can explain the observed excess of $\overline{d}$ anti-quarks compared to $\overline{u}$ anti-quarks in the proton sea~\cite{Thomas:2000ny,Chen:2001et,Chen:2001pva,Salamu:2014pka}. Interesting conclusions may also be extracted by comparing the quark contributions in the pion and kaon structure in relation to SU(3) flavor symmetry-breaking effects~\cite{Hutauruk:2016sug}.

In order to get an understanding of the strongly interacting system of quarks and gluons, distribution functions, such as parton distribution functions (PDFs), play an important role. PDFs provide essential information about the momentum distribution of quarks and gluons within their parent hadron, serving as fundamental tools for understanding the latter’s internal structure and dynamics. While extensive efforts have been devoted to constraining the proton PDF, the global analysis of PDFs within the pions and kaons has shown increasing interest in recent years~\cite{Barry:2018ort,Barry:2021osv,JeffersonLabAngularMomentumJAM:2022aix,Pasquini:2023aaf,Barry:2025wjx}. 
Several approaches are available, such as Dyson-Schwinger equations (DSE)~\cite{Chen:2016sno,Shi:2018mcb,Bednar:2018mtf}, light-front quantization~\cite{Lan:2019rba,Liu:2023yuj}, nonrelativistic constituent quark model~\cite{Wu:2022iiu}, statistical model~\cite{Bourrely:2022mjf}, QCD instanton vacuum~\cite{Kock:2020frx}, chiral constituent quark model~\cite{Watanabe:2017pvl}, and Bethe-Salpeter equation for the model of Nambu and Jona-Lasinio~\cite{Hutauruk:2016sug}. 
Nonetheless, light-meson PDFs remain comparatively less constrained across the full $x$ range, making first-principles input especially valuable.
While some pion data exists from pion induced Drell-Yan (FNAL-E615 experiment)~\cite{Conway:1989fs} analyzed in Ref.~\cite{Aicher:2010cb}, knowledge of the kaon is even more limited with some early data on $\bar{u}_{K^-}(x)/\bar{u}_{\pi^-}(x)$~\cite{Badier:1983mj}.

PDFs are defined by correlation functions of non-local operators containing light-like separated parton fields. As such, their direct extraction from a Euclidean lattice is not permitted. The main avenue of obtaining information on PDFs has been from their Mellin moments, $\langle x^n \rangle$, which have been explored in lattice QCD~\cite{Brommel:2006zz,Brommel:2006ww,Bali:2013gya,Abdel-Rehim:2015owa,Alexandrou:2017blh,Oehm:2018jvm,Alexandrou:2020gxs,Alexandrou:2021mmi,Alexandrou:2021ztx,ExtendedTwistedMass:2024kjf}. However, obtaining the $x$ dependence from reconstructing the PDF from its Mellin moments remains challenging due to the difficulties calculating higher moments. In practice, the inversion from a limited set of noisy moments is ill-posed: higher moments suffer from rapidly deteriorating signal-to-noise ratios, operator mixing, and power divergences, as well as increasing sensitivity to discretization and renormalization systematics. An $x$-dependence reconstruction for the pion and kaon can be found in Ref.~\cite{Alexandrou:2021mmi}, demonstrating the difficulty of constraining $\langle x^3 \rangle$, as well as the parametrization of PDFs. A more recent development that utilizes gradient flow enables the determination of moments of parton distribution functions of any order~\cite{Shindler:2023xpd}. This approach has been implemented very recently for the pion up to $\langle x^5 \rangle$~\cite{Francis:2025rya}.

In recent years, the development of alternative approaches for accessing the $x$-dependence of PDFs has ignited a new research program in lattice QCD. 
In particular, some calculations of the $x$-dependence of the pion and kaon PDFs became available in the last few years using methods like the Large-Momentum Effective Theory (quasi-distributions)~\cite{Ji:2013dva,Ji:2020ect}, short-distance factorization (pseudo-distributions)~\cite{Radyushkin:2017cyf,Radyushkin:2019mye}, and current-current correlators~\cite{Ma:2014jla,Ma:2014jga,Ma:2017pxb}. 
In particular, Refs.~\cite{Zhang:2018nsy,Izubuchi:2019lyk,Gao:2022iex,Holligan:2024umc} present a calculation on the pion PDF, and Ref.~\cite{Lin:2020ssv} on both the pion and the kaon case; the aforementioned approaches use LaMET for the analysis. The pion PDF was studied within the pseudo-distributions method~\cite{Joo:2019bzr} and the current-current correlators~\cite{Sufian:2019bol,Sufian:2020vzb}. A combined study of both approaches for the pion can be found in Ref.~\cite{Gao:2020ito}.
Comprehensive reviews, covering various methods for obtaining $x$-dependent distribution functions, can be found in Refs.\cite{Cichy:2018mum,Ji:2020ect,Constantinou:2020pek,Cichy:2021lih,Cichy:2021ewm}. It is notable that the framework of Large-Momentum Effective Theory (LaMET) and the short-distance factorization (SDF) are the two most widely used methods. What sets these approaches apart from other methods is that they rely on the same lattice correlation functions and, thus, are complementary. Both are formulated in terms of equal-time matrix elements of non-local quark bilinears with straight Wilson lines; they differ in how the factorization is organized, either by boosting the hadron to large momentum or by exploiting the short-distance behavior of the correlator expressed in terms of the Ioffe-time.
Here, we explore both methods and assess systematic uncertainties in each analysis. We note that this is the first combined study for the kaon PDF case.

The remainder of the paper is organized as follows: Section~\ref{sec:setup} describes the lattice setup and the analysis of the matrix elements. We also include subsections on the SDF (Sec.~\ref{sub:pseudo}) and LaMET (Sec.~\ref{sub:quasi}) methods. 
The specific computational setup is given in Section~\ref{sec:comp}. Sec.~\ref{sec:results} presents the results of the analysis including the general matrix elements, the results for SDF (Sec.~\ref{sec:results_pseudo}) and LaMET (Sec.~\ref{sec:results_quasi}). Finally, we conclude in Section~\ref{sec:conclusions} with a summary and outlook.

\newpage
\section{Lattice Calculation}
\label{sec:setup}

\subsection{Theoretical Setup}

The pion and kaon matrix elements of the non-local vector operator are defined as
\begin{equation}
{\mathcal F}_M^f(z, p_i, p_f)  \equiv \langle M(p_f) | \bar{\Psi} (0) \gamma_\mu \, {\cal W}(0, z)  \Psi (z) | M(p_i)  \rangle \,
\label{eq:def}
\end{equation} 
where $M$ represents the particle under study. The momentum of the initial and final state are indicated by $p_i$ and $p_f$, respectively. In general, the momenta $p_i$ and $p_f$ can be written as a combination of averaged momentum, $P=(p_i+p_f)/2$, and a momentum transfer, $\Delta \equiv p_f-p_i$. The third component of $\vec{p}_i$ and $\vec{p}_f$ has a momentum boost, $P_3$, so that it is along the direction of the Wilson line. Here, we are interested in the forward limit of Eq.\eqref{eq:def}, that is $\vec{p}_f=\vec{p}_i\equiv\vec{P}=(0,0,P_3)$, which is connected to the pion/kaon PDF. Regarding the operator inserted in the matrix element, the quark fields are separated in the third spatial direction and are connected through a Wilson line, $\mathcal{W}$, to establish gauge invariance. Also, the Dirac structure used is $\gamma^0$. Without loss of generality, we take the direction of the Wilson line to be $\hat{z}$. It should be noted that this choice is inspired by the light-cone correlation function, which uses $\gamma^+$. Another option is via an operator that aligns with the direction of the boost; in this case, $\gamma_3$. However, lattice QCD formulations that break chiral symmetry give rise to mixing between $\gamma_3$ and $\hat{1}$ operators~\cite{Constantinou:2017sej,Alexandrou:2017huk}. The mixing manifests in the renormalization procedure and is a pure lattice artifact; thus, it vanishes in the continuum limit. For twisted mass fermions, which is the formulation used in this work, the mixing in the twisted basis is between the $\gamma_3$ and $\gamma_5$ operators. This is advantageous, as the forward limit of the matrix element with operator $\gamma_5$ vanishes, and thus, the effect of mixing is simply noise contamination. This has been shown explicitly for the nucleon case in Ref.~\cite{Alexandrou:2019lfo}. Nevertheless, here, we use exclusively the operator $\gamma_0$, which is multiplicatively renormalized. 

The lattice computation is performed in coordinate space, that is, one calculates the matrix element of Eq.~\eqref{eq:def} for different values of $z$ between $\pm L/(2a)$, where $L/a$ is the number of lattice points in the spatial directions for the ensemble chosen. 
For a particular quark flavor $f$, the ground-state matrix element, $F^f_M$, is related to the unpolarized light-cone PDF, $q_M^f$.   
In particular, the extraction of the latter is carried out using two distinct approaches that introduce intermediate quantities: the Ioffe-time pseudo-distribution (pseudo-ITD) and the quasi-parton distribution function (quasi-PDF). These two are computed independently, each following a separate analysis framework. The pseudo-ITD approach is based on short-distance factorization (SDF), whereas the quasi-PDF analysis relies on the large-momentum effective theory (LaMET). Both methods are detailed in Secs.~\ref{sub:pseudo} – \ref{sub:quasi}.
These approaches are complementary and, in principle, should yield the same light-cone distributions. Importantly, they both utilize the same matrix elements from Eq.~\eqref{eq:def}. In practice, the quasi-PDF analysis is performed at each momentum $P_3$ separately, and the results can be used to check convergence in the final estimates. In contrast, the pseudo-ITD approach uses data across multiple momenta simultaneously, resulting in a higher computational cost.

\subsubsection{quasi-distributions approach}
\label{sub:quasi}

The first approach we use to reconstruct the $x$-dependence of the spin-0 unpolarized PDFs, $q_M^f$, is the quasi-distribution method, which utilizes large-momentum effective theory (LaMET). 
In this approach, we calculate the matrix elements for each individual momentum boost, $P_3$, and multiple values of $z$. 
The LaMET formalism states that, as $|P_3|$ increases, we approach the light-cone counterparts faster once the matching is applied.
The separate analysis of each matrix element requires that the renormalization of the operator under study be considered.
Thus, we define the renormalized matrix element,
\begin{equation}
   F_M^{f,\,{\cal S}}(z, p_i, p_f,\mu) = Z^{\cal S}_{\gamma^{0}}(z,\mu)\, F^f_M(z, p_i, p_f)\,,
\end{equation}
where $Z^{\cal S}_{\gamma^0}$ is the multiplicative renormalization function determined at a general scale $\mu$ and a scheme ${\cal S}$.
The renormalization of the vector operator is obtained non-perturbatively in the RI scheme~\cite{Martinelli:1994ty}, using the momentum source method~\cite{Gockeler:1998ye,Alexandrou:2015sea}. 
This scheme imposes the condition
\bea
\label{eq:renorm}
Z^{-1}_{\gamma_0}(z,\mu_R) =\left( {\rm Tr} \left[S^{-1}\, S^{\rm tree} \right] \right)^{-1}  {\rm Tr} \left[ {\cal V}^{\gamma_0}_V(p,z) \left({\cal V}_V^{{\gamma_0}, \rm tree}(p,z)\right)^{-1}\right]_{p^2{=}\mu_R^2} \,,
\eea
where ${\cal V}(p,z)$ ($S{\equiv} S(p)$) is the amputated vertex function of the operator (fermion propagator), and ${\cal V}^{{\rm tree}}$ ($S^{{\rm tree}}(p)$) is its tree-level value. 
The first trace is the fermion field renormalization.
$Z_{\gamma_0}$ is calculated on various pion mass ensembles, and a chiral extrapolation is required to extract the mass-independent ${Z}^{\rm RI}_{\gamma_0,0}(z,\mu_R)$. We use the fit
\begin{equation}
\label{eq:Zchiral_fit}
Z^{\rm RI}_{\gamma_0}(z,\mu_R,m_\pi) = Z^{\rm RI}_{\gamma_0,0}(z,\mu_R) + m_\pi^2 \,Z^{\rm RI}_{\gamma_0,1}(z,\mu_R) \,,
\end{equation}
which eliminates the pion mass dependence in the renormalization functions~\cite{Alexandrou:2019lfo}.
The chirally extrapolated values are used to renormalize the matrix element prior to applying the matching kernel (see, Eqs.~\eqref{eq:matching} - \eqref{eq:kernel}).
Regardless of the scheme and scale of the quasi-distributions, the final light-cone PDFs are obtained in the ${\overline{\rm MS}}$ at a scale $\mu=2$ GeV.

Within the quasi-distributions approach, the renormalized quasi-PDFs, defined in the RI scheme at scale $\mu_R$, are transformed into their counterparts in momentum space. 
This is done within the Fourier transform, where, in continuum space, the coordinate variable is integrated within $\pm \infty$. 
 \begin{equation}
 F_M^{f,\rm RI}(x,P_3,\mu_R) = \int\frac{dz}{4\pi P_3} e^{-ixP_3z}F_M^{f,\rm RI}(z, P_3,\mu_R)\,. \label{e:convo}
   \end{equation} 
The discretization imposed by the lattice formulation and its periodicity restricts the coordinate space variable, $z$, up to half the lattice size. 
The integration of Eq.~\eqref{e:convo} results in a summation that is then truncated at a certain $z_{\rm max}$. 
Such a truncation poses an inverse problem in reconstructing the $x$-dependence, lacking a unique solution (for an in-depth discussion, see Ref.~\cite{Karpie:2019eiq}).
The simplest assumption is to set the coordinate-space quasi-distributions to zero beyond $z_{\rm max}$, but this introduces biases and systematic uncertainties in the final estimate.
The Backus–Gilbert (BG) estimator~\cite{BackusGilbert} trades bias for stability by minimizing the width of a resolution kernel under a normalization constraint. It should be noted that the reconstruction still has caveats, as it is limited by the number and noise of the coordinate-space data. Further implementation details can be found in Ref.~\cite{Alexandrou:2021bbo}.

The final component of extracting the light-cone distribution is the matching procedure, which eliminates the differences between quasi-PDFs and light-cone PDF in the ultraviolet regime, which can be written as
\begin{equation}
F_M^{f, {\rm RI}}(x,P_3,\mu_R) = \int_{-1}^1 \frac{dy}{|y|} \,C_{\gamma_0}^{{\rm RI}, \MSb}\left(\frac{x}{y},\frac{(\mu_R)^2}{(p^z_R)^2},\frac{yP_3}{\mu},\frac{yP_3}{p^z_R}\right) \,q^\MSb(y,\mu) \,\,+\, \mathcal{O}\left(\frac{m^2}{P_3^2},\frac{\Lambda_{\rm QCD}^2}{x^2P_{3_{\phantom{L}}}^2},\frac{\Lambda^2_{\rm QCD}}{(1-x)^2P_3^2}\right)\,. 
\label{eq:matching}
\end{equation}

\noindent We recall that $P_3$ is the momentum boost of the hadronic state, which is chosen in the $z$ direction. Also, $p^z_R$ is the $z$ component of $\mu_R$, where $p^2=\mu_R^2$, as shown in Eq.~\eqref{eq:renorm}. The kernel, $C_{\gamma_0}$, is calculated order by order in perturbation theory and, at the one-loop level, it reads 
\begin{equation}
C^{{\rm RI,}\MSb}_{\gamma_0}\left(x{,}\frac{(\mu_R)^2}{(p^z_R)^2}{,}\frac{y P_3}{\mu}{,}\frac{y P_3}{p^z_R}\right) = \delta(1{-}x) + f_1\left(\gamma_0,x,\dfrac{y P_3}{\mu}\right)_{+}{-}\left[ \left|\dfrac{y P_3}{p^z_R}\right|f_{2\slashed{p}}\left(\gamma_0, \dfrac{y P_3}{p^z_R}(x{-}1)+1,\frac{(\mu_R)^2}{(p^z_{R})^2}\right) \right]_{+} {+} \,\mathcal{O}(\alpha_s^2),
\label{eq:kernel}
\end{equation}
in which one utilizes the plus prescription in the quasi-distribution framework,
\begin{equation}
\int_{-\infty}^{\infty}dx~[h(x)]_+g(x) = \int_{-\infty}^{\infty}dx~h(x)[g(x) - g(1)] \,.
\end{equation}

\noindent The numerical expression is taken from Ref.~\cite{Liu:2019urm}, which is written for GPDs. In the PDFs limit  ($\xi \rightarrow 0$ in the expressions of Ref.~\cite{Liu:2019urm}), one finds the following for $f_1$ and $f_{2\slashed{p}}$ using the notation of Ref.~\cite{Liu:2019urm}
\begin{eqnarray*}
    \label{eq:kernel_f1}
f_1\left(\gamma_0,x,0,\frac{y P_3}{\mu}\right) = \frac{\alpha_sC_F}{2\pi} \left\{
        \begin{array}{ll}
           -\left(\frac{1+x^2}{1-x}\right)\ln{\left(\frac{x}{x-1}\right)-1}   & \quad  x < 0 \,, \\
           \frac{1+x^2}{1-x}\ln{\left(\frac{4x(1-x)(y P_3)^2}{\mu^2}\right)}-\frac{x(1+x)}{1-x} & \quad 0< x < 1 \,,\\
           \frac{1+x^2}{1-x}\ln{\left(\frac{x}{x-1}\right)+1}   & \quad x > 1\,.
        \end{array}
    \right.
\end{eqnarray*}
\begin{eqnarray*}
    \label{eq:kernel_f2}
f_{2\slashed{p}}\left(\gamma_0,x,r\right) = \frac{\alpha_sC_F}{2\pi} \left\{
    \begin{array}{ll}
        \frac{3-3r-2x}{2(r-1)(x-1)} + \frac{4rx-8x^2+8x^3}{(r-4x+4x^2)^2} + \frac{2-2r-rx+2x^2}{(r-1)^{3/2}(x-1)}\tan^{-1}\left({\frac{\sqrt{r-1}}{2x-1}}\right) & \quad x > 1 \,, \\
        \frac{3-3r-2x+4x^2}{2(r-1)(1-x)} + \frac{-2+2r+rx-2x^2}{(r-1)^{3/2}(1-x)}\tan^{-1}\left(\sqrt{r-1}\right) & \quad 0 < x < 1 \,, \\
        -\frac{3-3r-2x}{2(r-1)(x-1)}-\frac{4rx-8x^2+8x^3}{(r-4x+4x^2)^2}-\frac{2-2r-rx+2x^2}{(r-1)^{3/2}(x-1)}\tan^{-1}\left(\frac{\sqrt{r-1}}{2x-1}\right) & \quad x < 0 \,.
    \end{array}
    \right.
\end{eqnarray*}

We remind that, the above definition of $C_{\gamma_0}$ connects the quasi-distributions evaluated in the RI scheme at a scale $\mu_R$, which is introduced via the renormalization function, to the light-cone PDF in the standard $\overline{\rm MS}$ scheme, through $f_{2\slashed{p}}$, where $r = (\mu_R/p^z_R)^2$.
As mentioned above, the final renormalization scale for the light-cone PDFs is chosen to be $\mu$=2 GeV.

\subsubsection{pseudo-distributions approach}
\label{sub:pseudo}

An alternative approach to analyzing the lattice data and reconstructing the $x$-dependence of the pion and kaon PDFs is the Ioffe-time pseudo-distribution method.
It is useful to express the matrix elements of Eq.~\eqref{eq:def} as a function of the Lorentz invariant Ioffe time, $\nu=z\cdot P$, where $z_\mu=(0,0,0,z)$ and $P_\mu=(P_0,0,0,P_3)$. 
This enables one to combine different values of $z$ and $P$ in $F^f_M(\nu,z^2)$ because the latter contains physical information even at low values of $P_3$. 
While obtaining the matrix elements for multiple values of $P_3$ increases the computational cost, it offers a denser range of $\nu$, which is useful for reconstructing the $x$-dependence of the Ioffe-time distributions. The caution for this approach is that $z$ has to be small.
We construct the so-called reduced-ITD by taking the following double ratio of matrix elements,
\begin{equation}
    \mathcal{M}(\nu,z^2) = \frac{F^f_M(\nu,z^2)/F^f_M(\nu,0)}{F^f_M(0,z^2)/F^f_M(0,0)}\,.
    \label{eqn:DR}
\end{equation}
This combination serves as a renormalization scheme, which is effective for multiplicative renormalizable operators such as $\gamma_0$; it would have failed to completely capture all divergences for the case of the $\gamma_3$ operator. 
A benefit of the double ratio is the suppression of higher-twist contamination~\cite{Orginos:2017kos}. 

Once $\mathcal{M}(\nu,z^2)$ is constructed for multiple $P_3$ and $z$ values, it is useful to fit its $\nu$ dependence. 
At a fixed $z^2$ and following the $\nu$ symmetries of  $\mathcal{M}(\nu,z^2)$, the real and imaginary parts of the double ratio are fitted to a power expansion given by
\begin{eqnarray}
\label{eq:fit_re}
{\rm Re}[\mathcal{M}(\nu,z^2)] = 1 +  c_2{\footnotesize{(z^2)}} \,\nu^2 + c_4(z^2) \,\nu^4 + c_6(z^2) \,\nu^6\,+ ...\,, \\[2.5ex] 
\label{eq:fit_im}
{\rm Im}[\mathcal{M}(\nu,z^2)] = c_1(z^2) \,\nu^{\phantom{1}} + c_3(z^2) \,\nu^3 + c_5(z^2) \,\nu^5\,+ ...\,,
\end{eqnarray}
where $c_i$ are fit parameters. As can be seen, the real (imaginary) part is symmetric (antisymmetric), so the fit contains only even (odd) powers of $\nu$.

The fitted function $\mathcal{M}(\nu,z^2)$ is then evolved to a common scale, $\mu$; here, we chose $\mu=2$ GeV as the final scale. The evolved function, $\mathcal{M}'(\nu,z^2,\mu^2)$ is then matched to its light-cone counterpart, $\mathcal{Q}(\nu,\mu^2)$, through a short-distance factorization~\cite{Radyushkin:2017lvu}. 
The above procedure is implemented numerically, and to one-loop level, it can be written as
\begin{equation}
    \mathcal{M}(\nu,z^2) = \mathcal{Q}(\nu,\mu^2) + \frac{\alpha_sC_f}{2\pi} \int_0^1 du~\left[\mathrm{ln}\left(z^2\mu^2 \frac{e^{2\gamma_E + 1}}{4} \right)B(u) + L(u)\right]\mathcal{Q}(u\nu,\mu^2)\,.
    \label{eqn:EvolveMatch}
\end{equation}
The functions $B(u)$ and $L(u)$ represent the evolution and matching components, respectively. $B(u)$ is given by
\begin{equation}
    B(u) = \left[\frac{1+u^2}{u-1} \right]_+\,,
    \label{eqn:Evolver}
\end{equation}
and the matching function $L(u)$, which is given by
\begin{equation}
    L(u) = \left[4\frac{\mathrm{ln}(1-u)}{u-1}-2(u-1) \right]_+\,.
    \label{eqn:Matcher}
\end{equation}
The final light-cone ITD are given in $\overline{\mathrm{MS}}$ scheme at 2 GeV. 
Eqs.~\eqref{eqn:Evolver} - \eqref{eqn:Matcher} utilize the plus-prescription, that is
\begin{equation}
    \int_0^1 du~[f(u)]_+\mathcal{Q}(u\nu) = \int_0^1 du~f(u)[\mathcal{Q}(u\nu)-\mathcal{Q}(\nu)]\,.
\end{equation}
Upon inverting Eq.~\eqref{eqn:EvolveMatch} we can separate the evolved pseudo-ITD ($\mathcal{M}'$) and matched light-cone ITD ($\mathcal{Q}$) by
\begin{equation}
    \mathcal{M}'(\nu,z^2,\mu^2) = \mathcal{M}(\nu,z^2) - \frac{\alpha_sC_F}{2\pi}\int_0^1 du~ \mathrm{ln}\left(z^2\mu^2 \frac{e^{2\gamma_E + 1}}{4} \right)B(u)\mathcal{M}(u\nu,z^2)\,,
    \label{eqn:Evolved_ITD}
\end{equation}
and
\begin{equation}
    \mathcal{Q}(\nu,\mu^2) = \mathcal{M}'(\nu,z^2,\mu^2) - \frac{\alpha_sC_F}{2\pi} \int_0^1 du~ L(u)\mathcal{M}(u\nu,z^2)\,.
    \label{eqn:Matched_ITD}
\end{equation}
Eq. (\ref{eqn:Evolved_ITD}) is dependent upon the Ioffe time $\nu$, the common scale $\mu$, and the initial scale $z$. As stated earlier, fitting the $\nu$ dependence is useful to access $\mathcal{M}(u\nu,z^2)$ entering the matching.
Since $\mathcal{Q}(\nu,\mu^2)$ is constructed from multiple $P_3$ and $z$ values, some correspond to the same value of $\nu$.
Thus, the matched-ITD is averaged for common values of $\nu$.
We also check consistency between different pairs of $(P_3,z)$ that have the same $\nu$ value.

$\mathcal{Q}(\nu,\mu^2)$ is related to the PDF, $q(x,\mu^2)$, via a Fourier transform in Ioffe time
\begin{equation}
\label{eq:PDF2ITD}
\mathcal{Q}(\nu,\mu^2) =\int_{-1}^1 dx \, e^{i\nu x} q(x,\mu^2)\,.
\end{equation}
The crossing symmetries for the vector case depict that the antiquark distribution for positive $x$ is $\bar{q}(x)=-q(-x)$, which allows one to associate them to the real and imaginary parts of $\mathcal{Q}$, via
\begin{eqnarray}
\label{eq:ReQ}
{\rm Re}\,\mathcal{Q}(\nu,\mu^2) =\int_0^1 dx \cos(\nu x) \big(q(x,\mu^2)-\bar{q}(x,\mu^2)\big) =  \int_0^1 dx \cos(\nu x) q_v(x,\mu^2),
\end{eqnarray}
\begin{eqnarray}
\label{eq:ImQ}
{\rm Im}\,\mathcal{Q}(\nu,\mu^2) = \int_0^1 dx \sin(\nu x) \big(q(x,\mu^2) + \bar{q}(x,\mu^2)\big) =  \int_0^1 dx \sin(\nu x) q_{v2s}(x,\mu^2)\,.
\end{eqnarray}
In particular, for the pion, the valence distribution, $q^{\pi^u}_v(x)=q^{\pi^u}(x)-q^{\pi^{\bar{u}}}(x)$ with support $x \in [0,1]$, becomes
\begin{equation}
\label{eq:valence_pion}
q^{\pi^u}_v(x)=q^{\pi^u}(x)+q^{\pi^u}(-x)\,,\quad x \in [0,1]\,,
\end{equation} 
and is related to the real part of the ITDs.
Similarly, the combination $q^{\pi^u}_{v2s}\equiv q_v(x) + 2q^{\pi^{\bar{u}}}(x)$
becomes 
\begin{equation}
\label{eq:v2s_pion}
    q^{\pi^u}_{v2s}(x)=q^{\pi^u}(x)-q^{\pi^u}(-x)\,,\quad x \in [0,1]\,,
\end{equation}
and is related to the imaginary part of the ITDs. 
For the kaon, we use the following combinations for each flavor, $f=\{u,\,s\}$, that is 
\begin{eqnarray}
\label{eq:valence_kaon}
q^{K^f}_{v}(x)=q^{K^f}(x)-q^{K^{\bar{f}}}(x) = q^{K^f}(x)+q^{K^f}(-x)\,,\quad x \in [0,1]\,,\\[0.5ex]
\label{eq:v2s_kaon}
q^{K^f}_{v2s}(x)=q^{K^f}(x)+q^{K^{\bar{f}}}(x) = q^{K^f}(x)-q^{K^f}(-x)\,,\quad x \in [0,1]\,.
\end{eqnarray}

The left-hand side of Eqs.~\eqref{eq:PDF2ITD} - \eqref{eq:ImQ} is the reduced-ITDs computed on the lattice, and an inversion is needed to extract the PDF. 
This poses the so-called inverse problem as the inverse equations are ill-defined~\cite{Karpie:2018zaz}.
In particular, the lattice data are restricted to finite values of $z$ and $P_3$ that only cover a finite range of Ioffe time, while the integral over $\nu$ assumes continuous values up to infinity.
To reconstruct the pion and kaon PDFs, we will follow a reconstruction technique in which one assumes a functional form for a fitting ansatz for the light-cone PDF. 
This is motivated by the phenomenological fits of experimental data sets to extract the PDFs.
Here, we will use a functional form that has the expected low- and high-$x$ behavior in the range $x\in(0,1)$, that is
\begin{equation}
\label{eq:ansatz}
q(x) = N x^a (1-x)^b\,,
\end{equation}
where the exponents $a,\,b$ are fitting parameters.
$N$ is a normalization constant, which for $q_v$ ensures charge conservation. 
Thus, it is fixed to $1/B(a+1,b+1)$, where $B(x,y)$ is the Euler beta function related to the gamma function via $B(x,y)=\Gamma(x)\Gamma(y)/\Gamma(x+y)$. 
For $q_{v2s}$, $N$ is kept as a fit parameter, similar to $a$ and $b$.
The fits on the lattice data are performed by minimizing the $\chi^2$ function defined as
\begin{equation}
\label{eq:chi2}
\chi^2=\sum_{\nu=0}^{\nu_{\rm max}}\frac{[\mathcal{Q}(\nu,\mu^2)-\mathcal{Q}_f(\nu,\mu^2)]^2}{\sigma_\mathcal{Q}^2(\nu,\mu^2)}\,,
\end{equation}
where $\sigma_\mathcal{Q}^2(\nu,\mu^2)$ is the statistical uncertainty of $\mathcal{Q}(\nu,\mu^2)$.
$\mathcal{Q}_f(\nu,\mu^2)$ is given by the cosine and the sine Fourier transform of the fitting ansatz of Eq.\eqref{eq:ansatz} for fits of the real and the imaginary part of ITDs, respectively.
Obtaining a continuous fit function is advantageous, as the Fourier transform is not subject to any inverse problem. Nevertheless, there are systematic uncertainties associated with the choice of the fit function.
It should be noted that the fits depend on the maximum Ioffe time, $\nu_{\rm max}$, and we will investigate different choices of this parameter and the sensitivity of the final PDF results to this choice.
We remark that the $x$-dependence reconstruction used in the SDF analysis relies on a physics-motivated ansatz, while in LaMET we use BG reconstructions. This different approach in the two methodologies is guided by the fact that using BG in the SDF method requires a large value of $z_{max}$, much larger than the validity of the SDF formalism. Thus, attempts of using BG have not been found to be beneficial (see, e.g., Fig. 12 of Ref.~\cite{Bhat:2020ktg}).

\subsection{Computational Setup}
\label{sec:comp}

In this work, we focus on the connected contributions to the pion and kaon unpolarized PDFs, which have been calculated on the \texttt{cA211.30.32} ensemble of twisted-mass clover fermions and Iwasaki improved gluons. 
The ensemble has degenerate light quarks, and strange and charm quarks in the sea ($N_f=2+1+1$). 
The gauge configurations have been produced by the Extended Twisted Mass Collaboration (ETMC). More details on the ensemble and the quark-mass tuning can be found in Refs.~\cite{Alexandrou:2018egz,ExtendedTwistedMass:2021gbo}. 
In Table~\ref{tab:params}, we summarize the main parameters of the ensemble. 
\begin{table}[h!]
\centering
\renewcommand{\arraystretch}{1.2}
\renewcommand{\tabcolsep}{6pt}
\begin{tabular}{| l| c | c | c | c | c  | c | c |}
    \hline
    \multicolumn{8}{|c|}{Parameters} \\
    \hline
    Ensemble   & $\beta$ & $a$ [fm] & volume $L^3\times T$ & $N_f$ & $m_\pi$ [MeV] &
    $L m_\pi$ & $L$ [fm]\\
    \hline
    cA211.30.32 & 1.726 & 0.0934  & $32^3\times 64$  & 2+1+1 & 260
    & 4 & 3.0 \\
    \hline
\end{tabular}
\caption{\small{Parameters of the ensemble used in this work.}}
\label{tab:params}
\end{table}

The components needed to extract the matrix element of Eq.~\eqref{eq:def} are the two-point and three-point correlation functions. The former is defined as
\begin{equation}
C^{\rm 2pt}_M(\vec{p};t) = \sum_{\vec{x}} J_M(t,\vec{x}) J^\dag_M(0,0) e^{-i\vec{p}\cdot \vec{x}}\,,
\end{equation} 
The three-point function for quark flavor $f$ and particle $M$, ${\mathcal C}^f_M$, reads 
\begin{equation}
{ \mathcal C}^f_M(\vec{p};z,t_s,t) = \sum_{\vec{x}_s,\vec{x}} J_M(t_s,\vec{x}_s){\mathcal O}^f_{{\gamma_0}}(t,\vec{x};z)J^\dag_M(0,0) e^{-i\vec{p}\cdot\vec{x}_s}\,,
\end{equation} 
The matrix element that will be extracted from Eq.~\eqref{eq:def} corresponds to the special case of the forward limit, that is, the momentum of the initial and final states is the same ($p$). Also, $t$, and $t_s$ are the insertion and sink Euclidean times, respectively. The corresponding spatial coordinates of the current insertion and sink are $\vec{x}$, $\vec{x}_s$. 
Without loss of generality, we have taken the source to be at the point $(0,\vec{0})$, and, therefore, $t_s$ is the source-sink separation.
In the calculation of three-point functions, we use the fixed-sink sequential-inversion approach, which allows us to couple various operators at minimal additional cost.
We employ momentum smearing~\cite{Bali:2016lva}, which improves the overlap with the ground state of the pion/kaon and also reduces gauge noise. 
It has been demonstrated that the noise in the matrix elements of non-local operators decreases significantly~\cite{Alexandrou:2016jqi}.

To further decrease gauge noise, we apply five steps of stout smearing~\cite{Morningstar:2003gk} with smearing parameter $\rho=0.15$. 
We have previously demonstrated that renormalized matrix elements are independent of the number of stout smearing steps. 
As discussed in Sec.~\ref{sub:pseudo} and Sec.~\ref{sub:quasi}, the matrix elements require a momentum boost indicated by $P_3$. 
In particular, the pseudo-distributions approach utilizes several values of $P_3$ within the same analysis, and in this work, we implement $|P_3|=0,~0.41,~0.83,~1.25,~1.66$, and 2.07 GeV; the quasi-distributions method may also be applied to multiple values of $P_3$ to check for convergence in the final results as its value increases.
The upper value of $P_3$ employed is restricted from the increase of gauge noise.
To handle the increased noise-to-signal ratio as $P_3$ increases, for the pion, we obtain the momenta with $P_3<1$ GeV at $t_s=12a=1.12\,{\rm fm}$, while the higher momenta correspond to $t_s=10a\sim 0.93$ fm. For the kaon, momenta up to $P_3=1.25$ GeV are at $t_s=12a$.
To further reduce noise in the matrix elements, we increase the number of source positions as needed with increasing $P_3$. 
In addition, for the case of the pion, which is a lighter particle and susceptible to increased noise, we further extend the statistics for $|P_3|=1.25,\,1.66,\,2.07$ GeV compared to the kaon case.
In particular, at the highest momentum, we reach up to 400 and 300 source positions for the pion and kaon, respectively.
The statistics for each momentum is listed in Table \ref{tab:stats}.

\begin{table}[h!]
\begin{center}
\renewcommand{\arraystretch}{1.9}
\begin{tabular}{l|cccccc}
\hline
$P_3$ [GeV] & $\quad$0$\quad$ & $\quad$$\pm$0.41$\quad$ & $\quad$$\pm$0.83$\quad$ & $\quad$$\pm$1.25$\quad$ & $\quad$$\pm$1.66$\quad$ & $\quad$$\pm$2.07$\quad$  \\ \hline
$t_s/a$ & 12 & 12 & 12 & 10$^\star$, 12$^\dagger$ & 10 & 10 \\\hline
$N_{\rm confs}$ & 1,198 & 1,198 & 1,198 & 1,198 & 1,198 & 1,198 \\\hline
$N^{\pi}_{\rm src}$ & 1 &  8 &  8 &  56 &  84 & 400  \\\hline
$N^{K}_{\rm src}$ & 1 &  8 &  8 &  8 &  48 & 300  \\\hline
$N^{\pi}_{\rm tot}$ & 1,198 & 9,584 & 9,584 & 67,088  & 100,632  & 479,200 \\\hline
$N^{K}_{\rm tot}$ & 1,198 & 9,584 & 9,584 & 9,584  & 57,504  & 359,400 \\
\hline
\end{tabular}
\caption{\small Statistics for the pion and kaon matrix elements at different values of momentum $P_3$. $N_{\rm confs}$, $N^{\pi}_{\rm src}$, $N^K_{\rm src}$, $N^{\pi}_{\rm total}$ \\[0.5ex] and $N^K_{\rm total}$ are the number of configurations, source positions per configuration, and total statistics, respectively. \\[1ex] $\star$: pion; $\dagger$: kaon.}
\label{tab:stats}
\end{center}
\end{table}

Calculating the matrix elements for multiple values of $P_3$ allows us to implement both the quasi-distribution method and the pseudo-distribution approach. 
We remind the reader that the quasi-PDF method analyzes a single value of the momentum boost, whereas the pseudo-ITD method combines multiple values of $P_3$. 

\medskip
For the interpolating fields, $J_M$, of the mesons under study, we use
\begin{equation}
J_{\pi^+} = \overline{d}\gamma_5 u\,, \quad
J_{K^+} = \overline{s}\gamma_5 u\,,
\end{equation}
which satisfy the quantum numbers of $\pi^+$ and $K^+$.
The pseudoscalar structure of the pion and the $\gamma_5$-hermiticity relation of the twisted mass quark propagators
\begin{equation}
    G_{u}(x,x')=\gamma_5 G_{d}^\dagger(x',x)\gamma_5\,,
\end{equation}
have a useful consequence, which is that we only need to calculate the up-quark contribution to the pion three-point functions. 
The pion and kaon interpolating fields are smeared using Gaussian smearing at both the source and sink. 
The smearing parameters are tuned separately for the pion and kaon. 
We use the same value of $\alpha_G$ but varying the number of smear iterations $N_G$ for the light and strange quarks. 
An optimal choice for $N_G$ is based on the criterion that the root mean squared radius of the smeared source reproduces the experimental radius of the pion~\cite{AMENDOLIA1984116} for the light quarks and the experimental radius of the kaon~\cite{AMENDOLIA1986435} for the strange quarks. 
In this work we obtain $(\alpha_{G},N_{G})=(0.2,50)$ for the light quarks and $(\alpha_{G},N_{G})=(0.2,40)$ for the strange quark.  
APE smearing is applied on the gauge links that enter the Gaussian smearing with parameters ($\alpha_{APE},N_{APE})=(0.5,50)$.
In addition to the standard Gaussian smearing we apply momentum smearing. 

The matrix element of Eq.~\eqref{eq:def} is related to the two- and three-point correlation functions via an appropriate ratio that cancels unknown overlap factors. 
In the forward limit, the ratio takes the simple form

\begin{equation}
R^f_M(z,p;t_s,t) = \frac{{\mathcal C}^f_M(z,p;t_s,t)}{C^{2pt}_{M}(p;t_s)} \,.
    \label{eq:Ratio}
\end{equation}

\noindent For the two-point function, we exploit the symmetry properties of the spin-0 particles to symmetrize the correlator corresponding to $t$ and $T-t$, for  $ t \in [0,T/2]$, i.e., the value at $t$ has been averaged with their corresponding value at $T-t$; $T$ is the temporal extent of the lattice. As a test, we also implemented the two-state fit given by

\begin{equation}
    C^{\rm 2pt}_{M,\,\rm fit}(t) = c_0 \left(e^{-E_0 t}+e^{-E_0 (T-t)}\right) + c_1 \left(e^{-E_1 t}+e^{-E_1 (T-t)}\right)\,.
    \label{eq:tsf_twop}
\end{equation}

\noindent The amplitudes $c_0$ and $c_1$, as well as the ground and first excited state energies $E_0$, $E_1$ are fit parameters.
Using this parametrization and following Refs.~\cite{Alexandrou:2020gxs,Alexandrou:2021mmi,Alexandrou:2021ztx}, one may substitute the two-point function in the ratio with the fitted ground state, and Eq.~\eqref{eq:Ratio} becomes, up to contamination by excited states

\begin{equation}
R^f_M(z,p;t_s,t) = \frac{{\mathcal C}^f_M(z,p;t_s,t)}{c_0 \,e^{-E_0 t_s}} \,.
    \label{eq:Ratio2}
\end{equation}
\noindent We find that Eq.~\eqref{eq:Ratio2} produces compatible results as the two-state fit of Eq.~\eqref{eq:Ratio}. However, for large values of $P_3$, the fit is noisier. Thus, we continue with using the 2-point function of Eq.~\eqref{eq:Ratio}.

\smallskip
To identify the ground state, $F^f_M$, we seek a region of $t$ away from the source and sink time separations, we identify a plateau, indicating that excited states are suppressed

\begin{equation}
    R^f_M(z,p;t_s,t) \xrightarrow[t \gg a]{t_s-t \gg a} F^f_M(z,p;t_s)\,.
    \label{eq:Ratio3}
\end{equation}

\newpage
\section{Lattice Results}
\label{sec:results}

\begin{figure}[h!]
    \includegraphics[scale=0.5]{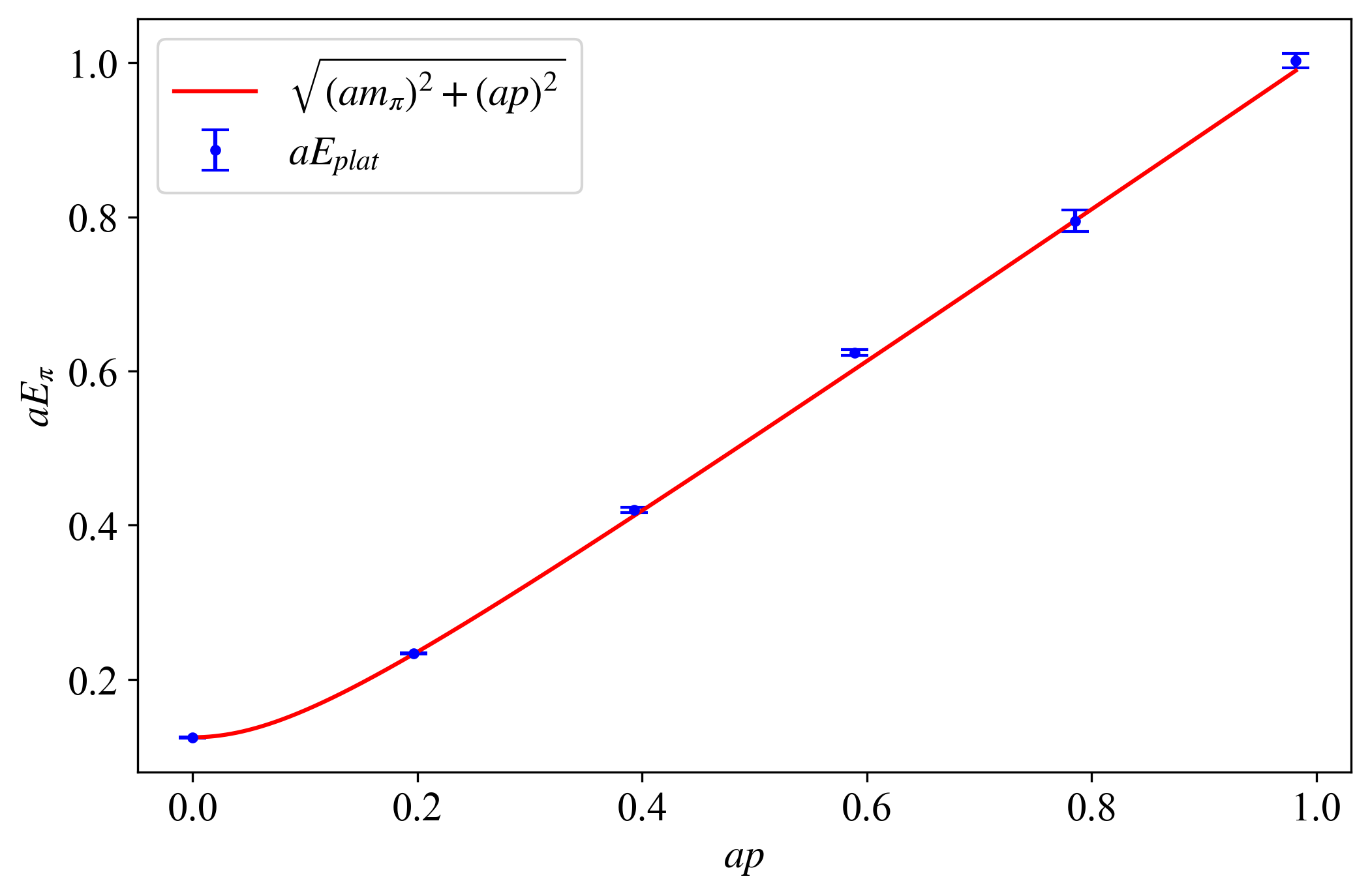}
    \includegraphics[scale=0.5]{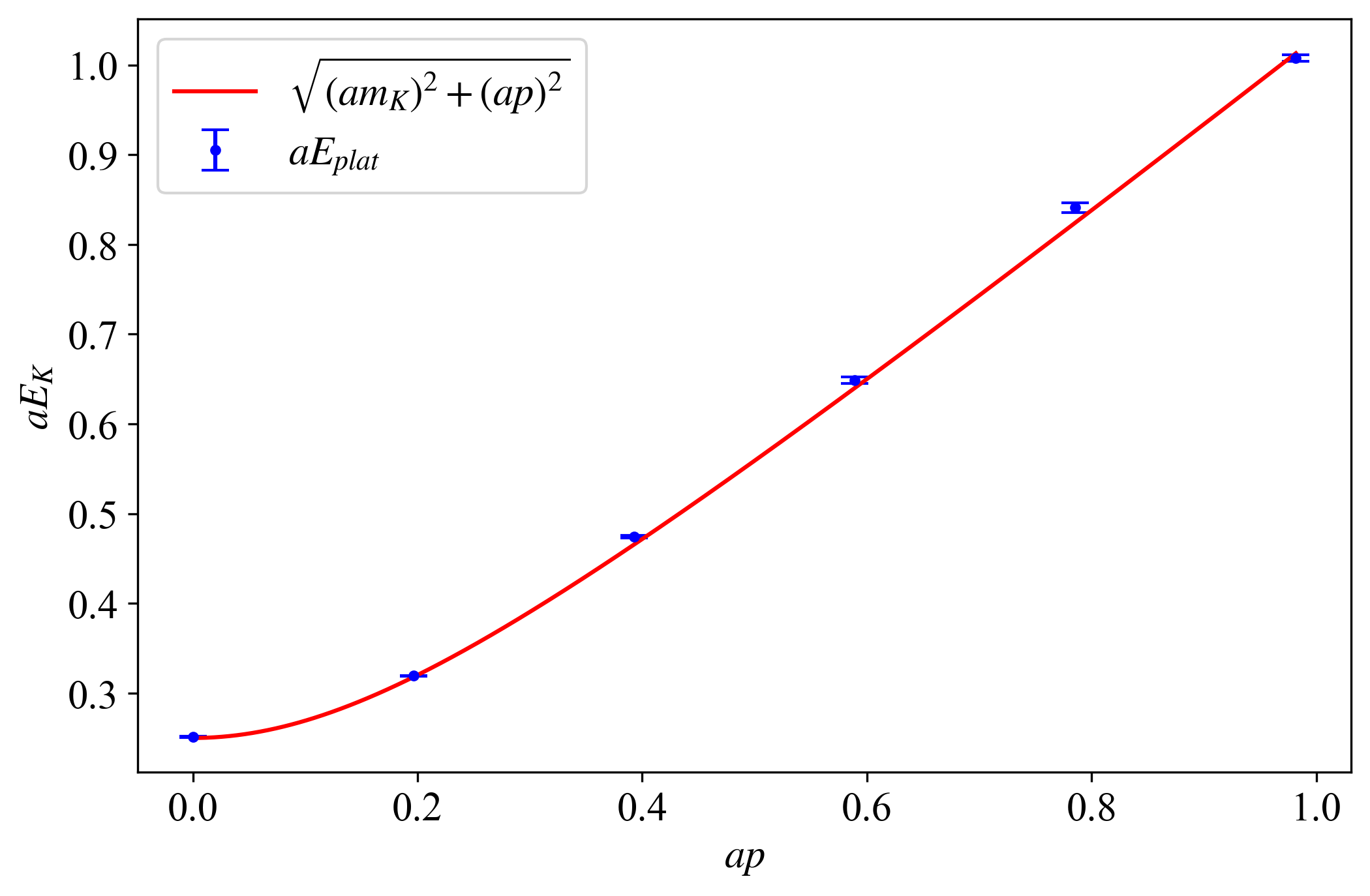}
    \vspace*{-0.4cm}
    \caption{\small{The ground-state energies for the pion (left) and kaon (right) at various values of the momentum boost, $a p$. The red curve corresponds to the dispersion relation, and the blue points correspond to the energy obtained from a plateau fit on the lattice data.}}
    \label{fig:dispersion_relation}
\end{figure}

Since our calculations involve matrix elements with momentum-boosted external states, we need to assess potential cutoff effects. To this end, we examine whether the extracted ground-state energies follow the continuum dispersion relation. This study is illustrated in Fig.~\ref{fig:dispersion_relation} for both the pion and the kaon for momenta $\vec{p}= \frac{2\pi}{L} (0,0,n)$ with $n$ in the range of $[0-5]$ (up to 2 GeV), where the continuum dispersion relation is shown along with the estimates from our calculation.
The results demonstrate excellent agreement with the dispersion relation for almost all the momenta used in this work. We also demonstrate the effect of increasing statistics by using multiple source positions per configuration (see Table~\ref{tab:stats}). We provide the values of the energies and the relative errors in Table~\ref{tab:energy}. As can be seen, the errors in the energy remains well controlled, even at the highest momentum values. Having performed this test, we now proceed with the next steps of the analysis.
\begin{table}[h!]
\begin{center}
\renewcommand{\arraystretch}{1.9}
\begin{tabular}{l|cccccc}
\hline
$P_3$ [GeV] & $\quad$0$\quad$ & $\quad$$\pm$0.41$\quad$ & $\quad$$\pm$0.83$\quad$ & $\quad$$\pm$1.25$\quad$ & $\quad$$\pm$1.66$\quad$ & $\quad$$\pm$2.07$\quad$  \\ \hline
$aE_\pi$ & 0.125 & 0.234 & 0.420 & 0.624 & 0.795 & 1.003 \\\hline
$aE_K$ & 0.251 & 0.319 & 0.474 & 0.649 & 0.841 & 1.008 \\\hline
$dE_\pi/E_\pi$ & 0.006 & 0.003 & 0.008 & 0.006 & 0.018 & 0.009  \\\hline
$dE_K/E_K$ & 0.003 & 0.001 & 0.003 & 0.005 & 0.007 & 0.003 \\  
\hline
\end{tabular}
\caption{\small Energy values and relative error
for each momentum boost $P_3$.}
\label{tab:energy}
\end{center}
\end{table}

\begin{figure}[h!] \hspace*{-0.35cm}
    \includegraphics[scale=0.175
]{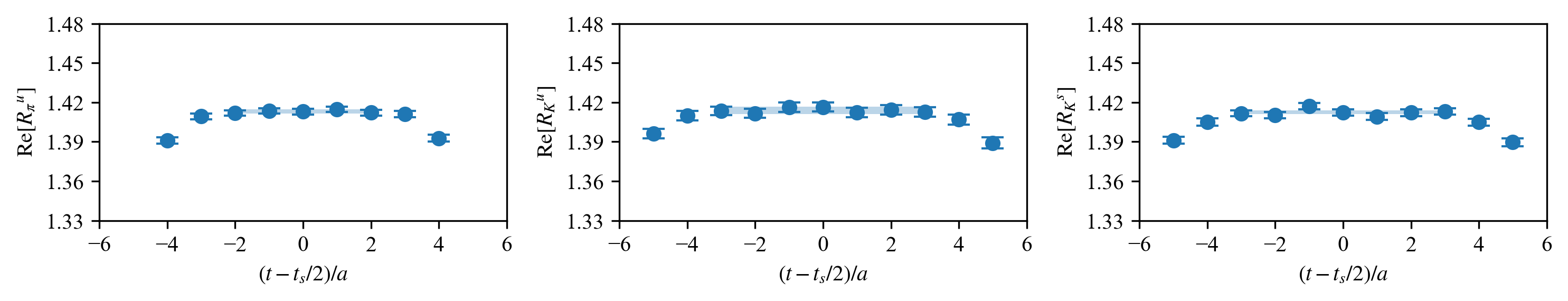}
    \vspace*{-0.4cm}
\caption{\small{The ratio of Eq.~\eqref{eq:Ratio} for $R_\pi^u$ (left), $R_K^u$ (center) and $R_K^s$ (right) for $z=0$ (no imaginary part) and $P_3=1.25$ GeV.}}
\label{fig:plateau_z0}
\end{figure}

\begin{figure}[h!]\hspace*{-0.35cm}
\includegraphics[scale=0.175]{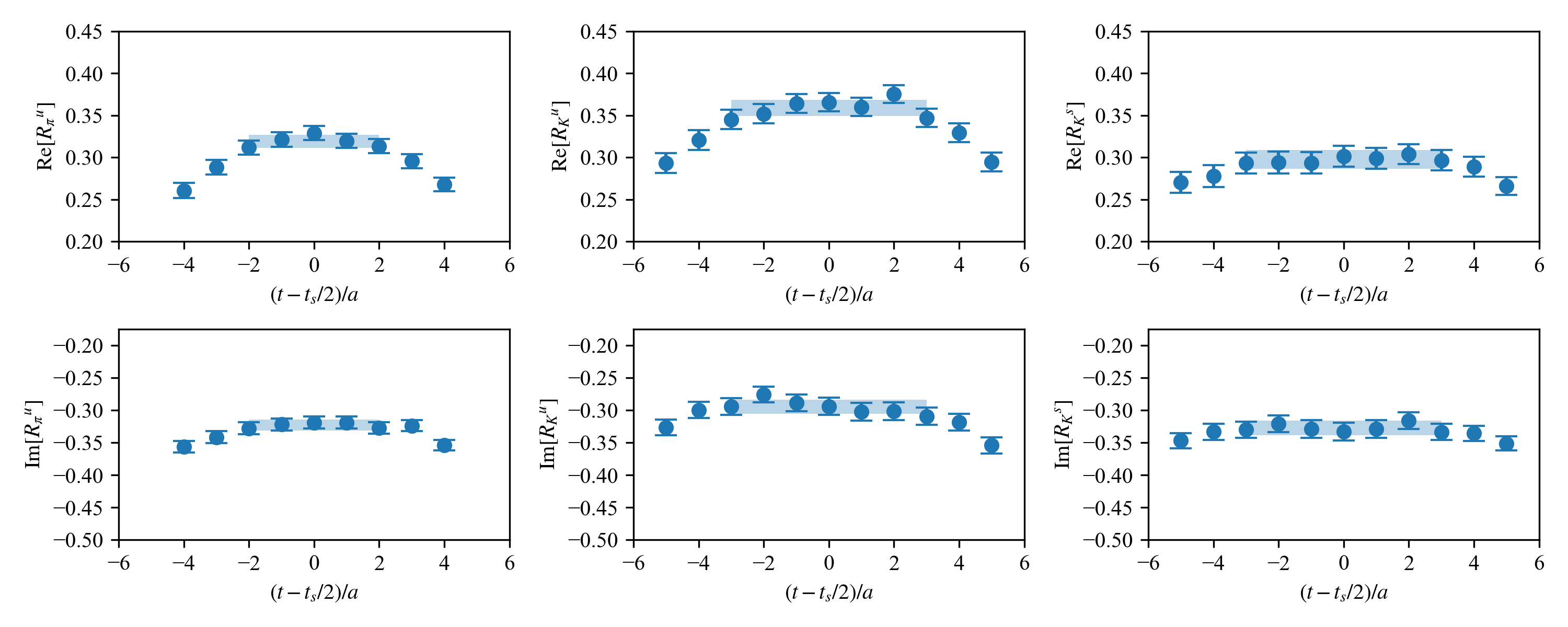}
    \vspace*{-0.4cm}
\caption{\small{The ratio of Eq.~\eqref{eq:Ratio} for $R_\pi^u$ (left), $R_K^u$ (center) and $R_K^s$ (right) for $z=5$ and $P_3=1.25$ GeV. 
The top (bottom) panel corresponds to the real (imaginary) part. }}
\label{fig:plateau_z5}
\end{figure}

The major component of the analysis of raw data is the extraction of the ground state contribution from the matrix element in Eq.~\eqref{eq:def}. 
In this work, we define the ground state via a plateau fit of the ratio in Eq.~\eqref{eq:Ratio3}, assuming single-state dominance for insertion timeslices away from the source and the sink.
For presentation purposes, we use $P=1.25$ GeV as a study case and display the ratio and plateau fit for $z=0$ (Fig.~\ref{fig:plateau_z0}) and $z=5$ (Fig.~\ref{fig:plateau_z5}). 
We remind the reader that $R_M^f$ is real at $z=0$. 
For a better comparison of the statistical uncertainties between the matrix elements of different particles/flavors, $R_\pi^u$, $R_K^u$, and $R_K^s$, we keep the same range for the $y$ axis. 
As can be seen in the above-mentioned figures, both the pion and kaon data have similar errors despite the lighter mass of the former. 
This was achieved by increasing the statistics for the pion at a higher number than the kaon data. 
Comparison of the up and strange components of the kaon shows a similar signal, with the ratio for the up quark having a slightly higher noise-to-signal ratio. 
More comparisons can be found below and in Table~\ref{tab:error}. Based on our analysis, we select $t \in [3a - 9a]$ when $t_s=12a$, and $t \in [3a - 7a]$ for $t_s=10a$, for the fit of the insertion time.
We note that the plateau fit is labeled by $F_M^f$ (Eq.~\eqref{eq:Ratio3}), and is applied separately in the real and imaginary parts and at each value of $z$. 

\begin{figure}[h!]
\hspace*{-0.2cm}    \includegraphics[scale=0.36]{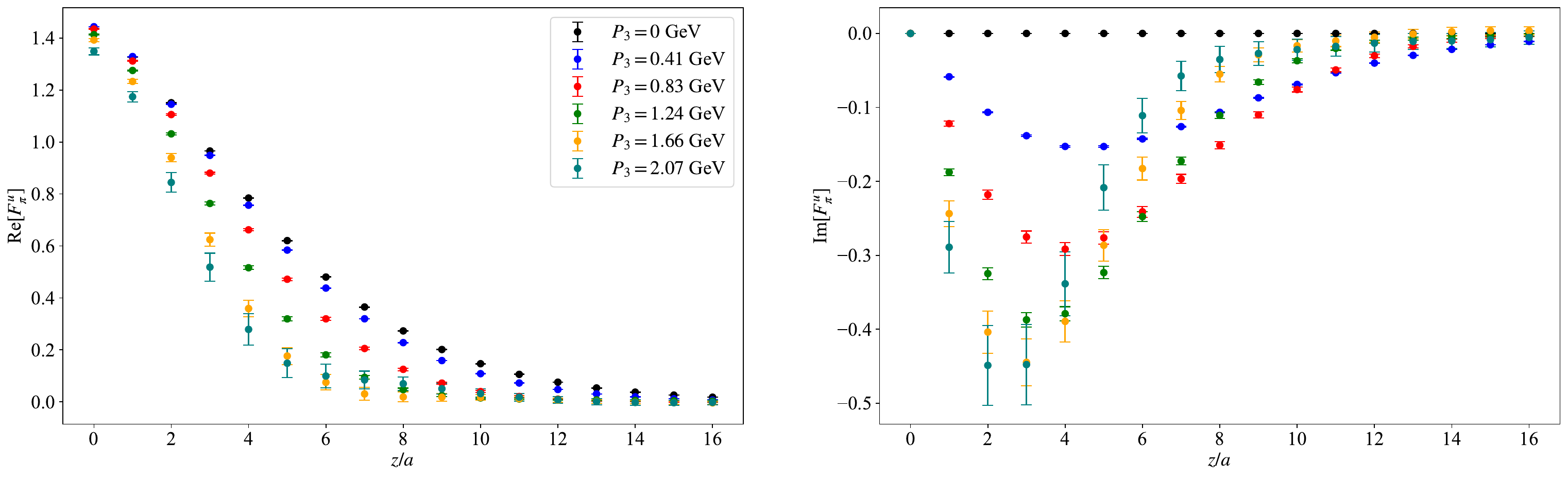}
    \caption{\small{Pion bare matrix element, $F_\pi^u$ for momentum boost $|P_3|=0,~0.41,~0.83,~1.25,~1.66,~2.07$ GeV. The real and imaginary components are shown in the left and right panels, respectively.\\}}
    \label{fig:g0_pion}
\end{figure}

\begin{figure}[h!]
\hspace*{-0.2cm}   \includegraphics[scale=0.36]{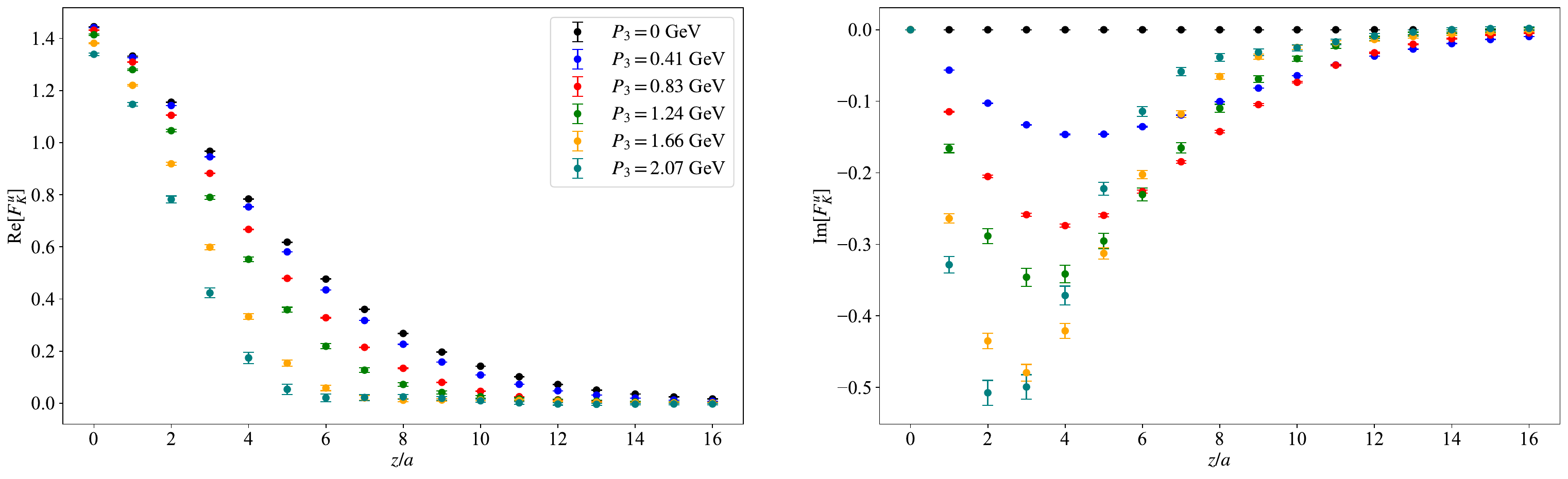}
    \caption{\small{Bare matrix elements for the kaon up flavor for momentum boost $|P_3|=0,~0.41,~0.83,~1.25,~1.66,~2.07$ GeV. The real and imaginary parts are shown in the left and and plots, respectively.\\}}
    \label{fig:g0_kaon_u}
\end{figure}

\begin{figure}[h!]
\hspace*{-0.2cm}    \includegraphics[scale=0.36]{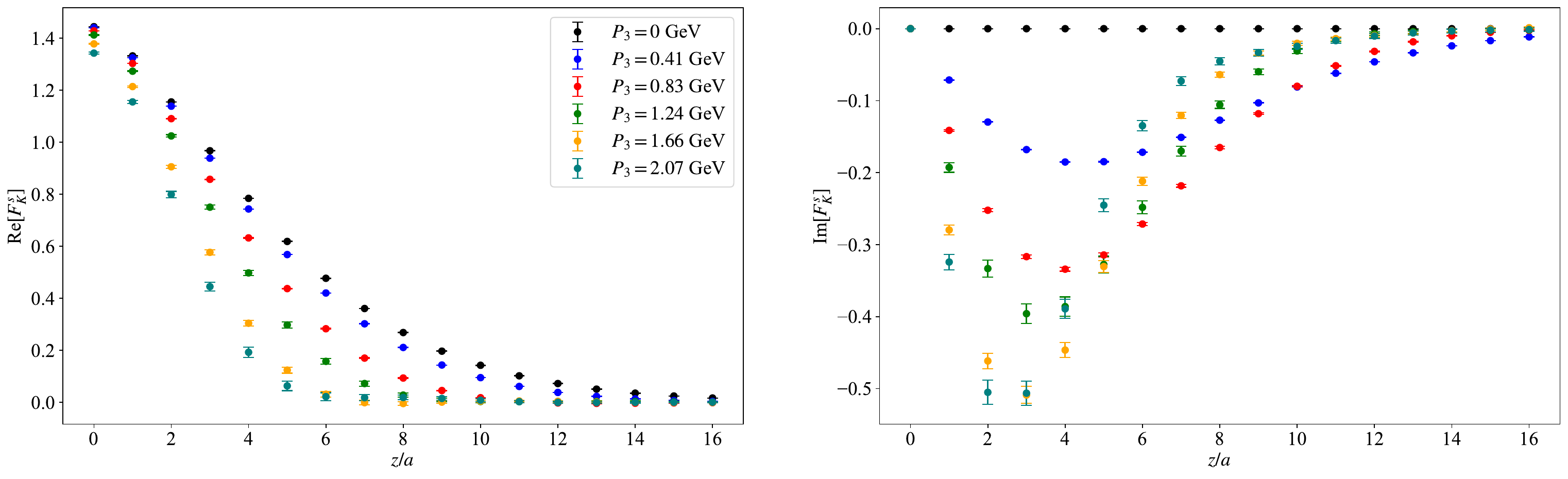}
    \caption{\small{Bare matrix elements for the kaon strange flavor for momentum boost $|P_3|=0,~0.41,~0.83,~1.25,~1.66,~2.07$ GeV. The real and imaginary parts are shown in the left and and plots, respectively. }}
    \label{fig:g0_kaon_s}
\end{figure}

\newpage
The ground state of the matrix elements, $F_M^f$, is shown in Fig.~\ref{fig:g0_pion} and Figs.~\ref{fig:g0_kaon_u} - \ref{fig:g0_kaon_s} for the pion and kaon, respectively. 
From these plots, we can also assess the $P_3$ dependence of the matrix elements.
Notably, the data exhibits a $P_3$ dependence that follows the anticipated behavior: As momentum increases, the real part of the matrix elements decay faster to zero (at smaller values of $z$), while the imaginary part enhances with $P_3$ increase, as at $P_3=0$ GeV it vanishes. 
Overall, the signal quality decreases for $P_3>1.25$ GeV despite the increase in statistics.
This is numerically demonstrated in Table~\ref{tab:error} for $z=0$ and $z=5$, where we give the noise-to-signal ratio at each momentum. 
Let us begin with the observations at $z=0$. For example, $P_3=0.41$ GeV and $P_3=1.25$ GeV have the same statistics for the kaon, and the errors increase by a factor of about 20 for the up quark and 17 for the strange quark. 

\vspace*{0.35cm}

\vspace*{0.35cm}

Let us now examine how the signal quality is affected by the Wilson-line length $z$ by comparing the relative errors at $P_{3}=1.25$~GeV and at the highest momentum, $P_{3}=2.07$~GeV, for both $z=0$ and $z=5a$.  
Focusing on the real part, for which the matrix element is nonzero at both separations, we observe the following behavior.  
For the pion, the relative error increases by a factor of approximately $8$ at $z=0$, while at $z=5a$ the increase is significantly larger, about a factor of $16$.  
Performing the same comparison for the up quark in the kaon, the relative error grows by a factor of $2$ at $z=0$ and by roughly $14$ at $z=5a$, when comparing $P_{3}=1.25$~GeV and $P_{3}=2.07$~GeV.  
For the strange-quark component of the kaon, the corresponding increases are a factor of $3$ at $z=0$ and a factor of about $8$ at $z=5a$.
It is important to appreciate these trends considering the substantial increase in statistics at the highest momentum, that is, by a factor of $7$ for the pion and by a factor of $38$ for the kaon.  
A further informative comparison concerns the relative errors of $F_{\pi}^{u}$, $F_{K}^{u}$, and $F_{K}^{s}$ at fixed momentum.  
At $z=0$ and $P_{3}=1.25$~GeV, the ratio of the relative error of $F_{\pi}^{u}$ to that of $F_{K}^{u}$ is about $0.6$, even though the pion has seven times more statistics.  
At $P_{3}=2.07$~GeV, this ratio increases to approximately $3$.  
A complete set of numerical comparisons for both $z=0$ and $z=5a$ can be found in Table~\ref{tab:error}.\\

\begin{table}[h!]
\begin{center}
\renewcommand{\arraystretch}{1.9}
\begin{tabular}{l|cccccc}
\hline
$\qquad\,\, P_3$ \,\, [GeV] & $\quad$0$\quad$ & $\quad$$\pm$0.41$\quad$ & $\quad$$\pm$0.83$\quad$ & $\quad$$\pm$1.25$\quad$ & $\quad$$\pm$1.66$\quad$ & $\quad$$\pm$2.07$\quad$  \\ 
\hline 
$\qquad\,\, N_{\rm tot}^{\pi} \,\, \qquad$  & 1,198 & 9,584 & 9,584 & 67,088  & 100,632  & 479,200 \\ \hline
$\qquad\,\, N_{\rm tot}^{K} \,\, \qquad$  & 1,198 & 9,584 & 9,584 & 9584  & 28,752  & 359,400 \\  \hline
{\rm Re}:\,\, $dF_\pi^u(z=0)\,/\,F_\pi^u(z=0) \quad$ & 0.00020 & 0.00019 & 0.0012 & 0.0012 & 0.0042 & 0.0100\\ \hline
{\rm Re}:\,\, $dF_K^u(z=0)\,/\,F_K^u(z=0)\quad$ & 0.00017 & 0.00010 & 0.0004 & 0.0020 & 0.0017 & 0.0033 \\ \hline
{\rm Re}:\,\, $dF_K^s(z=0)\,/\,F_K^s(z=0)\quad$ & 0.00012 & 0.00006 & 0.0002 & 0.0010 & 0.0014 & 0.0027\\  \hline
{\rm Re}:\,\, $dF_\pi^u(z=5a)\,/\,F_\pi^u(z=5a)\quad$ &0.00411  & 0.00390 & 0.02787 & 0.02401 & 0.18395 & 0.37702\\ \hline
{\rm Re}:\,\, $dF_K^u(z=5a)\,/\,F_K^u(z=5a)\quad$ &0.00392   & 0.00208 & 0.00846 & 0.02682 & 0.07396 & 0.36229\\ \hline
{\rm Re}:\,\, $dF_K^s(z=5a)\,/\,F_K^s(z=5a)\quad$&0.00391  & 0.00209 & 0.00914 & 0.03797 & 0.09462 & 0.29001\\ \hline
{\rm Im}:\,\, $dF_\pi^u(z=5a)\,/\,|F_\pi^u(z=5a)|\quad$ &N/A  & 0.00505 & 0.02397 & 0.02545 & 0.07464 & 0.14716\\ \hline
{\rm Im}:\,\, $dF_K^u(z=5a)\,/\,|F_K^u(z=5a)|\quad$ &N/A & 0.00276 & 0.00758 & 0.03742 & 0.02580 & 0.04073\\ \hline
{\rm Im}:\,\, $dF_K^s(z=5a)\,/\,|F_K^s(z=5a)|\quad$&  N/A & 0.00246 & 0.00685 & 0.03523 & 0.02402 & 0.03744\\  \hline
\end{tabular}
\caption{\small Signal to noise ratio for $F_M^f$ at $z=0$ and $z=5a$ for each momentum boost $P_3$. For comparison, we also include the total statistics for the pion ($N_{\rm tot}^{\pi}$) and kaon ($N_{\rm tot}^{K}$). }
\label{tab:error}
\end{center}
\end{table}

\newpage
\vspace*{0.5cm}
\subsection{Pseudo-distributions approach}
\label{sec:results_pseudo}

\begin{figure}[h!]
    \centering
    \includegraphics[scale=0.33]{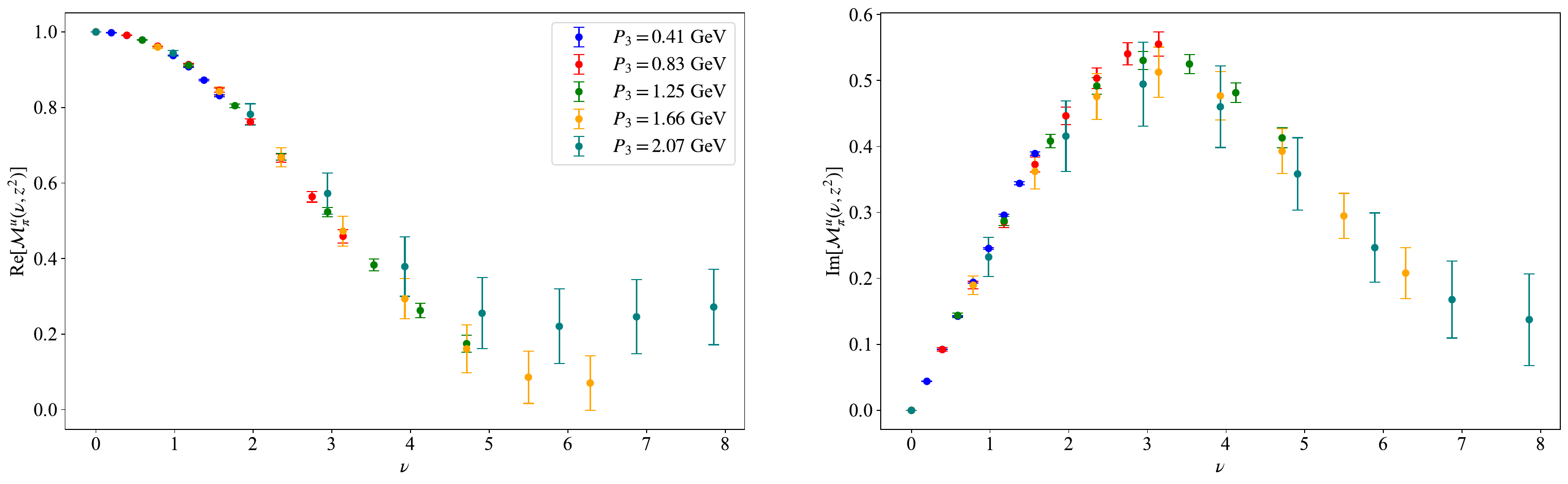}
    \vspace*{-0.4cm}
    \caption{\small{Reduced-ITD for the pion as a function of the Ioffe time $\nu=P\cdot z$. The data on individual momenta $P_3=0.41,\,0.83,\,1.25,\,1.66,\,2.07$ GeV are shown with blue, red, green, orange, and turquoise symbols, respectively. For every $P_3$, the plots go up to $z=8a$. The real (imaginary) part is shown in the left (right) plot.}}
    \label{fig:DR_pion}
\end{figure}

\begin{figure}[h!]
\hspace*{-0.6cm}
\includegraphics[scale=0.365]{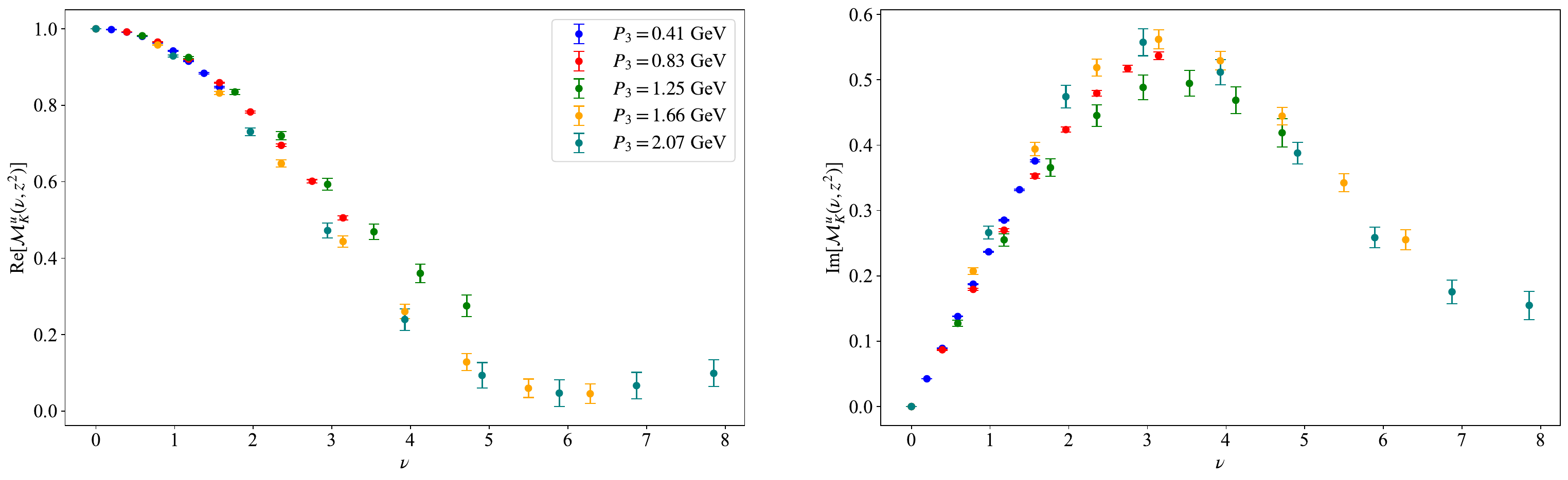}
        \vspace*{-0.35cm}
    \caption{\small{Reduced-ITD for the kaon up-quark as a function of the Ioffe time $\nu=P\cdot z$. The notation is the same as Fig.~\ref{fig:DR_pion}.}}
    \label{fig:DR_kaon_u}
\end{figure}

\begin{figure}[h!]
\hspace*{-0.6cm}
\includegraphics[scale=0.365]{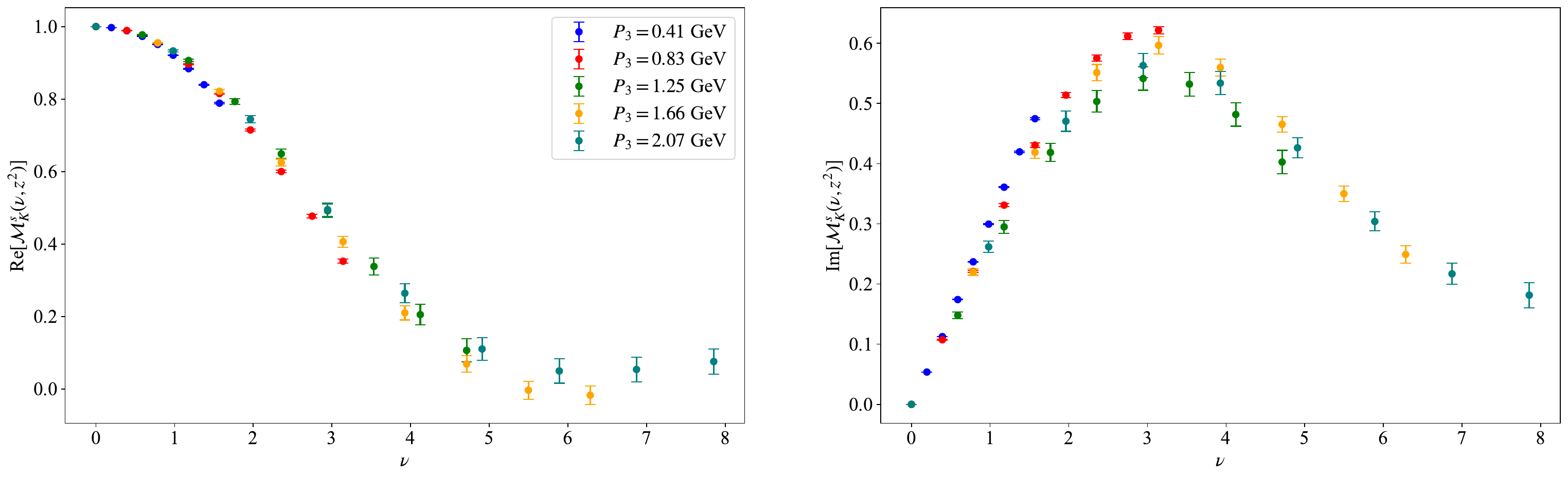} 
        \vspace*{-0.35cm}
    \caption{\small{Reduced-ITD for the kaon strange-quark as a function of the Ioffe time $\nu=P\cdot z$. The notation is the same as Fig.~\ref{fig:DR_pion}.}}
    \label{fig:DR_kaon_s}
\end{figure}

The matrix elements presented above, $F_M^f$, are used for the extraction of the reduced-ITD shown in Eq.~\eqref{eqn:DR}; the latter serve as the foundation of the pseudo-distributions analysis. 
The reduced-ITDs as a function of the Ioffe time are shown in Figs.~\ref{fig:DR_pion} - \ref{fig:DR_kaon_s} for both particles. The plots also indicate, through different colors, the momentum boost associated with each value of the Ioffe time.  
We show data up to $z = 8a$, which provides access to Ioffe times up to $\nu = 8$.  

\newpage

\medskip
It is interesting to observe that, for most values of $\nu$, the reduced-ITDs obtained from different $(P_{3}, z)$ pairs that correspond to the same Ioffe time are found to be statistically compatible, even though it is not expected at the level of the reduced ITDs.  
This empirical consistency may reflect the fact that ${\cal M}(\nu, z^2)$ depends on $\nu$ and $z^{2}$, but does not impose an a priori expectation that different momentum-separation combinations must yield identical values at fixed $\nu$.  
It is also noteworthy that as $P_3$ increases, the statistical uncertainties grow substantially in ${\cal M}(\nu, z^{2})$, particularly at large $\nu$, causing the corresponding points to be significantly less constraining.

\begin{figure}[h!]
\hspace*{-0.3cm} \includegraphics[scale=0.35]{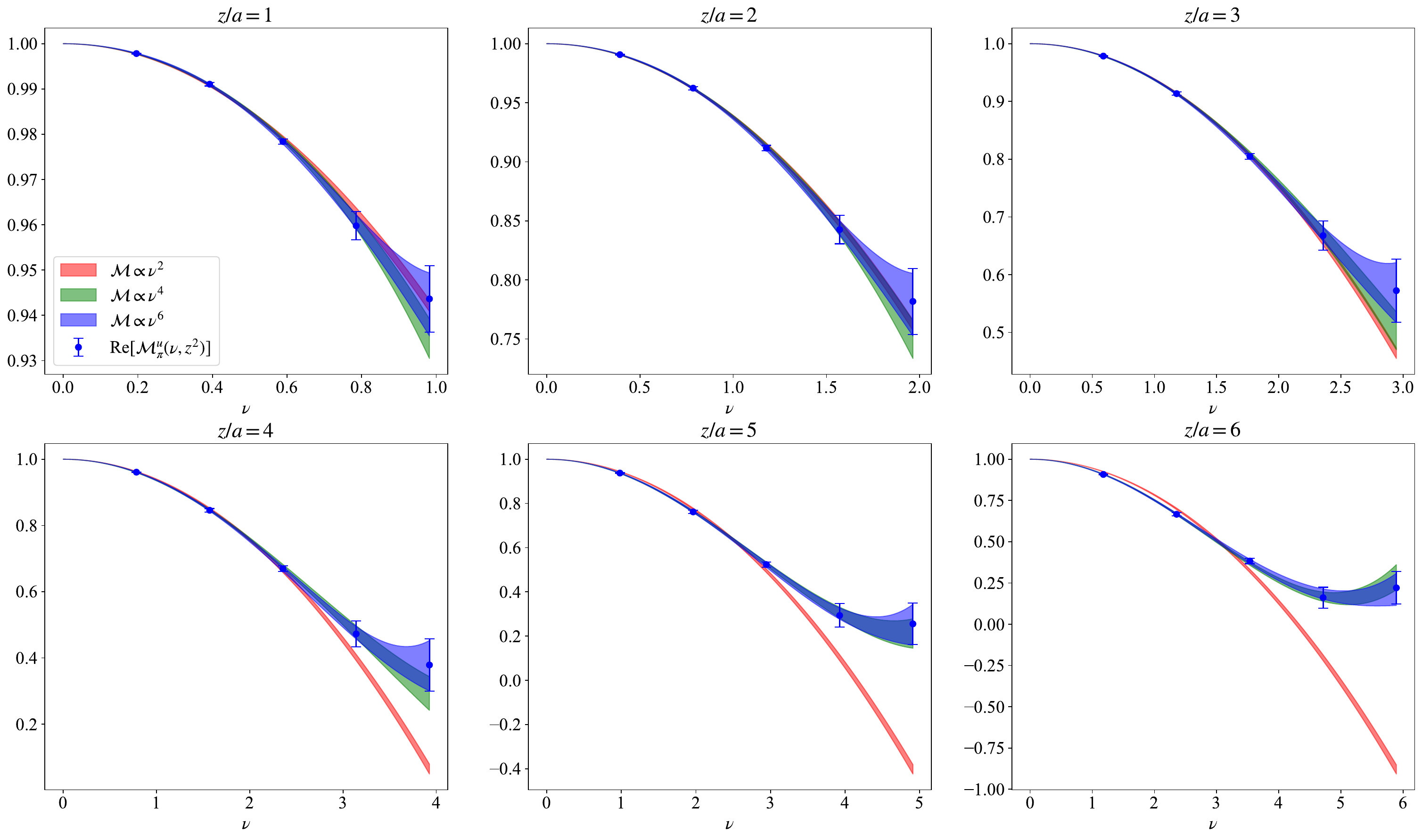}
\vspace*{-0.4cm}
\caption{ \small Interpolation of ${\rm Re}[\mathcal{M}_\pi^u]$ for $z/a \in [0,6]$. The 1-, 2-, and 3-parameter fits are shown with red, green, and blue bands. \\}
\label{fig:DR_re_pion_fit}
\end{figure}

\begin{figure}[h!]
\hspace*{-0.3cm} \includegraphics[scale=0.35]{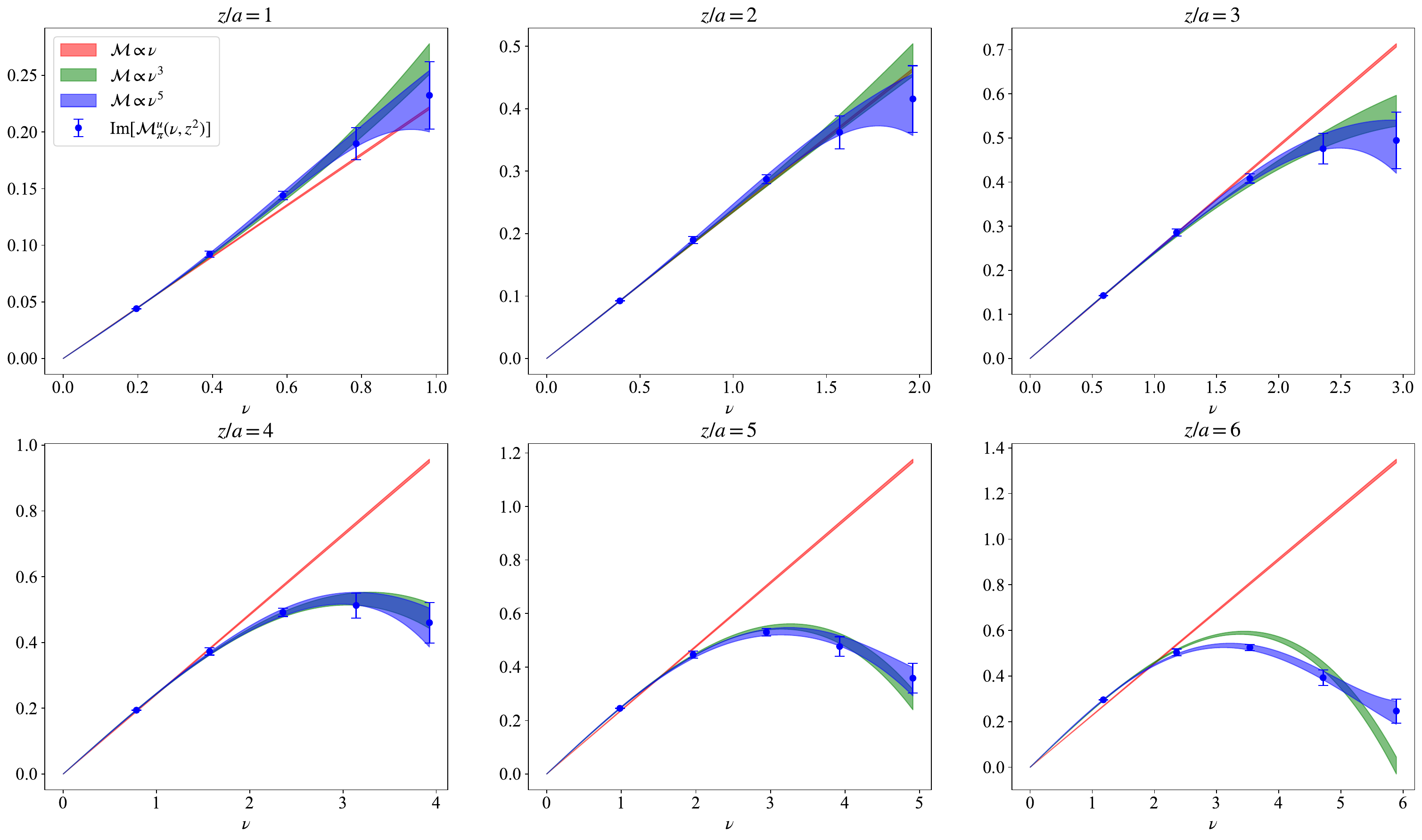}
\vspace*{-0.4cm}
    \caption{ \small Interpolation of ${\rm Im}[\mathcal{M}_\pi^u]$ for $z/a \in [0,6]$. The notation is the same as Fig.~\ref{fig:DR_re_pion_fit}.}
    \label{fig:DR_im_pion_fit}
\end{figure}

\begin{figure}[h!]
\hspace*{-0.6cm}
\includegraphics[scale=0.35]{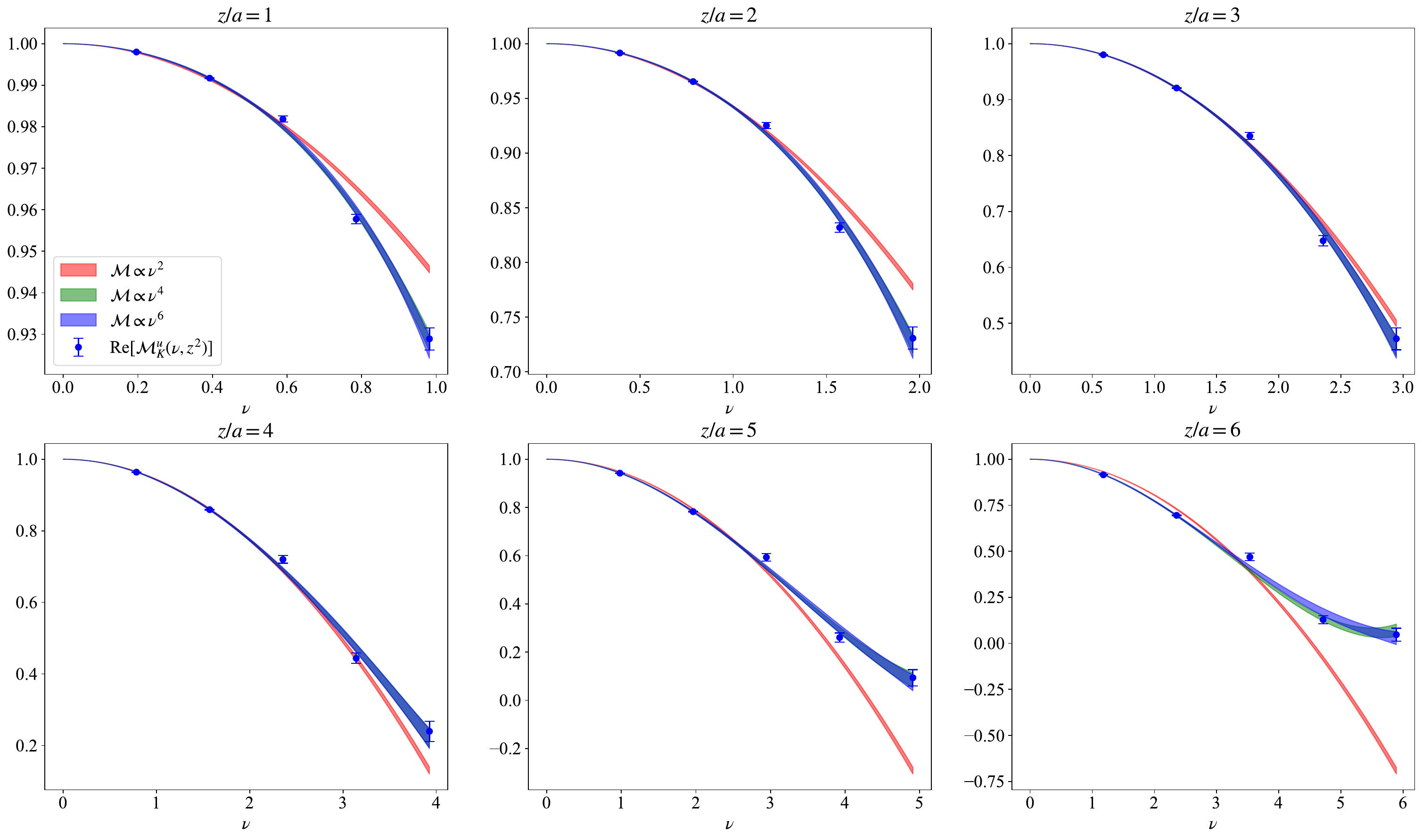}
\vspace*{-0.35cm}
    \caption{ \small Interpolation of ${\rm Re}[\mathcal{M}_K^u]$ for $z/a \in [0,6]$. The notation is the same as Fig.~\ref{fig:DR_re_pion_fit}.}
    \label{fig:DR_re_kaon_u_fit}
\end{figure}

\begin{figure}[h!]
\hspace*{-0.6cm}
\includegraphics[scale=0.35]{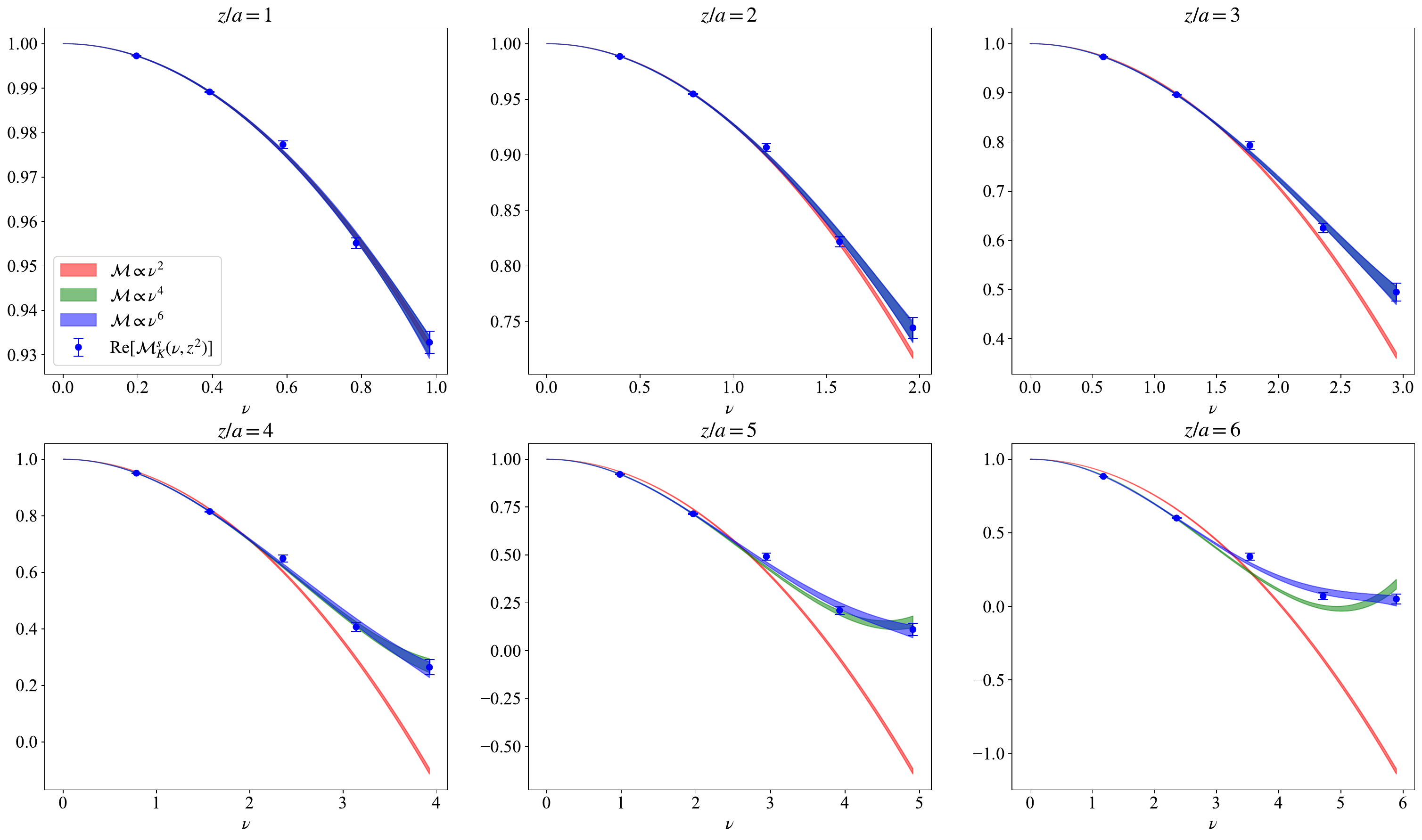}
\vspace*{-0.35cm}
    \caption{ \small Interpolation of ${\rm Re}[\mathcal{M}_K^s]$ for $z/a \in [0,6]$. The notation is the same as Fig.~\ref{fig:DR_re_pion_fit}.}
\label{fig:DR_re_kaon_s_fit}
\end{figure}

\begin{figure}[h!]
\hspace*{-0.6cm}
\includegraphics[scale=0.35]{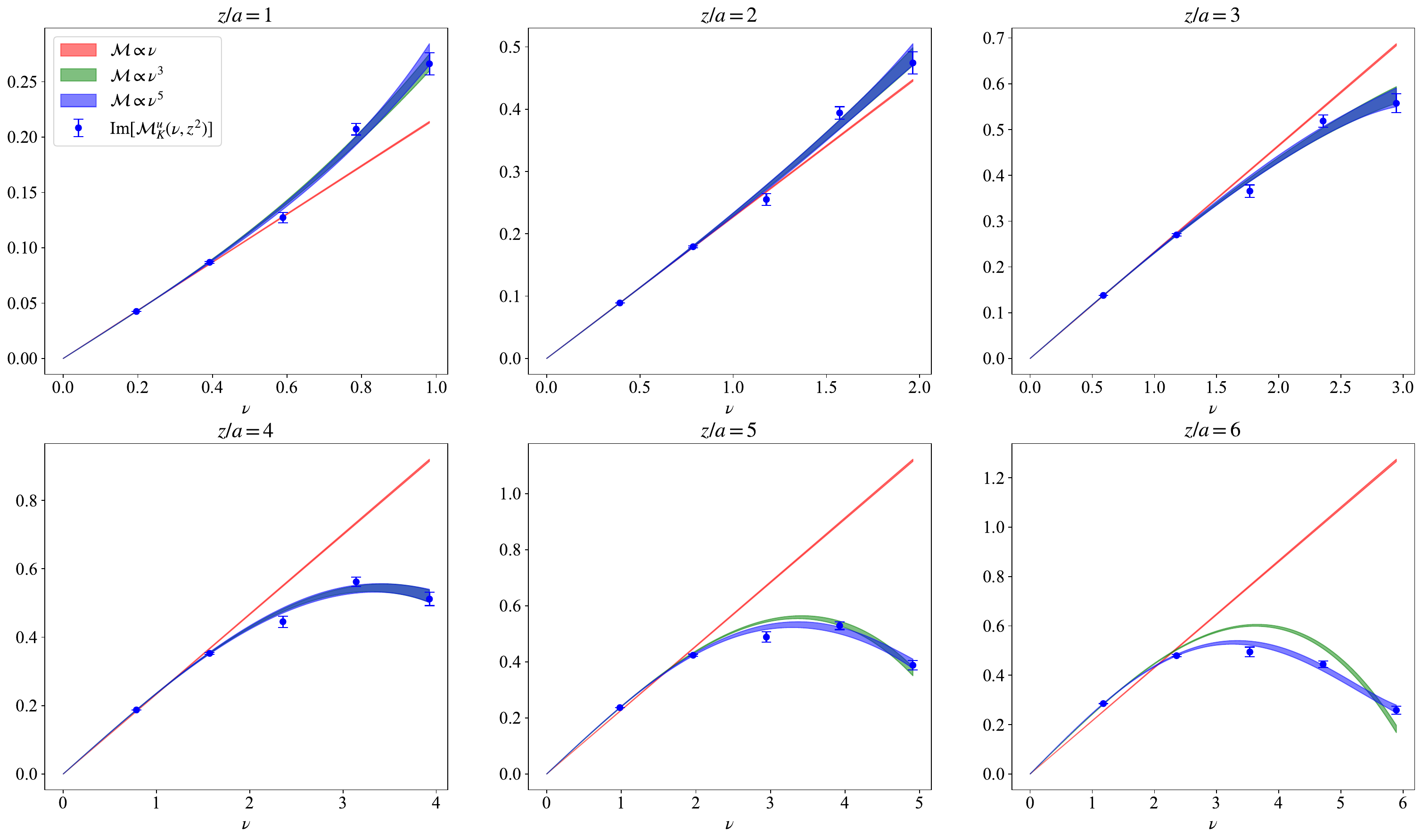}
\vspace*{-0.35cm}
    \caption{ \small Interpolation of ${\rm Im}[\mathcal{M}_K^u]$ for $z/a \in [0,6]$. The notation is the same as Fig.~\ref{fig:DR_re_pion_fit}.}
\label{fig:DR_im_kaon_u_fit}
\end{figure}

\begin{figure}[h!]
\hspace*{-0.6cm}
\includegraphics[scale=0.35]{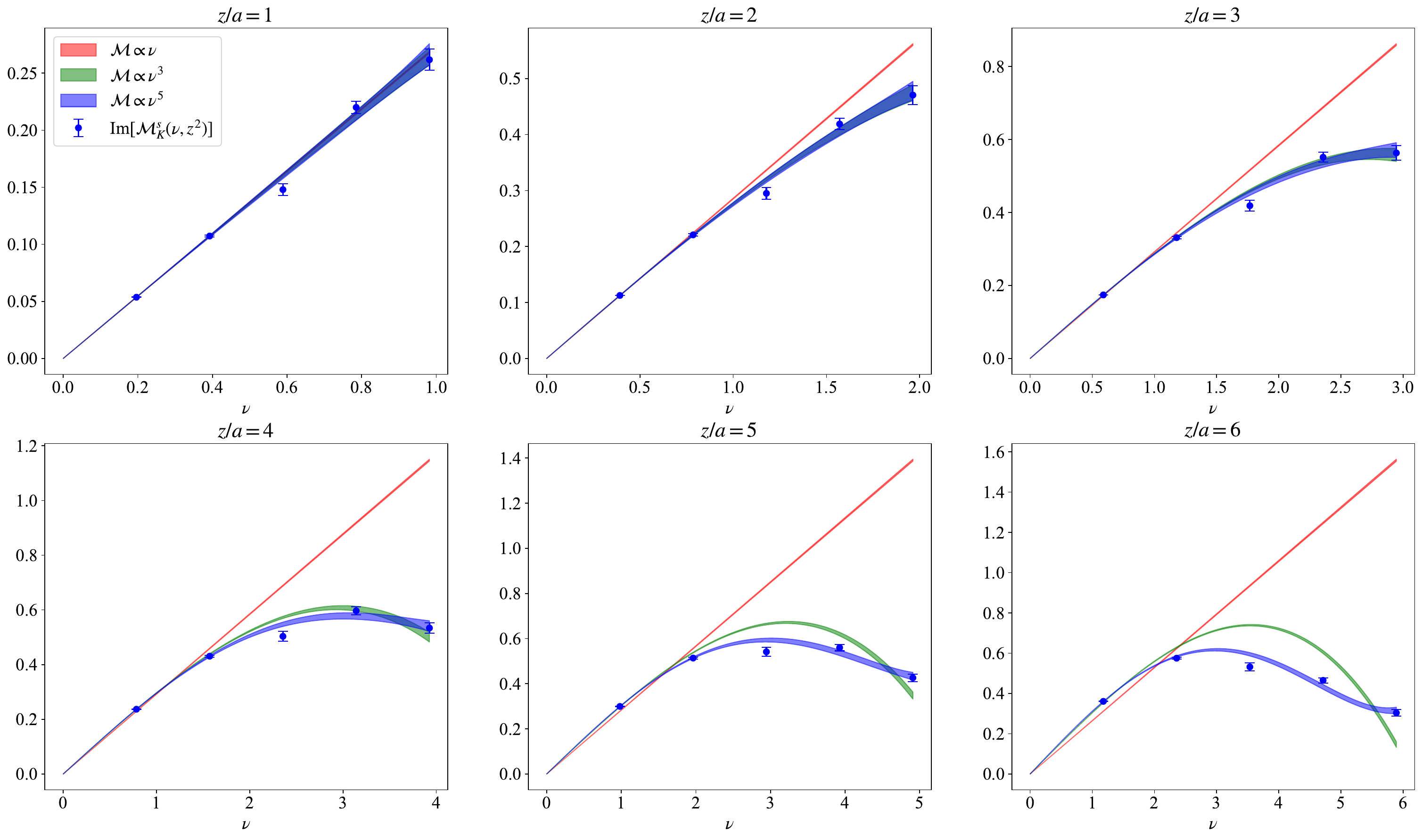}
\vspace*{-0.35cm}
    \caption{ \small Interpolation of ${\rm Im}[\mathcal{M}_K^s]$ for $z/a \in [0,6]$. The notation is the same as Fig.~\ref{fig:DR_re_pion_fit}.}
\label{fig:DR_im_kaon_s_fit}
\end{figure}

\bigskip
In order to implement the integration of Eq.~\eqref{eqn:Evolved_ITD} and Eq.~\eqref{eqn:Matched_ITD} for the evolution to the ITDs and scheme conversion, one needs to perform a fit on the data to obtain the reduced-ITD as a continuous function of the Ioffe time.
We implement an $n$-parameter polynomial fit in $\nu$ at fixed values of $z^2$, and we test $n=1,2,3$, namely a 1-, 2-, and 3-parameter fit, with the real part in even powers of $\nu$ and the imaginary part in odd powers as shown in Eqs.~\eqref{eq:fit_re} - \eqref{eq:fit_im}.
The results are shown in Figs.~\ref{fig:DR_re_pion_fit} - \ref{fig:DR_im_pion_fit} for the pion and Figs.~\ref{fig:DR_re_kaon_u_fit} - \ref{fig:DR_im_kaon_s_fit} for the kaon. 
The 1-parameter fit fails to describe the data, with the exception of the pion up to $z=2a$.
We find that the 2-parameter fit provides a very good description of ${\rm Re}[\mathcal{M}]$ and ${\rm Im}[\mathcal{M}]$ for most $z$ values in the pion, up to $z=4a-5a$ for the kaon, depending on the flavor and whether it is the real or imaginary part. However, in some cases, as $z$ increases, the 2-parameter fit seems unable to describe the data, particularly in the large $P_3$ region. On the contrary, the 3-parameter fit represents the data better as $z$ increases. Therefore, we implement the 3-parameter fit in our final analysis for all values of $z$.


\begin{figure}[h!]
    \centering
    \includegraphics[scale=0.345]{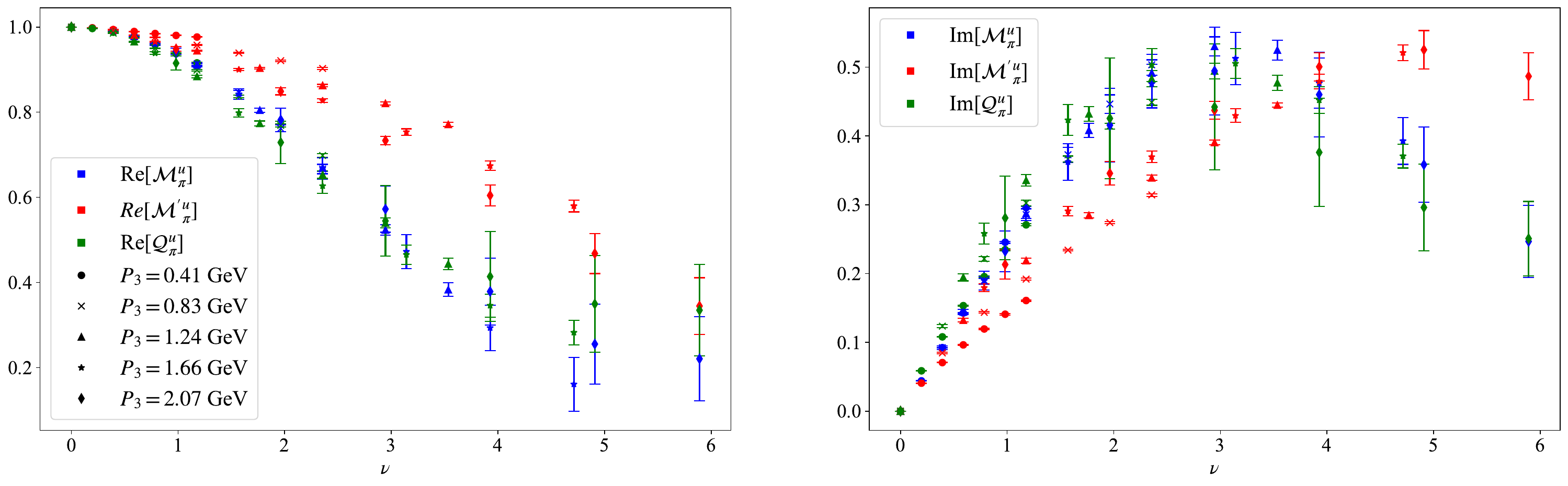}
    \vspace*{-0.4cm}
    \caption{\small{Reduced (blue), evolved (red), and matched (green) ITDs for $|P_3| = 0.41,~0.83,~1.25,~1.66,~2.07$ GeV, for the real (left) and imaginary (right) components.}}
\label{fig:DR_pion_evolved_matched}
\end{figure}

\begin{figure}[h!]
    \centering
\includegraphics[scale=0.345]{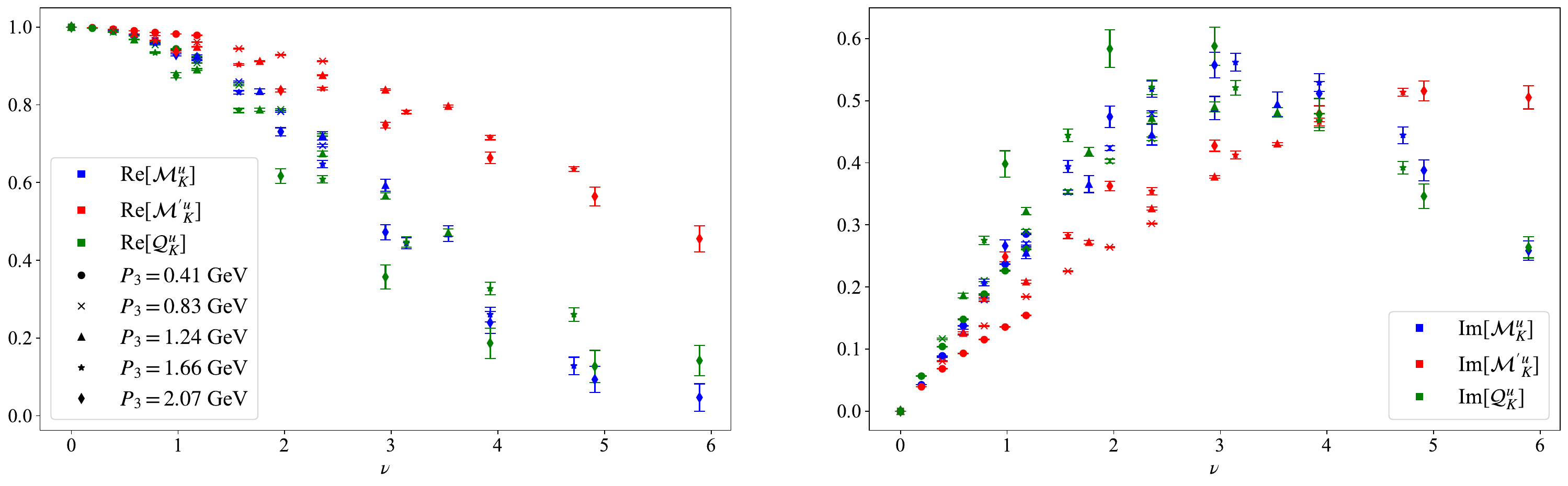}
    \vspace*{-0.4cm}
    \caption{\small{Same as Fig.~\ref{fig:DR_pion_evolved_matched}, but for the up-quark in the kaon.}}
\label{fig:DR_kaon_u_evolved_matched}
\end{figure}

\begin{figure}[h!]
    \centering
\includegraphics[scale=0.345]{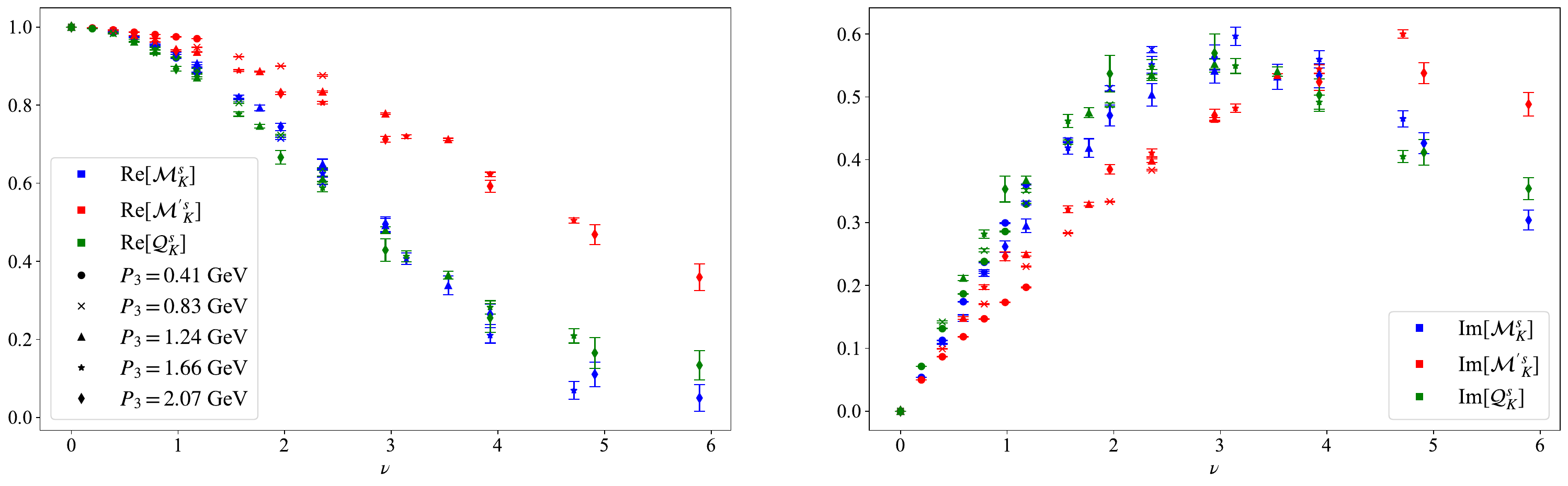}
    \vspace*{-0.35cm}
    \caption{\small{Same as Fig.~\ref{fig:DR_pion_evolved_matched}, but for the strange-quark in the kaon.}}
\label{fig:DR_kaon_s_evolved_matched}
\end{figure}

\newpage
The scale dependence of the ITDs at fixed Ioffe time is addressed via the evolution kernel $B(u)$ in Eq.~\eqref{eqn:Evolved_ITD}. All results are converted to the $\overline{\rm MS}$ scheme and evolved to a common scale of 2~GeV. To maintain consistency at one-loop order, we use the one-loop value of the strong coupling constant at 2~GeV, $\alpha_s/\pi \approx 0.129$.
Fig.~\ref{fig:DR_pion_evolved_matched} displays, for the pion, the reduced-ITDs $\mathcal{M}$ obtained from fits in $\nu$, the evolved ITDs $\mathcal{M}'$ from Eq.~\eqref{eqn:Evolved_ITD}, and the matched ITDs $\mathcal{Q}$ from Eq.~\eqref{eqn:Matched_ITD}. Corresponding results for the kaon are shown in Figs.~\ref{fig:DR_kaon_u_evolved_matched} -\ref{fig:DR_kaon_s_evolved_matched}. Each panel indicates the value of the boost momentum $P_3$ associated with the data.
Focusing on the pion, evolution to 2~GeV leads to an increase in the real part and a decrease in the imaginary part of $\mathcal{M}'$ relative to $\mathcal{M}$. Moreover, for both the real and imaginary components, differences emerge among data points corresponding to the same Ioffe time but originating from different combinations of $z$ and $P_3$. A similar pattern is seen for the kaon, with the added feature that the imaginary parts of $\mathcal{M}'_K{}^u$ and $\mathcal{M}'_K{}^s$ cross their reduced counterparts $\mathcal{M}_K{}^u$ and $\mathcal{M}_K{}^s$, respectively, for $\nu > 4$.
    \vspace*{-0.2cm}
    \vspace*{-0.2cm}

To obtain the final matched ITDs $\mathcal{Q}$, we apply the conversion from $\mathcal{M}'$ to the $\overline{\rm MS}$ scheme using Eq.~\eqref{eqn:Matched_ITD}. 
The numerical impact is also illustrated in Figs.~\ref{fig:DR_pion_evolved_matched} - \ref{fig:DR_kaon_s_evolved_matched}. 
Interestingly, the effects of evolution and scheme conversion are similar in magnitude but opposite in direction. 
As a result, the final matched ITDs are close to the original reduced-ITDs. This indicates that the one-loop matching has a relatively minor effect.
This observation aligns with Ref.~\cite{Bhat:2022zrw}, which shows that the difference between one- and two-loop matching is negligible for small $\nu$. 

Another important point concerns the comparison of data corresponding to the same Ioffe time $\nu$ but arising from different combinations of $z$ and $P_3$. Overall, there is improved agreement among such points compared to the reduced and evolved ITDs. However, there is a discrepancy in some cases, revealing non-negligible discretization effects.
As an example, let us focus on $\mathcal{Q}$ for the combinations $(n_3, z)=(2,6a),\,(3,4a),\,(4,3a)$, where $n_3$ is defined via $P_3 = \frac{2\pi}{L} \,n_3$; these correspond to $\nu\sim 2.4$. 
The data can be seen in Figs.~\ref{fig:DR_pion_evolved_matched} - \ref{fig:DR_kaon_s_evolved_matched} with different shapes: a cross ($n_3=2$), an up triangle ($n_3=3$), and a star ($n_3=4$).
For the pion, we find that in both the real and imaginary parts, the combination $(n_3, z)=(2,6a)$ differs from the other two cases. For the up-quark in the kaon, we observe differences between all three cases, while for the strange quark, we find compatibility. 
Such a difference in behavior is influenced by the different relative error of the data.  
We expect that, with sufficiently high statistics, the combination $(n_3, z)=(2, 6a)$ will be incompatible with the rest. Overall, similar observations for other values of $\nu$ reveal that, at small $z$, there are non-negligible discretization effects, which will affect the quality of the reconstruction for the $x$-dependence. To investigate this effect, we proceed with the averaging of the values of $\mathcal{Q}$ for the same $\nu$ values, and then perform a parametrization of the $\nu$ dependence. 

\begin{figure}[h!]
    \centering
\includegraphics[scale=0.35]{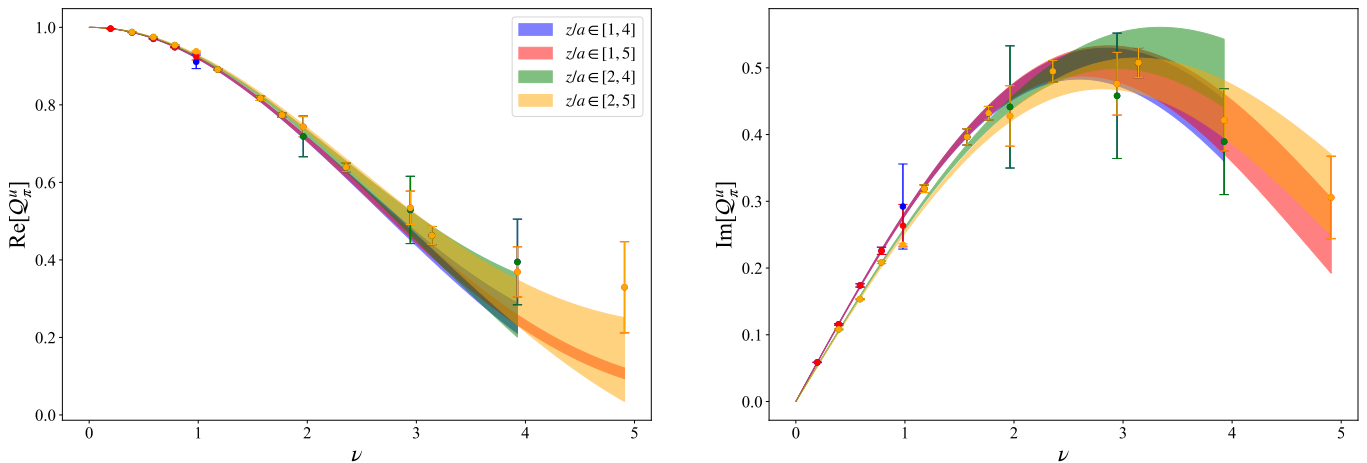}
    \vspace*{-0.35cm}
    \caption{\small{Parametrization of the real (left) and imaginary (right) parts of $\mathcal{Q}_\pi^u$  for various $z$ ranges. The data points correspond to the averaged $\mathcal{Q}$ over the $z$ and $P$ combinations at the same Ioffe time. The averaging excludes data outside the $z$ range of the fit under study.}}
    \label{fig:pion_zrange_fit}
\end{figure}

\begin{figure}[h!]
    \centering
\includegraphics[scale=0.35]{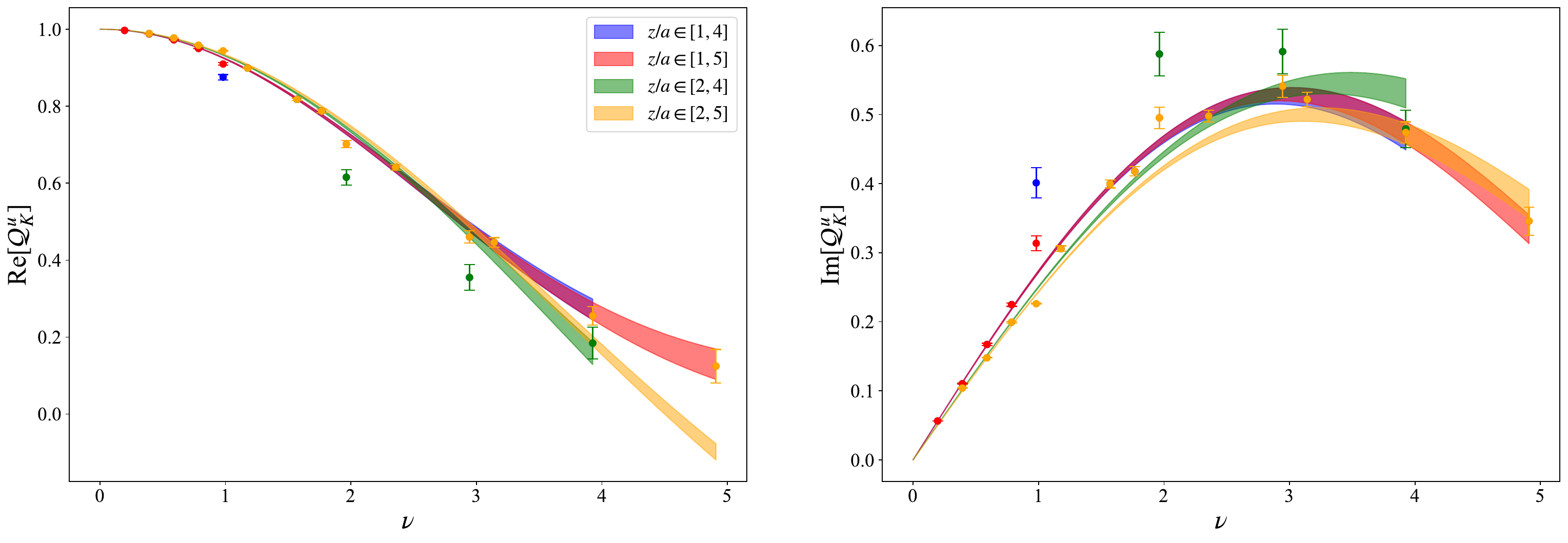}
    \vskip -0.35cm
    \caption{\small{Similar to Fig.~\ref{fig:pion_zrange_fit} but for $\mathcal{Q}_K^u$.}}
    \label{fig:kaon_u_zrange_fit}
\end{figure}

\begin{figure}[h!]
    \centering
\includegraphics[scale=0.35]{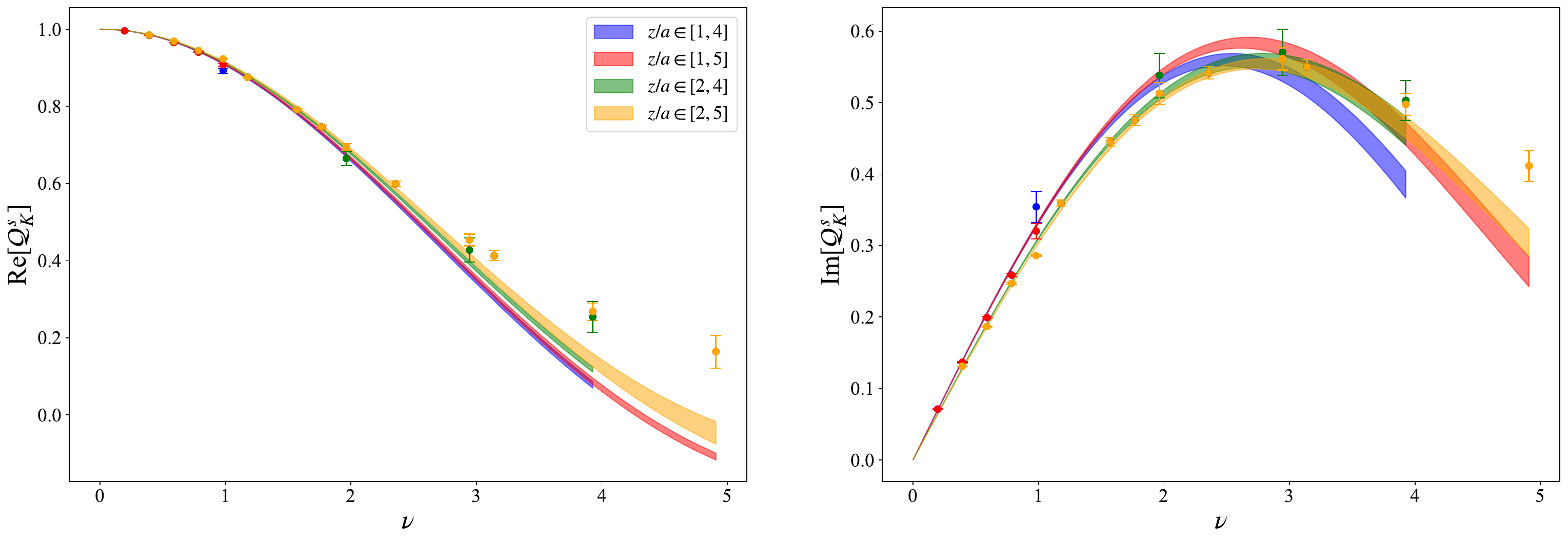}
    \vskip -0.35cm
\caption{\small{Similar to Fig.~\ref{fig:pion_zrange_fit} but for $\mathcal{Q}_K^s$.}}
    \label{fig:kaon_s_zrange_fit}
\end{figure}

Extracting the PDFs requires parametrizing the matched ITD ($\mathcal{Q}$) as a function of $\nu$. 
Considering the constraints mentioned above for the systematic uncertainties, we test a few options for the minimum and maximum values of $z$, named $z_{min}$ and $\zmax$, which are entered into the fits on $\mathcal{Q}$. Specifically, these are $z \in [1a, 4a],\,[2a, 4a],\,[1a,5a],\,[2a,5a]$ corresponding to physical distances $z_{\rm min} \in [0.09 - 0.18]$ fm, and $\zmax \in [0.38 - 0.47]$ fm.
These choices constrain the maximum accessible $\nu$ values to approximately $\nu_{\rm max}\sim4$ for $\zmax=4a$ and $\nu_{\rm max}\sim5$ for $\zmax=5a$.
The resulting matched ITD for each fit, along with their corresponding lattice data for $\mathcal{Q}$, are presented in Fig.~\ref{fig:pion_zrange_fit} for the pion and Figs.~\ref{fig:kaon_u_zrange_fit} - \ref{fig:kaon_s_zrange_fit} for the kaon. 
The data presented in these figures represent averages over multiple $(z,P_3)$ combinations that yield identical values of $\nu$. Only data points within the chosen fitting intervals ($[z_{\rm min},z_{\rm max}]$) are included in the fits. Consequently, the points shown in these figures that are at the same $\nu$ but from different fitting ranges (indicated by different colors) may not coincide for some $z\cdot P$.
Overall, all four fits produce similar results for the pion. However, for the kaon, differences between the fits of different ($[z_{\rm min},z_{\rm max}]$) become noticeable due to the higher statistical precision of the data compared to the pion. In particular, for $\mathcal{Q}_K^u$, fits using $z_{\rm max}=4a$ exhibit deviations in both the real and imaginary parts. For $\mathcal{Q}_K^s$, fits starting from $z_{\rm min}=2a$ do not adequately capture the data at larger values of $\nu$. Based on these findings, we select the fit corresponding to the interval $z \in [1a,5a]$ for the final analysis.

\newpage

\begin{figure}[h!]
    \centering
\includegraphics[scale=0.35]{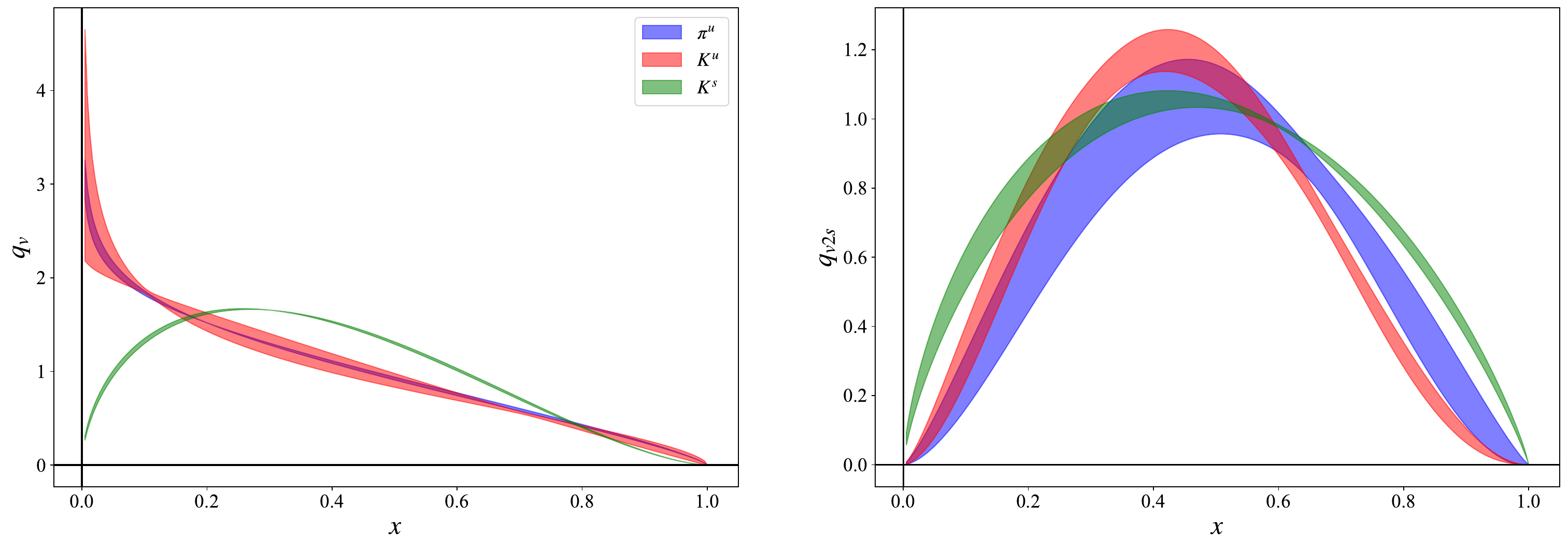}
\vskip -0.35cm
\caption{\small{The $x-$dependence for $q_v$ (left) and $q_{v2s}$ (right) distributions for the pion (blue) kaon up (red) and kaon strange (green) from the Ioffe-time distribution utilizing $z/a\in[1,5]$. Results are shown in $\MSb$ at a scale of $\mu = 2$ GeV. }}
    \label{fig:pseudo_x_dependence}
\end{figure}

\begin{figure}[h!]
    \centering
\includegraphics[scale=0.35]{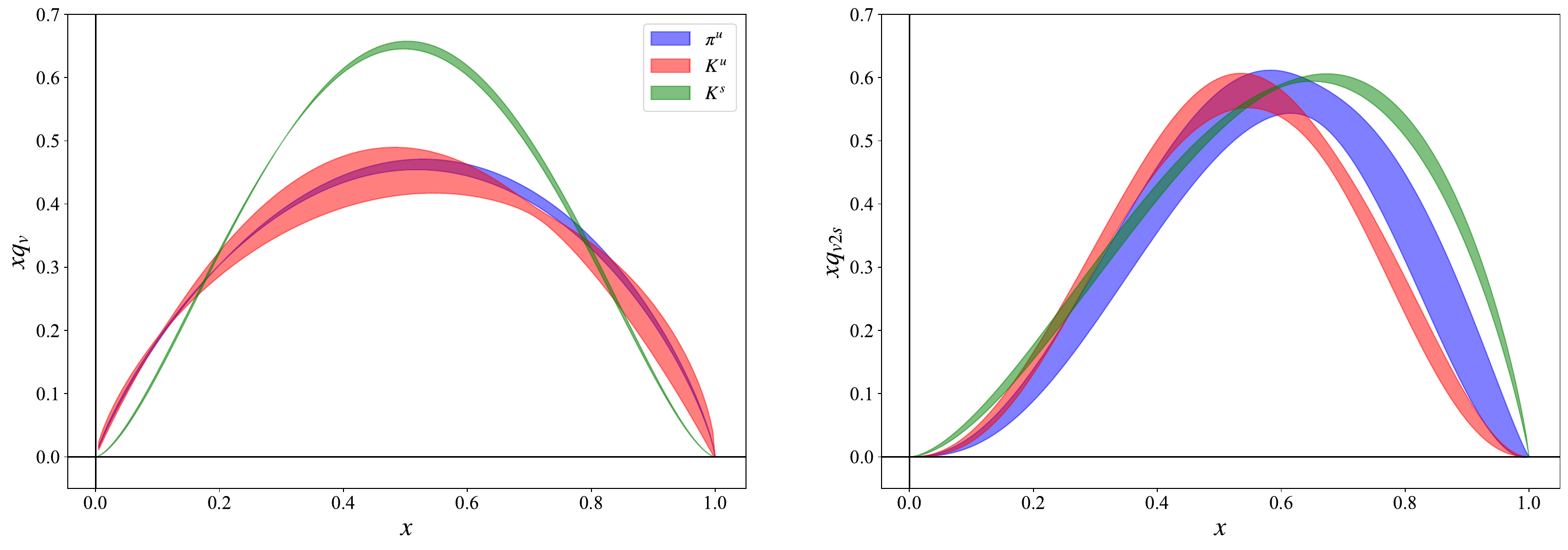}
\vskip -0.35cm
\caption{\small{Same as Fig.~\ref{fig:pseudo_x_dependence}, but for the rescaled distributions $xq_v$ (left) and $xq_{v2s}$ (right).}}
    \label{fig:pseudo_x_dependence_xq}
\end{figure}

\begin{figure}[h!]
    \centering
\includegraphics[scale=0.35]{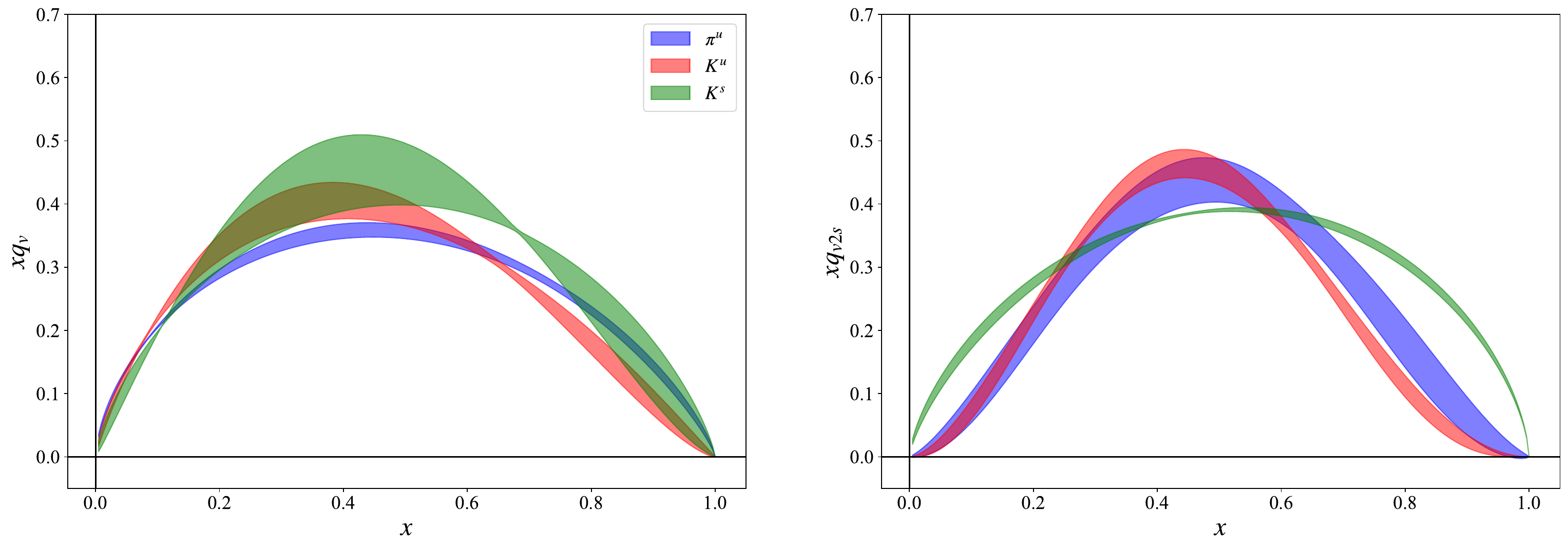}
\vskip -0.35cm
\caption{\small{Similar as Fig. \ref{fig:pseudo_x_dependence_xq} in the $\MSb$ at a scale of $\mu = 5.2$ GeV.}}
\label{fig:pseudo_x_dependence_xq_mu_5.2}
\end{figure}

The final step in the analysis is the extraction of the PDFs, $q(x)$, with their $x$-dependence obtained from $\mathcal Q$ via a Fourier transform in Ioffe time, as discussed in Sec.~\ref {sub:pseudo}. Separating the real and imaginary components of $\mathcal Q$ gives access to the valence (Eq.~\eqref{eq:ReQ}) and ${v2s}$ (Eq.~\eqref{eq:ImQ}) flavor combinations. The $x$-dependence of the $q_v$ and $q_{v2s}$ is given by the parametrization of Eq.~\eqref{eq:ansatz}. 
The results obtained from the Ioffe-time distributions within the range $z/a \in [1,5]$ are shown in Fig.~\ref{fig:pseudo_x_dependence} for $q_v(x)$ and $q_{v2s}(x)$ of the pion and the kaon. We remind the reader that the data presented are in the $\MSb$ at a scale of $\mu = 2$ GeV.
For clarity in the interpretation of results, we also show the distributions rescaled by $x$ in Fig.~\ref{fig:pseudo_x_dependence_xq} that avoids the small-$x$ behavior, which is unreliable due to the limited high-$\nu$ values, restricting fitting the large-$\nu$ behavior. Furthermore, for the comparison between the quark flavor/particles, we focus on the general features of the PDFs, rather than a quantitative comparison, as there are sources of systematic uncertainties to be explored. 
For the valence combination, $x q_v^{\pi^u}$ and $x q_v^{K^u}$ exhibit similar shapes and peaks around $x \sim 0.5$. 
Additionally, $x q_v^{K^u}$ shows a slightly broader profile compared to the pion.
$x q_v^{K^s}$ , however, is peaking at higher values around $x {\sim} 0.6$, indicating that the strange quark in the kaon carries, on average, a larger fraction of the hadron's momentum. 
Turning to the $xq_{v2s}$ combination, all three distributions have peaks that shifted to slightly lower $x$ values compared to the valence case, around $x \sim 0.4 - 0.5$. 
The distributions are somewhat the same within errors up to about $x\sim0.5$. For the intermediate- to large-$x$ regions, $xq_{v2s}^{K^u}$ decays to zero faster, followed by $xq_{v2s}^{\pi^u}$.
The $x q_v^{K^s}$ is the widest in this combination, and its peak is slightly shifted towards larger $x$ values relative to the pion and kaon up-quark distributions. As expected, all distributions fall to zero as $x$ approaches 1. 
It is also instructive to present the distributions at a scale $\mu=5.2$ GeV, which has been used in other lattice, $\chi$-PT, model calculations, as well as global analyses (see, e.g., Refs.~\cite{Aicher:2010cb,Joo:2019bzr,Sufian:2020vzb,Chen:2016sno,Bednar:2018mtf,Lan:2019rba,Watanabe:2017pvl,Alexandrou:2021mmi}). The comparison for $x q_v$ and $x q_{v2s}$ is shown in Fig.~\ref{fig:pseudo_x_dependence_xq_mu_5.2}.
As can be seen, the magnitude of the distributions decreases in all cases. $q_v^{K^s}$ and $q_{v2s}^{K^s}$ exhibits the largest changes in both shape and normalization. 
It is particularly intriguing to focus on the valence contributions, for which we have reconstructed the $x$ dependence from the Mellin moments, $\langle x^n \rangle$ (using $n=0,1,2,3$), on the same ensemble as this work (see Fig. 11 of Ref.~\cite{Alexandrou:2021mmi}). 
The two determinations are similar in magnitude and display the same pattern: the distributions agree for small to intermediate $x$; $q_v^{\pi^u}$ and $q_v^{K^u}$ remain close across the full $x$ range; and $q_v^{K^s}$ is dominant for intermediate- to large-$x$. 
\vspace*{-0.1cm}
\vspace*{-0.1cm}

\begin{figure}[h!]
\hspace*{-0.15cm}
    \includegraphics[scale=0.415]{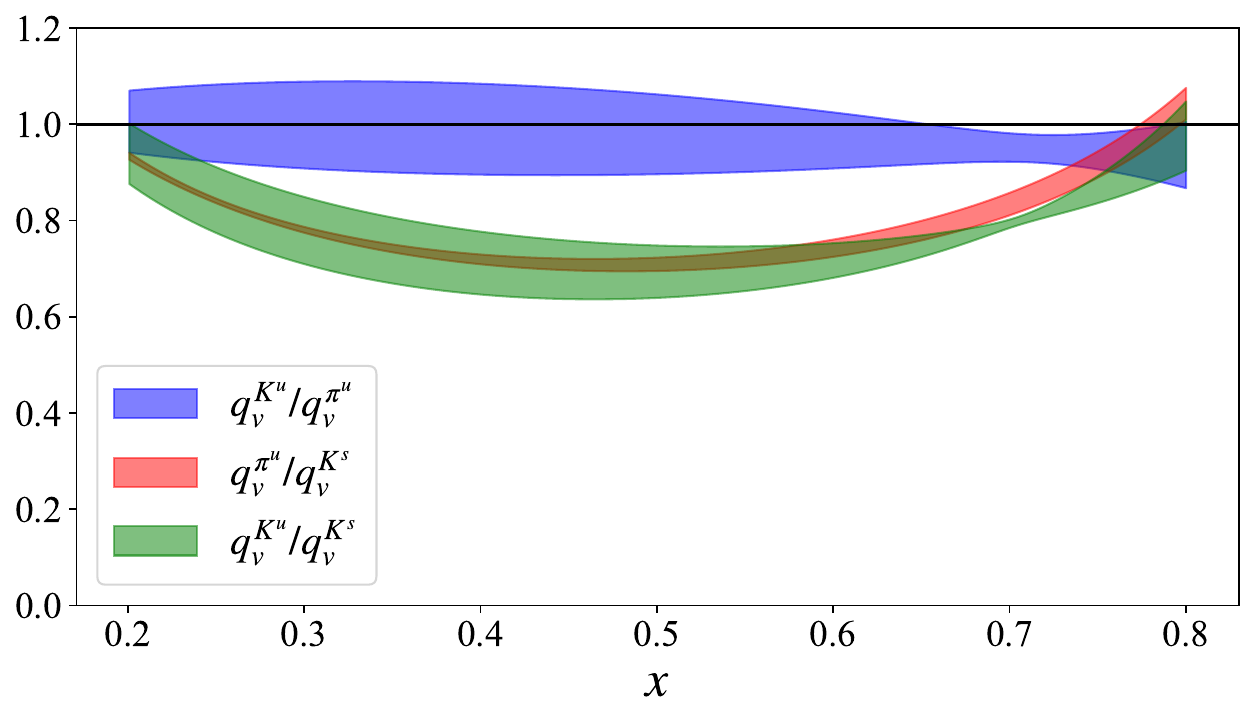}
    \includegraphics[scale=0.415]{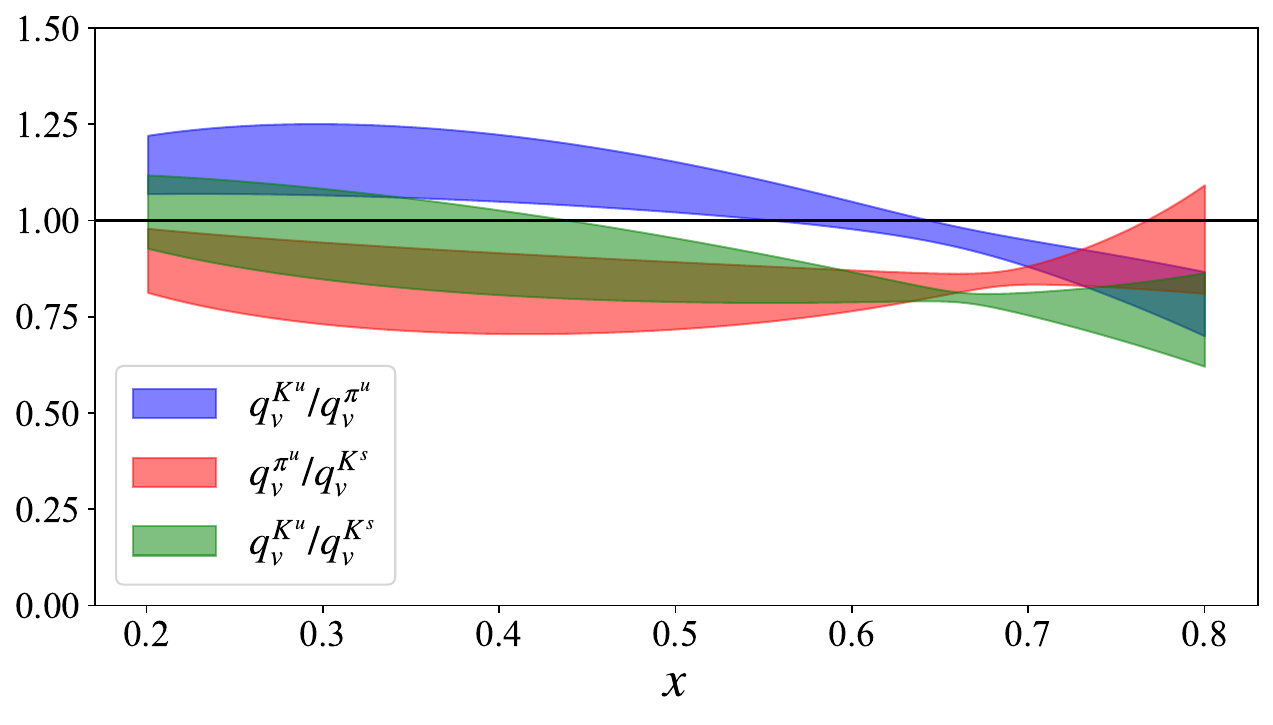}
\vskip -0.45cm
\caption{\small{SU(3) symmetry breaking from the ratios of $q_v^{K^u}/q_v^{\pi^u}$, $q_v^{\pi^u}/q_v^{K^s}$, and $q_v^{K^u}/q_v^{K^s}$ as found from the pseudo-PDF approach at scales $\mu=2$ GeV (left) and $\mu=5.2$ GeV (right).}}
    \label{fig:SU3_pseudo_scales}
\end{figure}

One of the motivations for comparing the pion and kaon PDFs is to investigate the effects of SU(3) flavor symmetry breaking. This arises due to the larger mass of the strange quark compared to the up and down quarks. 
This mass difference leads to the observed mass disparity between pions and kaons, illustrating the SU(3) flavor symmetry-breaking effect in Nambu-Goldstone bosons. 
To explore this, besides the comparison of Figs.~\ref{fig:pseudo_x_dependence} - \ref{fig:pseudo_x_dependence_xq_mu_5.2} show the ratio of the distributions for the valence case, as depicted in Fig.~\ref{fig:SU3_pseudo_scales} for a scale of 2 GeV and 5.2 GeV. 
For this comparison, we focus on the range $x \in [0.2, 0.8]$, as there are uncontrolled systematic effects in the small and large $x$ region due to higher-twist contaminations. 
At $\mu=2$ GeV, the pion and kaon have the same up-quark contribution, while at $\mu=5.2$ GeV, an approximate 10\% difference is permitted within uncertainties. The ratios $q_v^{\pi^u}/q_v^{K^s}$ and $q_v^{K^u}/q_v^{K^s}$ at $\mu=2$ GeV show that the up-quark contribution becomes up to 80$\%$ of $q_v^{K^s}$ in the intermediate $x$ region; close to $x=0.2$ and $x=0.8$ the aforementioned ratio becomes one. Some different behavior is observed at $\mu=5.2$ GeV, and we find that the uncertainties are larger. This is a result of the parametrization of ${\mathcal{Q}}$ and the constraint of the fit parameters $a$ and $b$ of Eq.~\eqref{eq:ansatz}. Interesting, similar conclusions as the $\mu=5.2$ GeV of Fig.~\ref{fig:SU3_pseudo_scales}, are obtained comparing the pion and kaon PDFs reconstructed from the Mellin moments on the same ensemble~\cite{Alexandrou:2021mmi}. The similar role of the up quark in the pion and kaon is found in the Mellin moments~\cite{Alexandrou:2020gxs,Alexandrou:2021mmi} and the form factors~\cite{Alexandrou:2021ztx} using the same ensemble. The ratio $q_v^{K^u}/q_v^{\pi^u}$ at $\mu=5.2$ GeV is similar to the analysis of Ref.~\cite{Bednar:2018mtf} (see Fig. 3), as well as the data of Ref.~\cite{Saclay-CERN-CollegedeFrance-EcolePoly-Orsay:1980fhh}. 

To conclude the discussion of the SDF approach, we emphasize that reconstructing the $x$-dependence of the PDFs requires careful selection and treatment of the input data used in the parametrization of $\mathcal{Q}$. We find that the shape of the distributions in terms of $x$ is influenced by the fit range, in particular, omitting the small $z$ data, as well as extending $\zmax$ beyond 0.45 fm. Thus, the reconstruction is sensitive to systematic uncertainties, which can limit the reliability of the final results. For this reason, analyzing the same matrix elements within the LaMET framework is essential, as it enables a comparison between methodologies and may help identify and quantify method-specific effects. We also note that the SDF approach has proven particularly effective in extracting Mellin moments of PDFs from non-local operators (see, e.g., Refs.~\cite{Bhattacharya:2023ays,Bhattacharya:2024wtg}).

\vspace*{0.75cm}
\subsection{Quasi-distributions approach}
\label{sec:results_quasi}

Similar to the pseudo-distribution approach, the matrix elements shown in Figs.~\ref{fig:g0_pion} - \ref{fig:g0_kaon_s} can be analyzed within the quasi-distribution framework to extract light-cone PDFs. As discussed in Section~\ref{sub:quasi}, the quasi-distribution method employs matrix elements evaluated at fixed values of the momentum boost $P_3$ and multiple values of the Wilson line length $z$.
The method requires that $P_3$ be sufficiently large for the matching formalism to effectively bring the lattice data to their light-cone counterparts. However, it does not impose a restriction that $z$ be small. In practice, both $P_3$ and $z$ introduce systematic effects and computational challenges that limit their usable ranges.
In particular, the signal-to-noise ratio deteriorates with increasing $P_3$ (see Table~\ref{tab:error}). For $z$, we typically retain data up to $\sim$1~fm, beyond which the bare matrix element decays to zero. Including such values ensures that the matrix element is fully captured over its relevant range (see, e.g., results for $P_3 > 1$~GeV in Figs.~\ref{fig:g0_pion} - \ref{fig:g0_kaon_s}).
Once the analysis is performed for each momentum, comparisons among different $P_3$ values provide a test of convergence toward the light-cone limit.

\begin{figure}[h!]
\hspace*{-0.15cm}
    \includegraphics[scale=0.29]{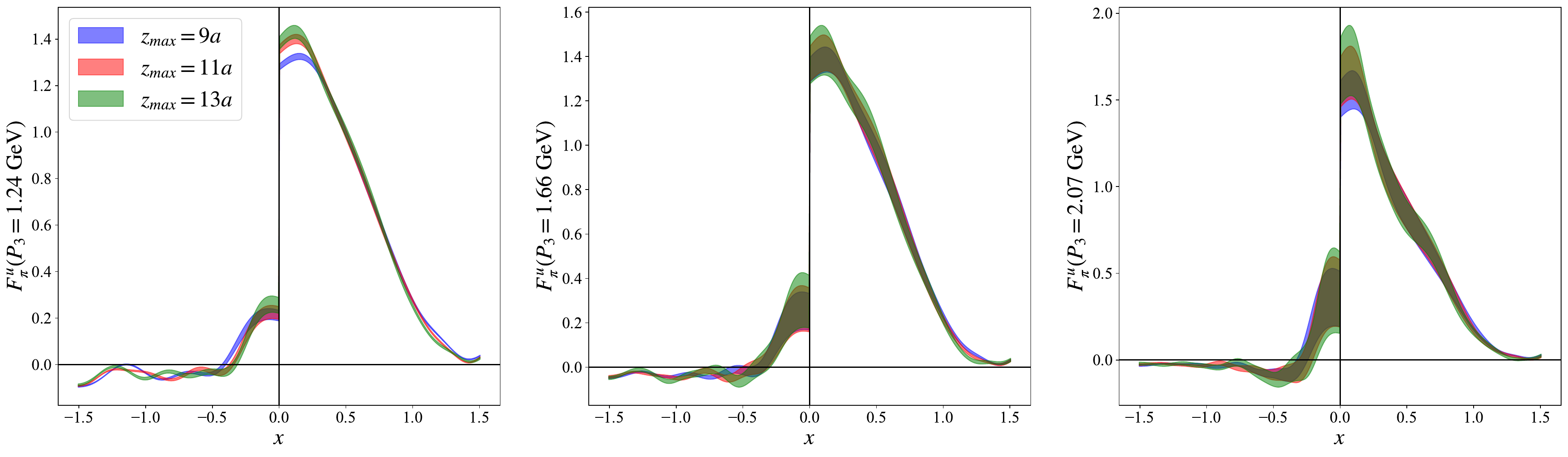}
    \vspace*{-0.65cm}
    \caption{\small{$x$-dependent quasi-PDF for the pion, $F_\pi^u$, for various values of $z_{\rm max}$ and using $P_3 = 1.25,~1.66,~2.07$ GeV in the left, middle, and right panels, respectively. Blue, red, and green bands correspond to $\zmax=9a,\,11a$, and $13a$, respectively.}}
    \label{fig:BG_pion_z_test}
\end{figure}

\begin{figure}[h!]
\hspace*{-0.15cm}
    \includegraphics[scale=0.29]{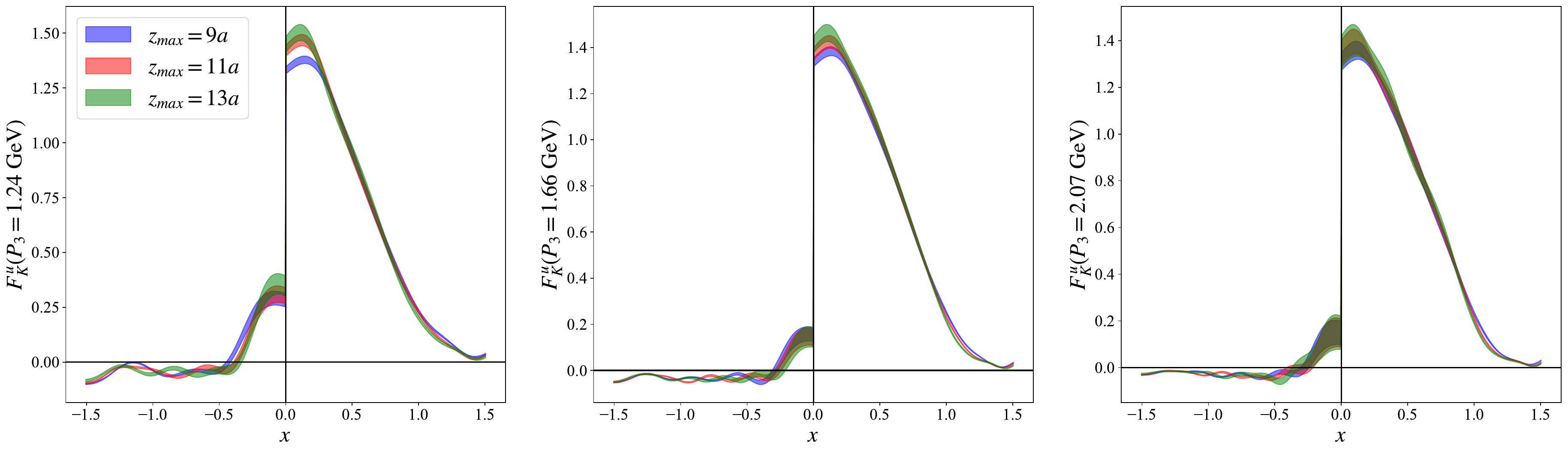}
        \vspace*{-0.65cm}
    \caption{\small{$x$-dependent quasi-PDF for the up quark in the kaon, $F_K^u$. The notation is the same as Fig.~\ref{fig:BG_pion_z_test}.}}
    \label{fig:BG_kaon_u_z_test}
\end{figure}

\begin{figure}[h!]
\hspace*{-0.15cm}
    \includegraphics[scale=0.29]{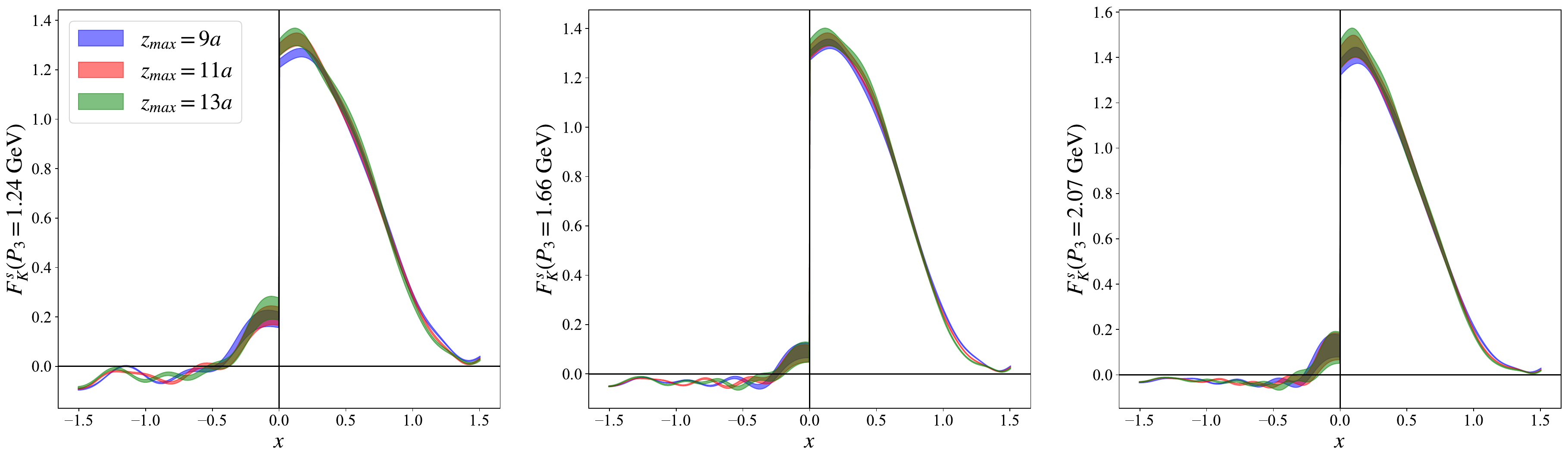}
        \vspace*{-0.65cm}
    \caption{\small{$x$-dependent quasi-PDF for the strange quark in the kaon, $F_K^s$. The notation is the same as Fig.~\ref{fig:BG_pion_z_test}.}}
    \label{fig:BG_kaon_s_z_test}
\end{figure}

Following the methodology outlined in Sec.~\ref{sub:quasi}, we reconstruct the $x$-dependence of the quasi-PDFs in momentum space using the Backus-Gilbert method~\cite{BackusGilbert}. A key parameter in this procedure is $\zmax$, which sets the maximal Wilson line length included in the analysis and influences the accuracy of the reconstruction.
To assess the impact of $\zmax$, we compare results for three representative values: $\zmax=9a,\,11a,\,13a$, corresponding to physical lengths of 0.84, 1.0, and 1.2~fm, respectively. The quasi-PDFs obtained at these $\zmax$ values, using the data at momentum boost above 1 GeV, are shown in Figs.~\ref{fig:BG_pion_z_test} - \ref{fig:BG_kaon_s_z_test}. These comparisons enable us to investigate how the choice of $\zmax$ impacts the shape and stability of the reconstructed distributions.
As shown in the plots, the quasi-PDFs for $\zmax=11a$ and $\zmax=13a$ are statistically compatible, while some differences are observed for $\zmax=9a$ at momentum 1.25 GeV, primarily in the small $x$ region. This discrepancy arises because, at $z=9a$, the matrix elements have not yet decayed to zero for the aforementioned momentum, which affects the stability of the inversion procedure.
We observe agreement between different values of $\zmax$ at the highest boosts, $P_3=1.66$ and 2.07~GeV, for both particles. Based on this analysis, we adopt $\zmax=11a$ as the preferred choice for proceeding with the analysis.
        \vspace*{-0.25cm}
        \vspace*{-0.25cm}

\begin{figure}[h!]
    \centering
    \includegraphics[scale=0.41]{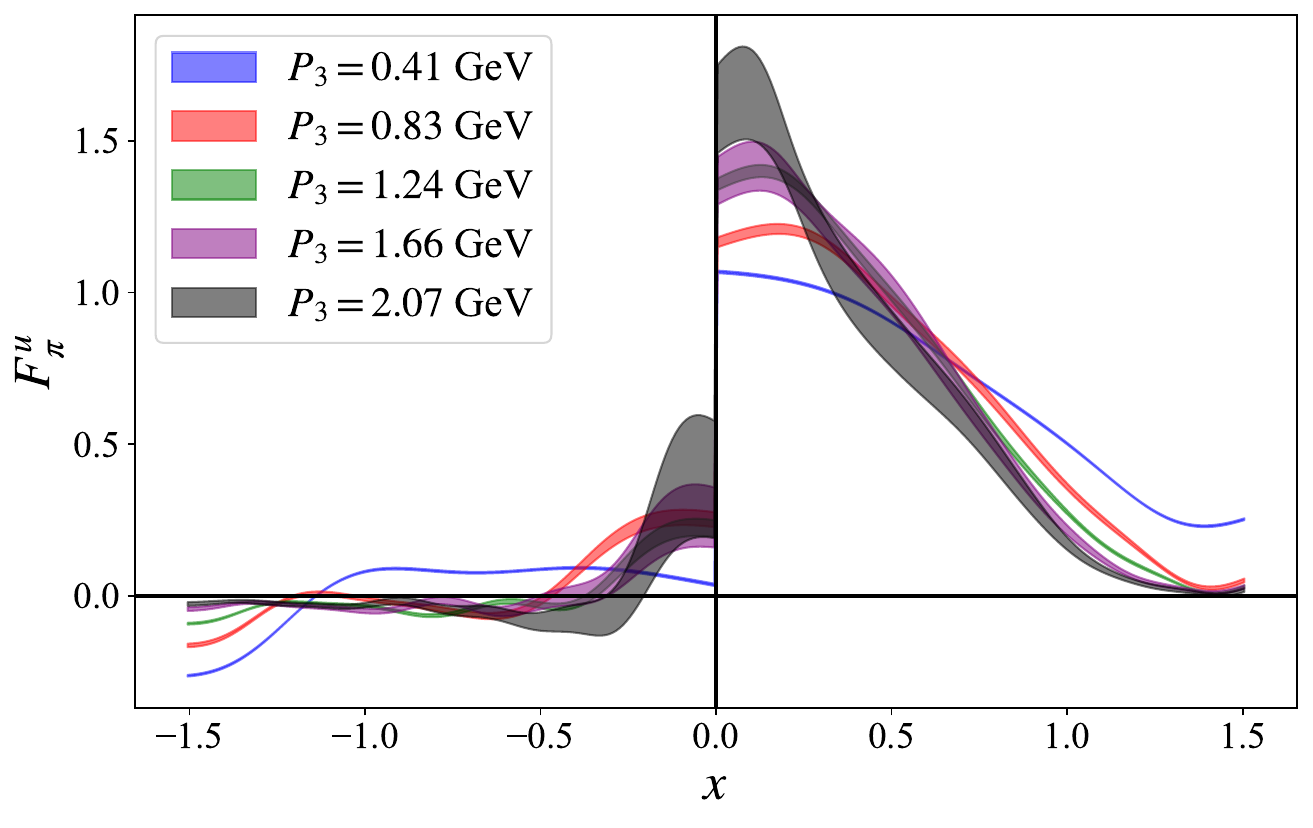}
        \vspace*{-0.4cm}
    \caption{\small{$x$-dependent pion quasi-PDF, $F_\pi^u$, using $\zmax=11a$ for various values of the momentum boost. The data for $|P_3| = 0.41,~0.83,~1.25,~1.66$, and $2.07$ GeV are shown with blue, red, green, magenta, and gray bands, respectively.}}
    \label{fig:BG_pion}
\end{figure}

\begin{figure}[h!]
    \includegraphics[scale=0.41]{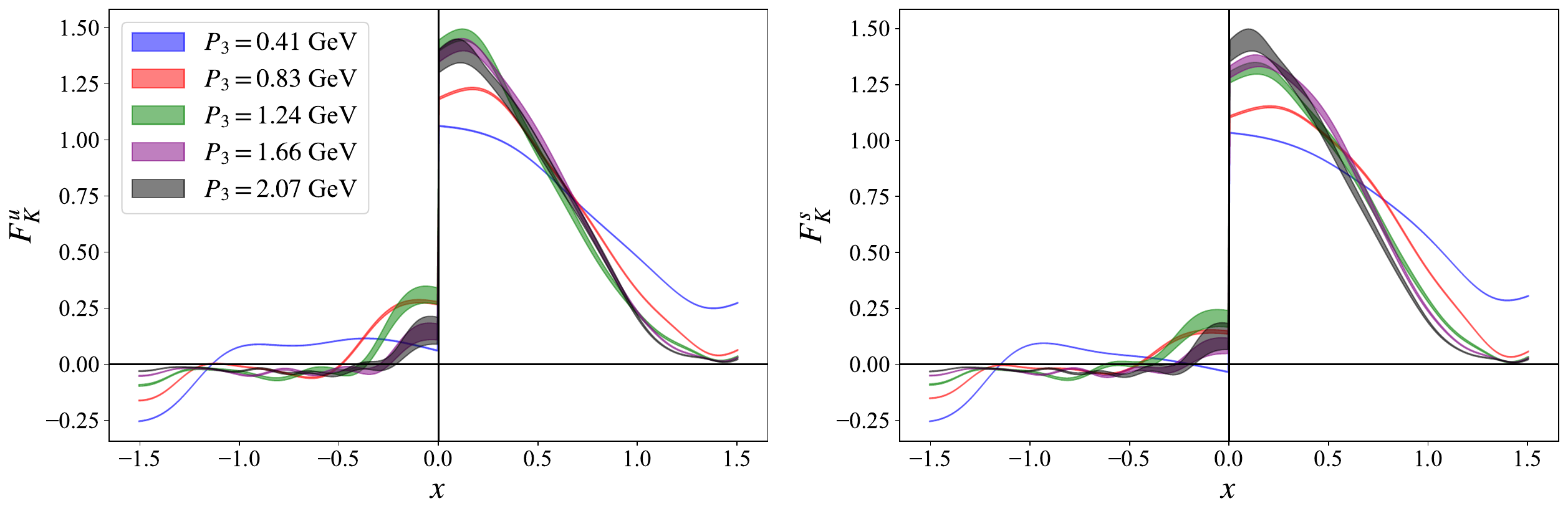}
        \vspace*{-0.4cm}
    \caption{\small{$x$-dependent of the $F_K^u$ (left) and $F_K^s$ (right) quasi-PDFs. The notation is the same as Fig.~\ref{fig:BG_pion}.}}
    \label{fig:BG_kaon}
\end{figure}

\newpage
An additional point of interest is the dependence of the quasi-PDFs on the momentum boost $P_3$. This dependence is inherited from the underlying matrix elements and is non-trivial. In principle, only at sufficiently large $P_3$ and after applying the matching procedure, the PDFs should exhibit independence from $P_3$.
Figs.~\ref{fig:BG_pion} - \ref{fig:BG_kaon} illustrate the $P_3$ dependence of the quasi-PDFs for the pion and kaon. We observe that the results for the three lowest momenta differ significantly from those at higher boosts, especially in the pion case. For $P_3 > 1$~GeV, the quasi-PDFs display similar behavior in the positive-$x$ region. In contrast, noticeable discrepancies remain in the negative-$x$ region, which is more sensitive to systematic uncertainties such as higher-twist contamination and truncation effects in the matrix elements.
Therefore, while the quasi-PDFs show encouraging consistency at higher momenta in the positive-$x$ region, the results in the negative-$x$ region must be interpreted with caution, and any conclusions drawn should be considered qualitative for the anti-quark region. The latter is also much smaller in magnitude than the quark contribution.


\begin{figure}[h!]
    \includegraphics[scale=0.28]{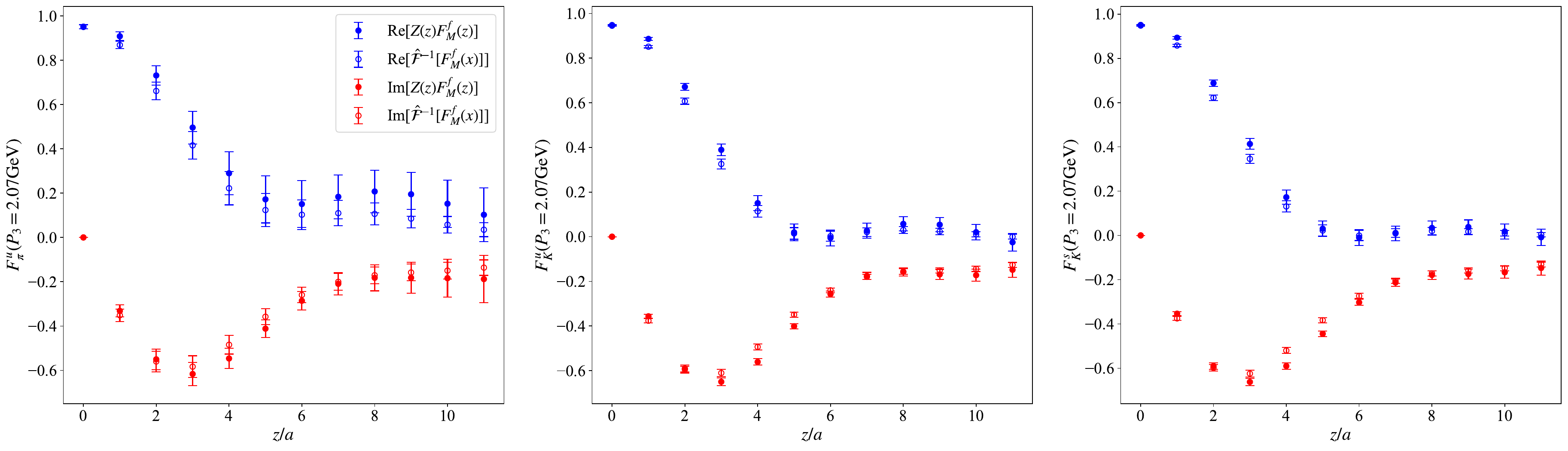}
        \vspace*{-0.28cm}
    \caption{\small{The real (blue) and imaginary (red) renormalized matrix element, $Z(z)F_M^f(z)$, with the inverse Fourier Transform of the Backus-Gilbert result, $\mathcal{F}^{-1}[F_M^f(x)]$ for the pion (left), kaon-up (middle), and kaon-strange (right) at $|P_3|=2.07$ GeV.}}
    \label{fig:BG_consistence_P5}
\end{figure}

Fig.~\ref{fig:BG_consistence_P5} provides a useful investigation for the reconstruction procedure. In particular, we compare the renormalized coordinate-space matrix elements, $Z(z)F^{f}_{M}(z)$, with the inverse Fourier transform of the Backus-Gilbert reconstructed quasi-PDFs, $\mathcal{F}^{-1}[F_M^f(x)]$ , at $P_{3} = 2.07$~GeV. For clarity of presentation, we display $\mathcal{F}^{-1}[F_M^f(x)]$ only at integer values of $z$, allowing for a direct comparison with the renormalized lattice data. This comparison provides insight into the BG reconstruction in momentum space. It should be noted that, even when the data appear consistent, the method remains subject to intrinsic limitations. The comparison is qualitative due to the larger statistical uncertainties observed at this momentum. For the pion, the reconstruction continues to follow the coordinate-space matrix elements reasonably well. In the kaon case, particularly for the up quark, differences between the two determinations become more pronounced at larger $z/a$. As the precision of lattice calculations continues to improve, the need for more robust, flexible reconstruction strategies becomes increasingly urgent. A first step toward such developments has already been taken by members of our group, as discussed in Ref.~\cite{Chu:2025jsi}.

\begin{figure}[h!]
    \centering
    \includegraphics[scale=0.40]{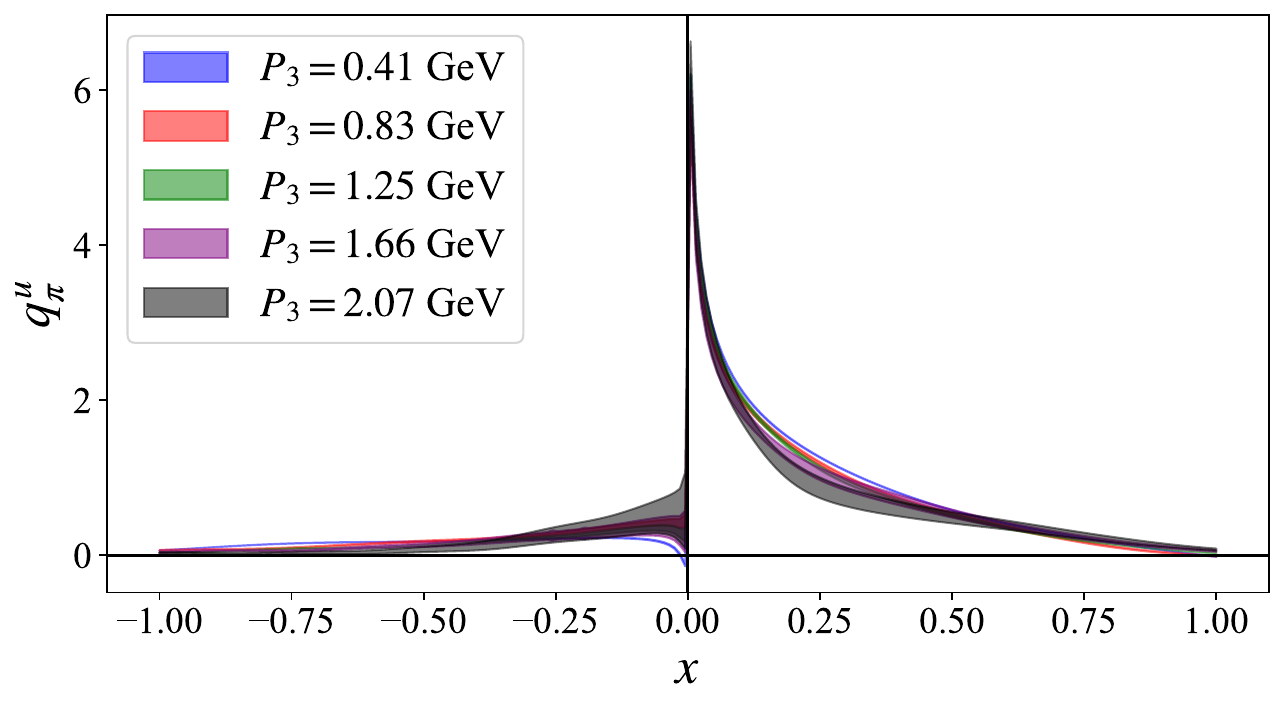}
\vspace*{-0.4cm}    
    \caption{\small{Light-cone pion PDF, $q_\pi^u$, for momenta $|P_3| = 0.41,~0.83,~1.25,~1.66,~2.07$ GeV as obtained from the LaMET approach. Results are shown in the $\MSb$ scheme at a scale of 2 GeV. }}
    \label{fig:lightcone_pion}
\end{figure}

\begin{figure}[h!]
    \centering
    \includegraphics[scale=0.37]{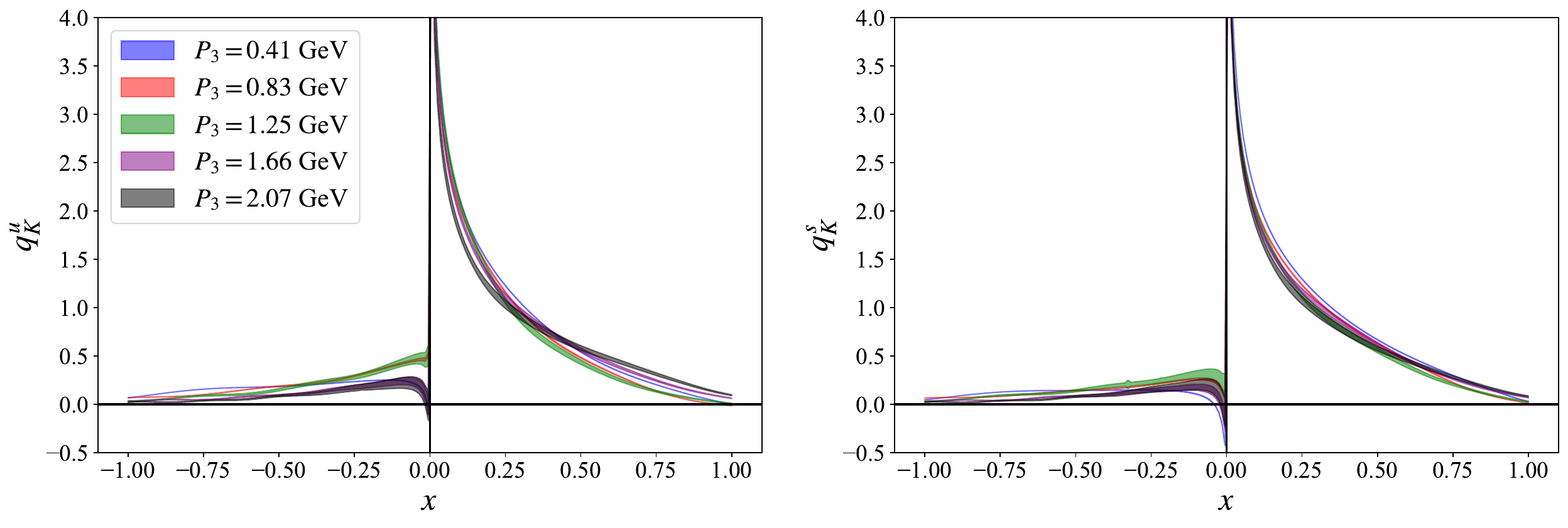}
\vspace*{-0.4cm}    
    \caption{\small{Same as Fig.~\ref{fig:lightcone_pion}, but for the light-cone kaon PDFs, $q_K^u$ (left) and $q_K^s$ (right).}}
    \label{fig:lightcone_kaon}
\end{figure}

\newpage
The final step in the quasi-distribution analysis is the matching to the light-cone PDFs, as described in Eq.~\eqref{eq:matching}. The resulting distributions are shown in Fig.~\ref{fig:lightcone_pion} for the pion and Fig.~\ref{fig:lightcone_kaon} for the kaon. It is informative to compare the $P_3$ dependence of the matched PDFs to that observed for the quasi-PDFs in Figs.~\ref{fig:BG_pion} - \ref{fig:BG_kaon}.
Notably, after matching, the results at lower momenta ($P_3 < 1$~GeV) are closer to those at higher momenta than in the quasi-PDF case, where a more pronounced separation was observed. As mentioned above, such a difference is expected prior to applying the matching kernel. Nonetheless, the low-momentum results in the final light-cone PDFs remain distinct and do not overlap with those at larger boosts. The intermediate momentum of 1.25 GeV yields PDFs that are broadly consistent with the high-momentum results, although some differences persist.
Focusing on the quark region ($x > 0$), we find excellent agreement between the results at $P_3 = 1.66$~GeV and $P_3 = 2.07$~GeV for both the pion and the strange-quark component of the kaon PDF. In contrast, some tension is observed in the up-quark distribution of the kaon PDF.
We emphasize that all quoted uncertainties are statistical only.

\begin{figure}[h!]
    \centering
    \includegraphics[scale=0.41]{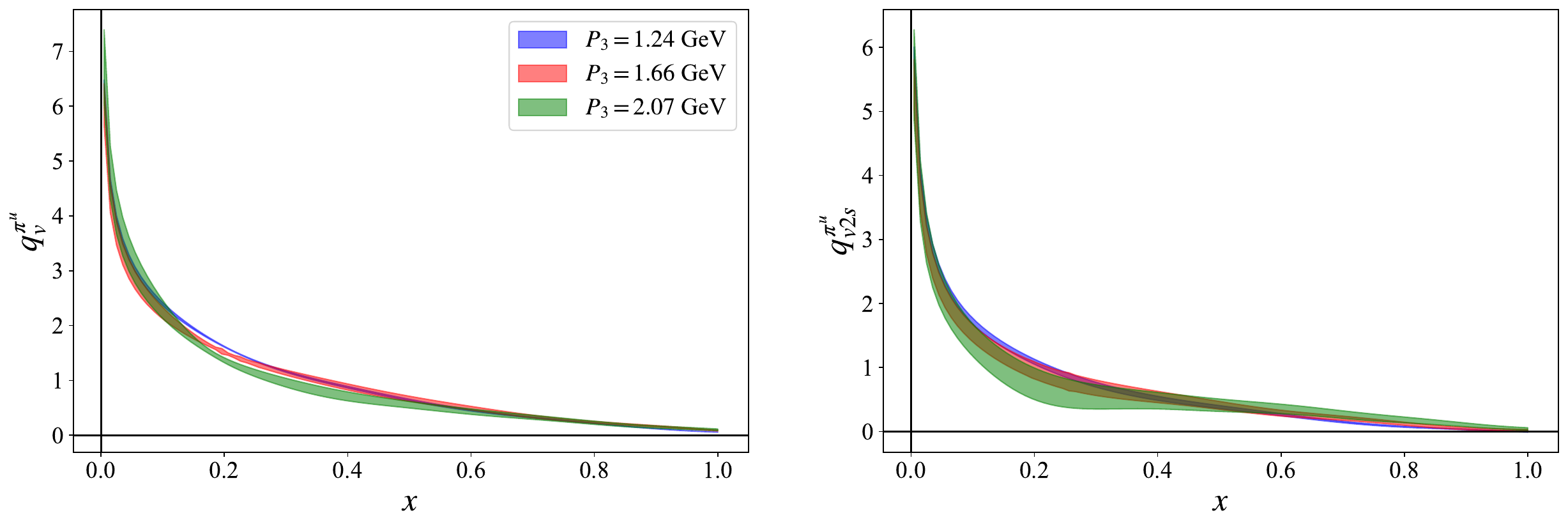}
\vspace*{-0.4cm}    
     \caption{\small{Light-cone pion PDFs,  $q^{\pi^u}_v$ (left) and $q^{\pi^u}_{v2s}$ (right), for momenta $|P_3| = 1.25,~1.66,~2.07$ GeV as obtained from the LaMET approach. Results are shown in the $\MSb$ scheme at a scale of 2 GeV.}}
    \label{fig:qv_qv2s_pion_quasi}
\end{figure}

\begin{figure}[h!]
    \centering
    \includegraphics[scale=0.41]{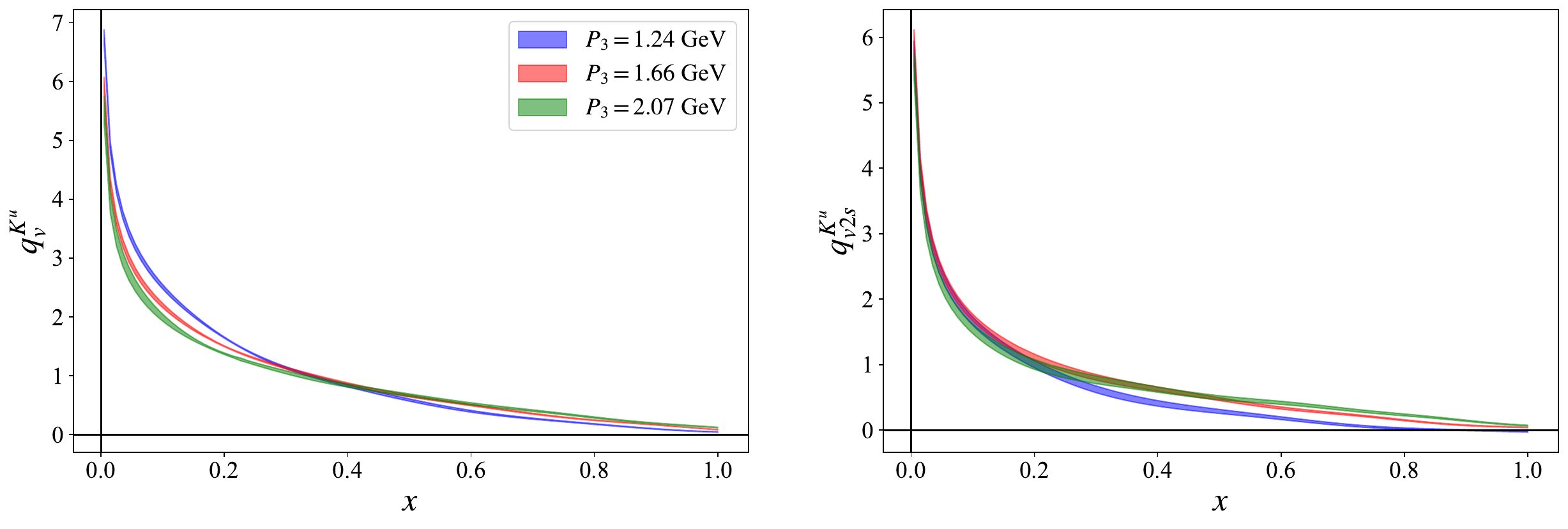}
\vspace*{-0.4cm}    
    \caption{\small{Light-cone up-quark kaon PDFs,  $q^{K^u}_v$ (left) and $q^{K^u}_{v2s}$ (right), for momenta $|P_3| = 1.25,~1.66,~2.07$ GeV as obtained from the LaMET approach. Results are shown in the $\MSb$ scheme at a scale of 2 GeV.}}
    \label{fig:qv_qv2s_kaon_u_quasi}
\end{figure}

\begin{figure}[h!]
    \centering
    \includegraphics[scale=0.41]{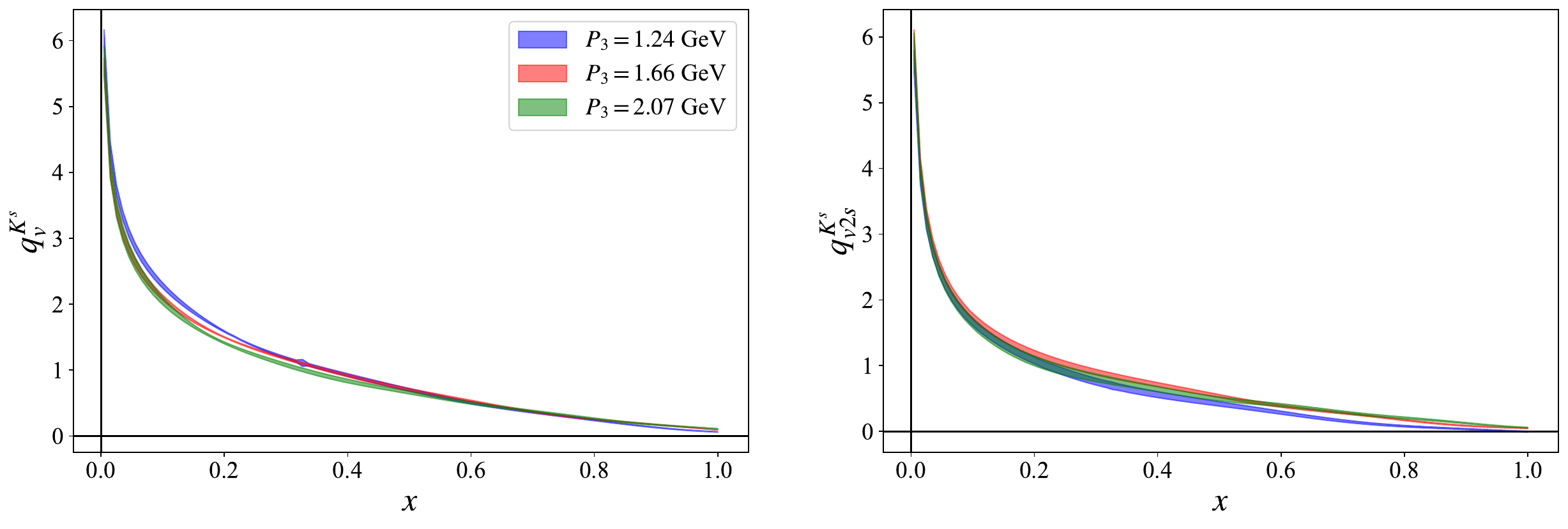}
\vspace*{-0.4cm}    
    \caption{\small{Light-cone strange-quark kaon PDFs,  $q^{K^s}_v$ (left) and $q^{K^s}_{v2s}$ (right), for momenta $|P_3| = 1.25,~1.66,~2.07$ GeV as obtained from the LaMET approach. Results are shown in the $\MSb$ scheme at a scale of 2 GeV.}}
    \label{fig:qv_qv2s_kaon_s_quasi}
\end{figure}

For completeness, we present the valence distributions, $q^{\pi^u}_v(x)$, $q^{K^u}_{v}(x)$, $q^{K^s}_{v}(x)$, defined in Eq.~\eqref{eq:valence_pion} and Eq.~\eqref{eq:valence_kaon} in Figs.~\ref{fig:qv_qv2s_pion_quasi} - \ref{fig:qv_qv2s_kaon_s_quasi}. The figures also include the combination $q^{\pi^u}_{v2s}$,  $q^{K^u}_{v2s}$, $q^{K^s}_{v2s}$, of Eq.~\eqref{eq:v2s_pion} and Eq.~\eqref{eq:v2s_kaon}.
For better clarity in the results interpretation, we only show the results for momentum above 1 GeV.
While we observe some differences in certain regions of $x$ as momentum changes, the overall trend of these functions aligns. The most notable differences between momenta are in the case of $q^{K^u}_{v}$ and $q^{K^u}_{v2s}$, a trend we have also seen in $q_K^u$ of Fig.~\ref{fig:lightcone_kaon}.
\vspace*{-0.25cm}   
\vspace*{-0.25cm} 

\begin{figure}[h!]
    \centering
    \includegraphics[scale=0.40]{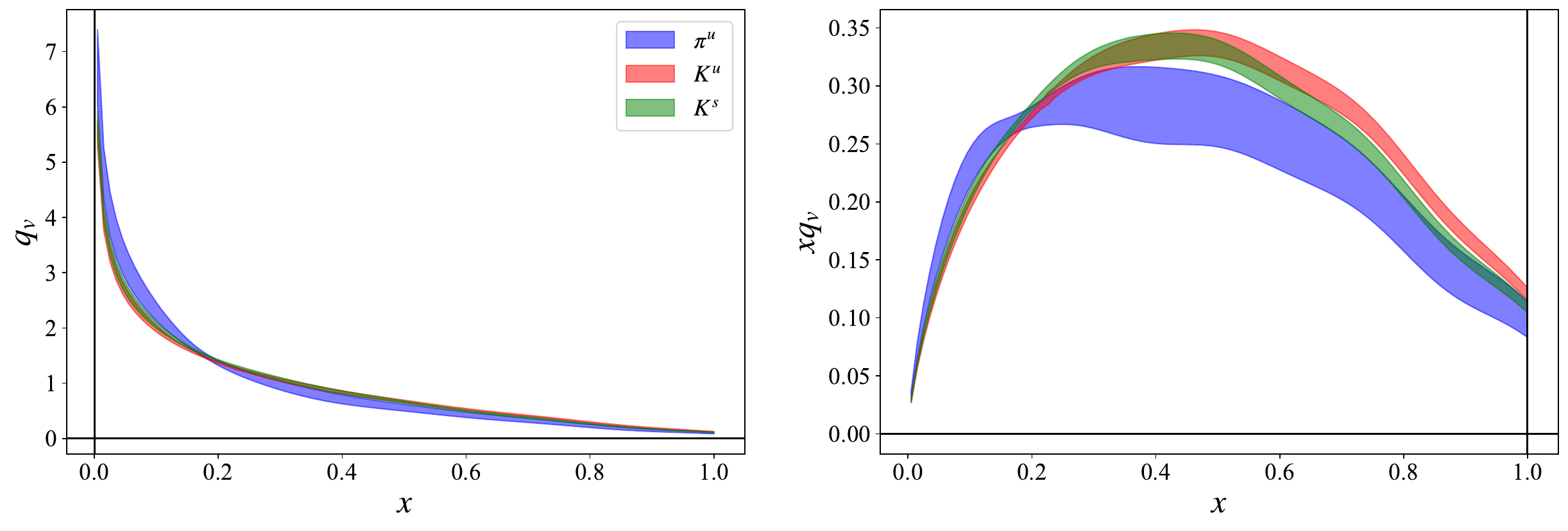}
\vspace*{-0.4cm}    
    \caption{\small{Comparison of $q_v$ (left) and $x q_v$ (right) for the pion and kaon at $P_3=2.07$ GeV. Results are shown in the $\MSb$ scheme at a scale of 2 GeV.}}
    \label{fig:qv_xqv_comparison_P5}
\end{figure}

\begin{figure}[h!]
    \centering
    \includegraphics[scale=0.40]{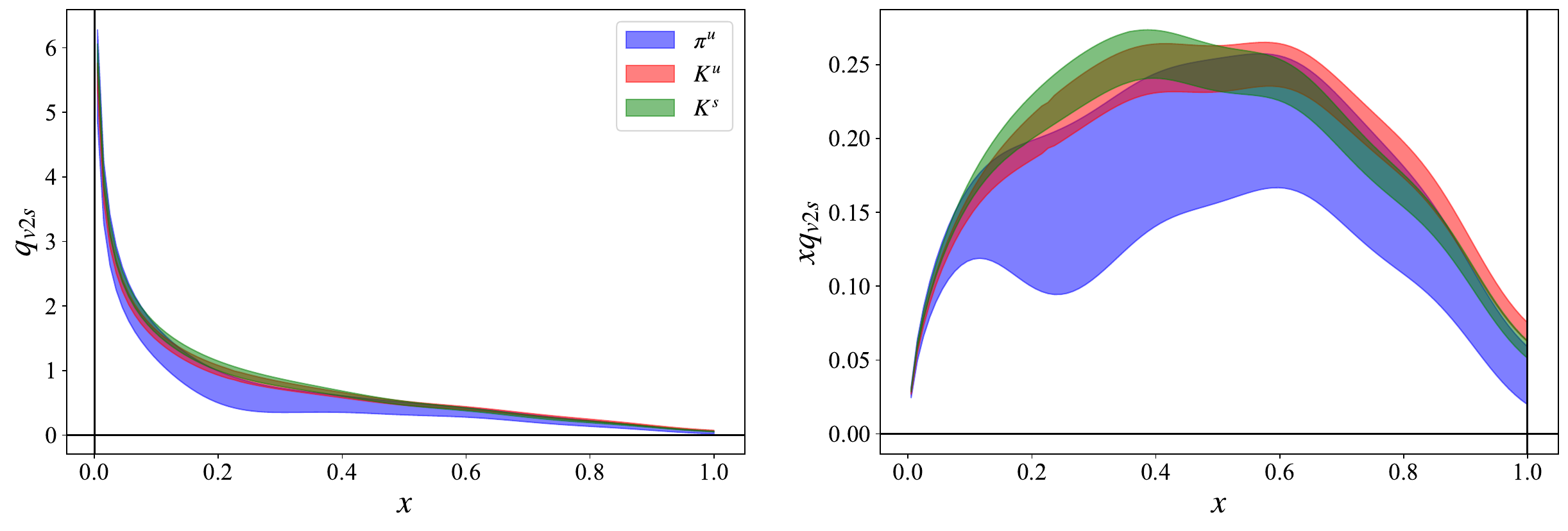}
\vspace*{-0.4cm}    
    \caption{\small{Comparison of $q_{v2s}$ (left) and $x q_{v2s}$ (right) for the pion and kaon at $P_3=2.07$ GeV. Results are shown in the $\MSb$ scheme at a scale of 2 GeV.}}
    \label{fig:qv2s_xqv2s_comparison_P5}
\end{figure}

\newpage
It is interesting to compare these distributions across different quark flavors and/or particles at a fixed momentum. 
Such a comparison is shown for the combination $q_v$ in Fig.~\ref{fig:qv_xqv_comparison_P5} for the highest momentum 2.07 GeV. 
Each figure shows panels for the valence PDFs, $q_v$, as well as their rescaled functions $x q_v$. 
The corresponding results for $q_{v2s}$ can be found in Fig.~\ref{fig:qv2s_xqv2s_comparison_P5} for $P_3=2.07$ GeV.
Focusing on the right panels where the distributions are rescaled by a factor of $x$, the main observation is that, for all cases, the distributions are of similar shape.
However, the pion is much noisier than the kaon despite the increased statistics. This effect is inherited by the matrix elements (see, e.g., comparisons in Figs.~\ref{fig:g0_pion} - \ref{fig:g0_kaon_s} and Table~\ref{tab:error}).
For both flavor combinations, the pion tends to be lower than the kaon PDFs, while $x q_{v2s}^{K^u}$ and $x q_{v2s}^{K^s}$ are very similar and overlap for extended values of $x$. 
More definite conclusions require an even higher statistics and addressing all possible systematic uncertainties. Due to such uncertainties, lattice results from different methodologies using the same ensemble may give results that have differences (see, e.g., Fig. 9 from Ref.~\cite{Constantinou:2020pek}). 

\begin{figure}[h!]
    \centering
    \includegraphics[scale=0.29]{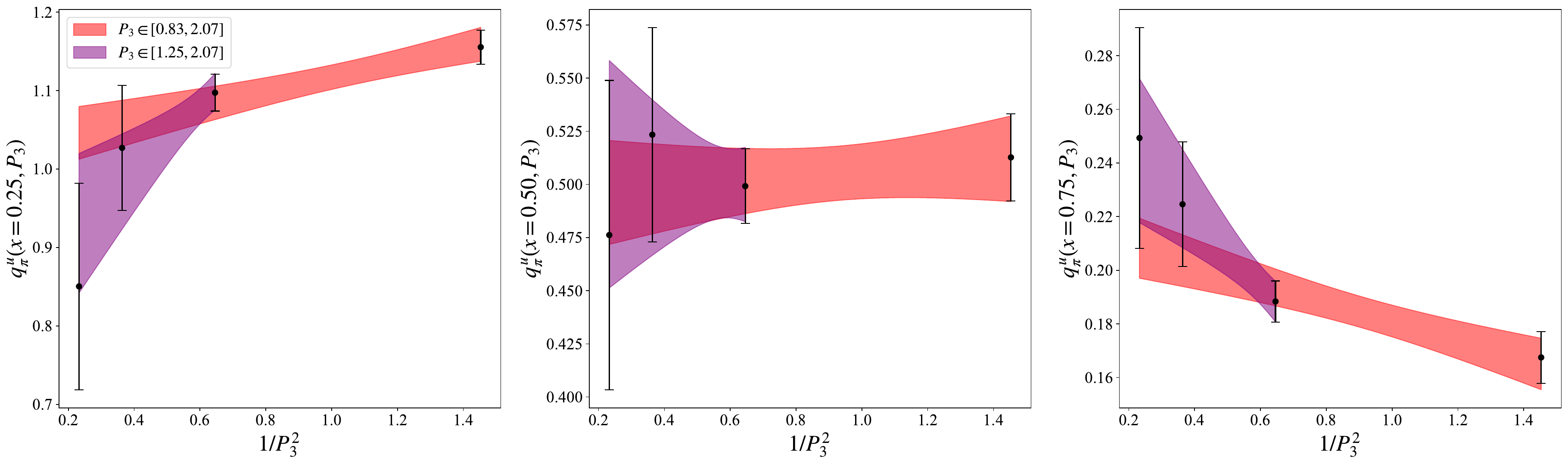}
\vspace*{-0.4cm}    
    \caption{\small{Fits on the pion PDF, $q_\pi^u$, at (left to right) $x=\{0.25,0.50,0.75\}$ for $P_3\in [0.83,2.07]$ GeV (red), and $P_3\in [1.25,2.07]$ GeV (purple).}}
    \label{fig:fixed_x_pion_inf_mom}
\end{figure}

\begin{figure}[h!]
    \centering
    \includegraphics[scale=0.29]{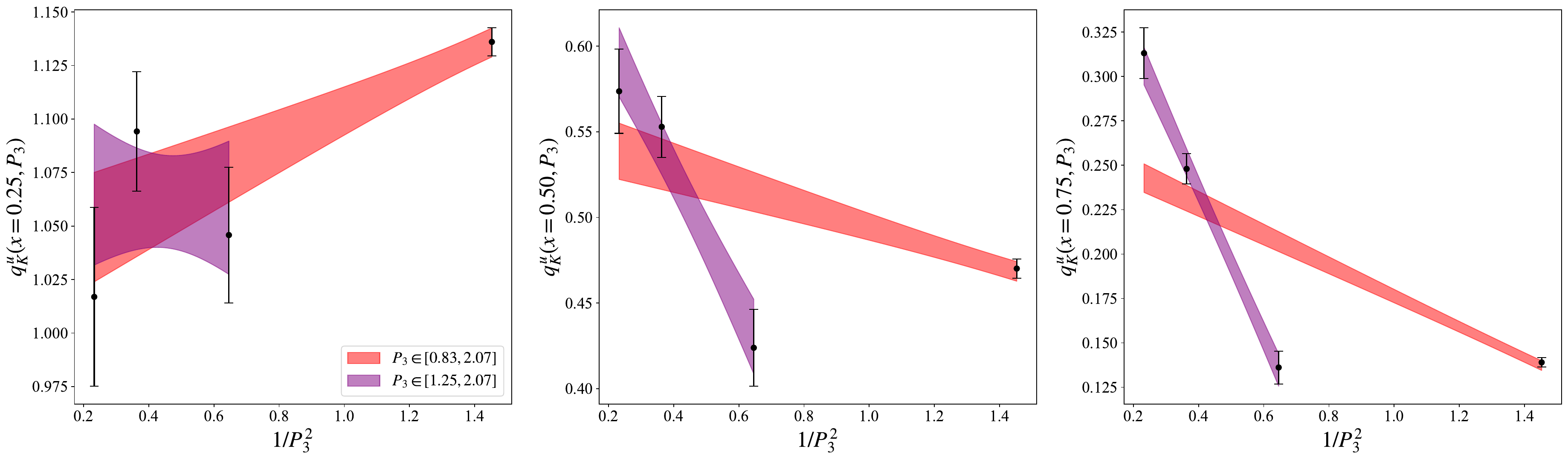}
\vspace*{-0.4cm}    
    \caption{\small{Fits on the kaon up-quark PDF, $q_K^u$, at (left to right) $x=\{0.25,0.50,0.75\}$ for $P_3\in [0.83,2.07]$ GeV (red), and $P_3\in [1.25,2.07]$ GeV (purple). }}
    \label{fig:fixed_x_kaon_u_inf_mom}
\end{figure}

\begin{figure}[h!]
    \centering
    \includegraphics[scale=0.29]{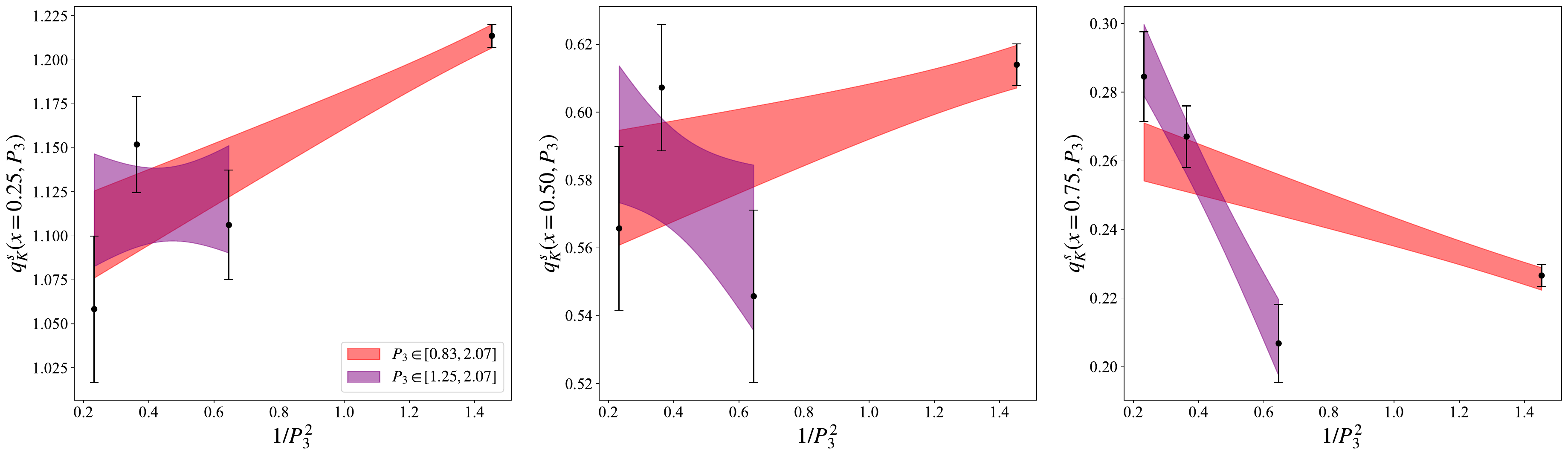}
\vspace*{-0.4cm}    
    \caption{\small{Fits on the kaon strange-quar PDF, $q_K^s$, at (left to right) $x=\{0.25,0.50,0.75\}$ for $P_3\in [0.83,2.07]$ GeV (red), and $P_3\in [1.25,2.07]$ GeV (purple). }}
    \label{fig:fixed_x_kaon_s_inf_mom}
\end{figure}

An additional component of this analysis is the investigation of the infinite-momentum limit for these distributions. To this end, we apply a parametrization of the form: 
\begin{eqnarray}
q(x) = q_\infty(x) + \frac{q_1(x)}{P_3^2}\,.
\label{eq:infty}
\end{eqnarray}
We have implemented this ansatz using the data up to 2.07 GeV, but with a different value for the lower momentum starting from $P_3=0.83$ GeV.
Here, we present the fits where $P_3 \in [0.83, 2.07]$ (\textit{fit 1}) and $P_3 \in [1.25, 2.07]$ (\textit{fit 2}). 
Figs.~\ref{fig:fixed_x_pion_inf_mom} - \ref{fig:fixed_x_kaon_s_inf_mom} display infinite-momentum fits to $q_\pi^u$, $q_K^u$, and $q_K^s$, respectively. For each meson, three representative values of the momentum fraction $x \in \{0.25, 0.50, 0.75\}$ are shown from left to right. The bands correspond to the fit uncertainty at infinite momentum, while the data points reflect the finite-momentum PDFs of Figs.~\ref{fig:lightcone_pion} - \ref{fig:lightcone_kaon} with their statistical errors. For the pion case (Fig.~\ref{fig:fixed_x_pion_inf_mom}), both fits describe the data reasonably well, though at low $x = 0.25$, \textit{fit 1} tends to overshoot the point at the highest momentum. In all $x$ values, \textit{fit 2} appears to best capture the $1/P_3^2$ behavior. Similar conclusions can be drawn from the kaon up-quark fits (Fig.~\ref{fig:fixed_x_kaon_u_inf_mom}), where the \textit{fit 1} fails to describe the $P_3$ behavior at large $x$. The kaon strange-quark distribution(Fig.~\ref{fig:fixed_x_kaon_s_inf_mom}) shows a comparable trend: while all fits perform well, \textit{fit 2} provides a better description of the data at higher $P_3$, in particular at high $x$. Based on the above, we choose \textit{fit 2} ($P_3 \in [1.25, 2.07]$~GeV) for the final data, despite the enhanced uncertainties. This is also motivated by the fact that a large $P_3$ is an important element of LaMET. 

\begin{figure}[h!]
\hspace*{-2.5cm}    \includegraphics[scale=0.34]{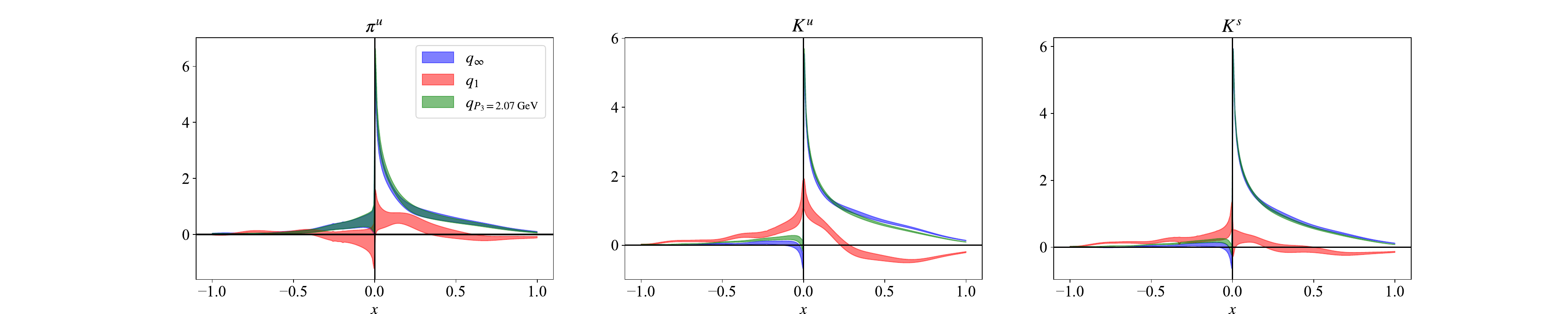}
    \caption{Comparison of $q_{\infty}(x)$ (blue), $q_1(x)$ (red), and $q(x; P_3=2.07~\mathrm{GeV})$ (green) for the pion (left), kaon up (middle), and kaon strange (right) distributions}
    \label{fig:qinf_q1_q5}
\end{figure}

\begin{figure}[h!]
\hspace*{-2.5cm}    \includegraphics[scale=0.34]{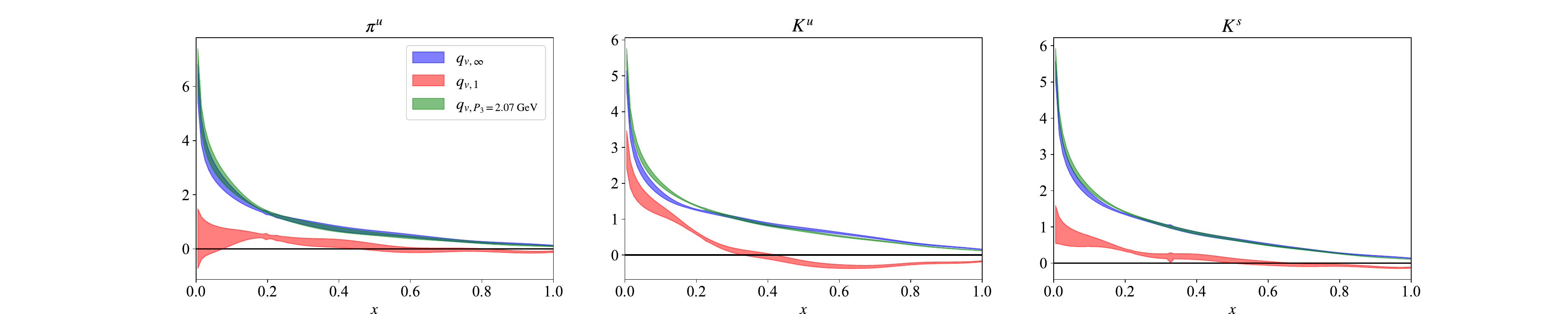}
    \caption{Comparison of $q_{v,\infty}(x)$ (blue), $q_{v,1}(x)$ (red), and $q_{v}(x;P_3=2.07~\mathrm{GeV})$ (green) for the pion (left), kaon up (middle), and kaon strange (right) distributions. }
    \label{fig:qvinf_qv1_qv5}
\end{figure}

\begin{figure}[h!]
\hspace*{-2.5cm}    \includegraphics[scale=0.34]{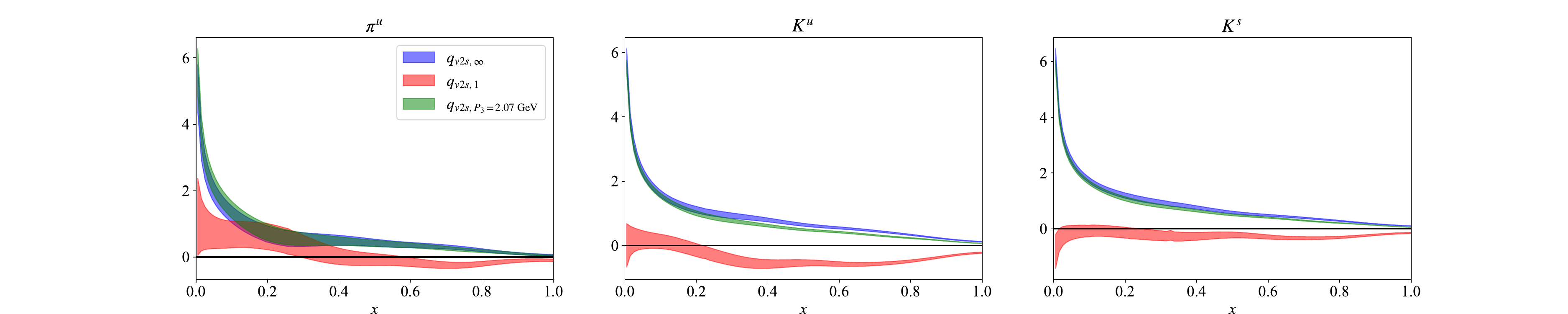}
    \caption{Comparison of $q_{v2s,\infty}(x)$ (blue), $q_{v2s,1}(x)$ (red), and $q_{v2s}(x;P_3=2.07~\mathrm{GeV})$ (green) for the pion (left), kaon up (middle), and kaon strange (right) distributions.}
    \label{fig:qv2sinf_qv2s1_qv2s5}
\end{figure}

It is instructive to examine the residual function $q_1(x)$ obtained from the fit of Eq.~\eqref{eq:infty}. 
Figs.~\ref{fig:qinf_q1_q5} - \ref{fig:qv2sinf_qv2s1_qv2s5} compare the infinite-momentum extrapolated distribution $q_\infty(x)$ with both the distribution extracted at the highest momentum, $P=2.07$~GeV, and the corresponding residual term $q_1(x)$. 
For completeness, we present the results for the individual quark flavors, as well as for the valence and $v2s$ flavor combinations.
Overall, we observe that $q_\infty(x)$ follows the distribution obtained at the largest momentum, indicating that the highest-boost data already approximate the asymptotic behavior within uncertainties. 
The residual contribution $q_1(x)$ is generally subleading, although it is not strictly negligible. 
In particular, its magnitude becomes more noticeable in the intermediate-to-large $x$ region, providing indication of non-negligible power corrections.
We note that the fit of Eq.~\eqref{eq:infty} is performed independently at each value of $x$, and therefore $q_1(x)$ is not constrained to form a continuous function. 
Despite this, the extracted $q_1(x)$ exhibits a smooth behavior across most of the $x$ range, with the exception of the $x=0$ point in Fig.~\ref{fig:qinf_q1_q5}, which is not an area accessible from lattice QCD.

\begin{figure}[h!]
    \centering
    \includegraphics[scale=0.395]{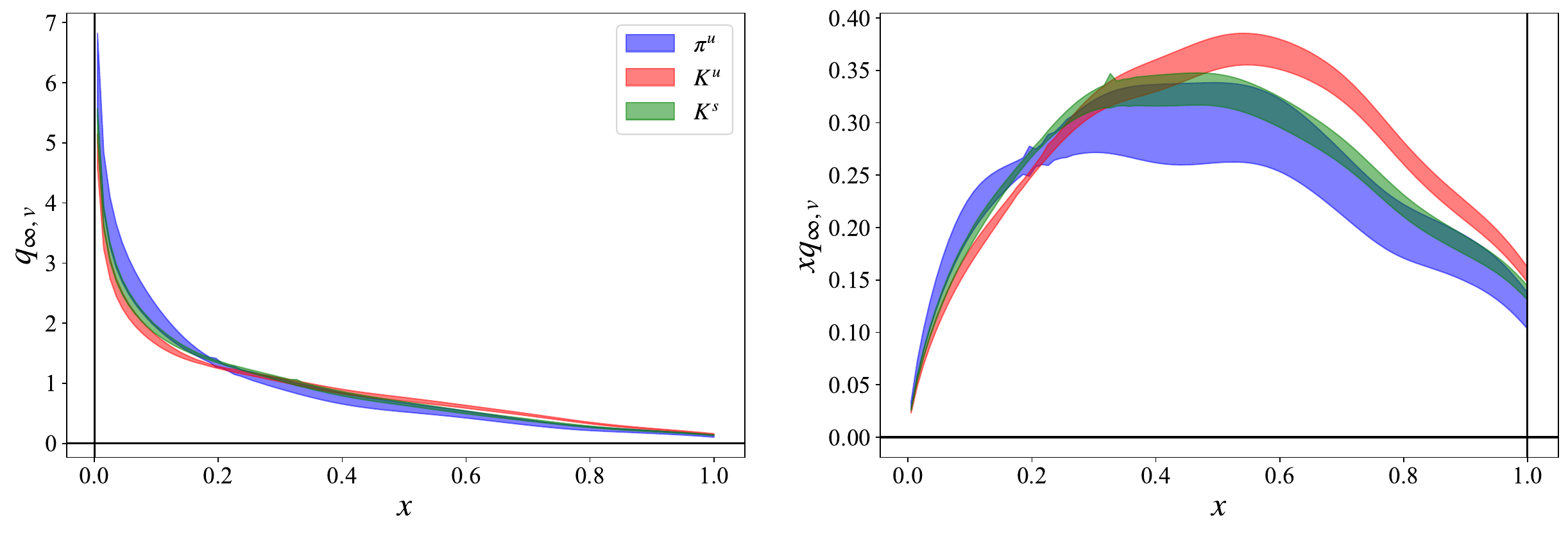}
\vspace*{-0.4cm}    
    \caption{\small{Comparison of pion (blue), kaon up-quark (red), and kaon strange-quark (green) $q_{\infty,\,v}$ (left) and $xq_{\infty,\,v}$ (right). Results are shown in the $\MSb$ scheme at a scale of 2 GeV. }}
    \label{fig:inf_mom_qv_pion_kaon}
\end{figure}

\begin{figure}[h!]
    \centering
    \includegraphics[scale=0.395]{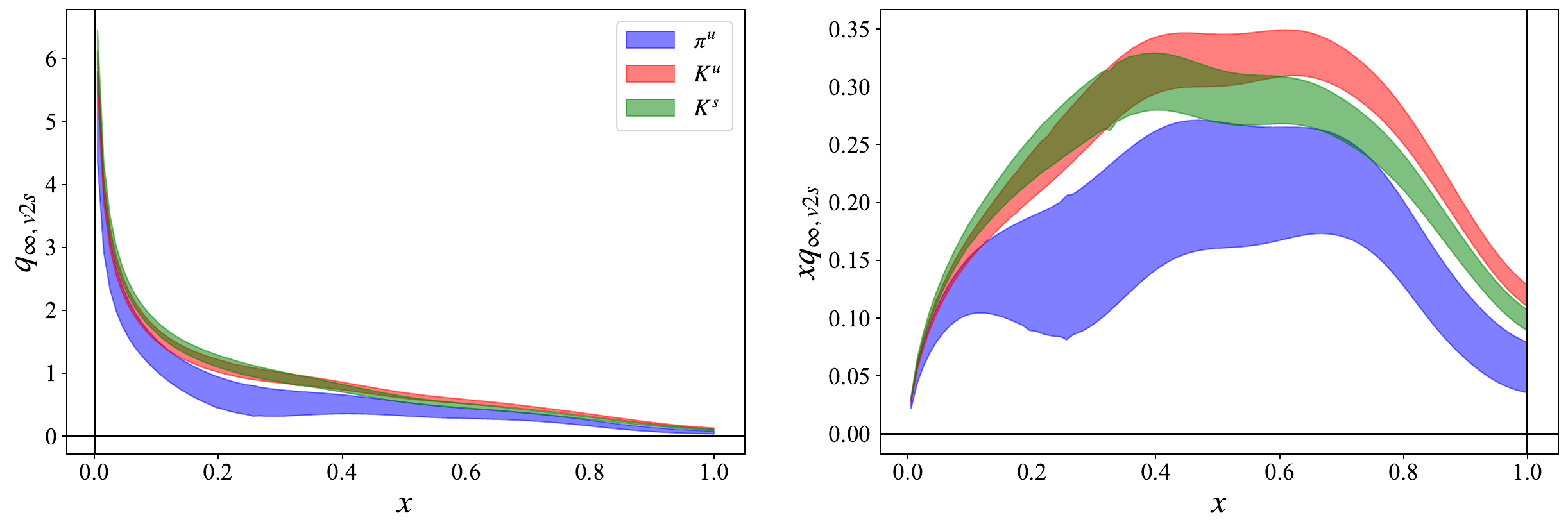}
\vspace*{-0.4cm}    
    \caption{\small{Comparison of pion (blue), kaon up-quark (red), and kaon strange-quark (green) $q_{\infty,\,v2s}$ (left) and $xq_{\infty,\,v2s}$ (right). Results are shown in the $\MSb$ scheme at a scale of 2 GeV.}}
    \label{fig:inf_mom_qv2s_pion_kaon}
\end{figure}

To complement the $P_3 \to \infty$ extrapolations shown above, we now present a comparison of the pion and kaon distributions at infinite momentum. At finite momentum, such comparisons are shown in Fig.~\ref{fig:qv2s_xqv2s_comparison_P5} for $P_3 = 2.07$~GeV. 
The observations of the finite-momentum comparison are reinforced by the extrapolated $P_3 \to \infty$ results shown in Figs.~\ref{fig:inf_mom_qv_pion_kaon} - \ref{fig:inf_mom_qv2s_pion_kaon}. 
Overall, the distributions have a non-vanishing tail at $x\to1$, which is a caveat of the inverse problem in the reconstruction of the $x$ dependence. With this in mind, it is not instructive to compare the distributions quantitatively. However, they have good qualitative features, with the pion having the largest statistical uncertainties, and tends to be lower than the kaon. A more direct comparison requires even higher statistics, as well as other reconstruction methods, such as neural network methods~\cite{Karpie:2019eiq}.
For the same reason, comparing the PDFs for the SDF and LaMET methods cannot be direct. In general, we have seen that each method is susceptible to different systematic uncertainties, leading to differences in some cases.
Further investigation at even higher statistics is required to clarify this behavior.

\section{Summary and Outlook}
\label{sec:conclusions}

In this work, we present a detailed lattice QCD determination of the unpolarized twist-2 parton distribution functions (PDFs) for the pion and kaon. The calculation is done on an $N_f=2+1+1$ ensemble of maximally twisted mass fermions with a clover term. The ensemble reproduces a pion mass of 260 MeV, a kaon mass of 530 MeV, and a lattice spacing of 0.0934 fm. Utilizing meson boosts up to $|P_3|=2.07$ GeV, we analyzed non-local operators containing Wilson lines within both the Large-Momentum Effective Theory (LaMET) and Short-Distance Factorization (SDF) frameworks.

The use of both LaMET and SDF allows us to analyze the same lattice matrix elements through two distinct factorization formalisms. In the LaMET framework, the extraction is performed at fixed momentum boosts and probes convergence in the large-momentum expansion, including sensitivity to power corrections in $1/P_3$. In contrast, the SDF approach exploits short-distance factorization in Ioffe time and combines multiple momentum–separation pairs, introducing a different reconstruction strategy and sensitivity to short-distance behavior. The comparison between the two approaches can, therefore, provide a nontrivial cross-check of methodology-driven systematics, including power corrections, matching effects, and reconstruction assumptions. Once systematic uncertainties are quantified (some requiring multi-ensemble datasets), comparison between the two frameworks can help assess the reliability of the lattice data. In this sense, the dual analysis can strengthen the robustness of the conclusions beyond what could be inferred from either method alone.

Interestingly, the light-cone PDFs reconstructed via LaMET exhibit small, but non-negligible, momentum boost dependence even at $P_3$ below 1 GeV. To this end, we perform an infinite-momentum extrapolation, focusing on the momenta above 1 GeV.
The quoted uncertainties in this work are purely statistical, as the data are obtained from a single lattice ensemble. A controlled assessment of these effects requires additional ensembles.

In addition to the standard PDFs, we also obtain the valence and $v2s$ flavor combinations. The light-cone PDFs are presented in the $\MSb$ scheme at a scale $\mu$ of 2 GeV; selected results are shown at $\mu=5.2$ GeV.By comparing the different flavors and particles, we discuss the role of the up-quark in the pion and kaon, as well as the differences between the up and strange quarks in the kaon. At $\mu=5.2$ GeV, we find that the up-quark contribution is similar in the pion and kaon, and they are up to 80\% of the strange-quark contribution in the kaon; this is consistent with other studies as discussed in Sec.~\ref{sec:results_pseudo}.

A promising direction for future work is the systematic investigation of methodology-related uncertainties. This includes exploring less-parametric reconstruction strategies, such as neural-network–based fits, as well as examining alternative renormalization schemes, consideration of fitting the large-$z$ region in LaMET, and implementing next-to-next-to-leading-order matching. Our preliminary analysis on the hybrid renormalization~\cite{Ji:2020brr} shows similar behavior as the RI-type renormalization used in this work. In addition, implementing a fit of the form  $Ae^{-m\nu}/|\nu|^b$, where $A,~m,~$ and $b$ are fitting parameters at large $\nu$~\cite{Ji:2020brr,Ding:2024saz,Ji:2026vir} is under exploration.
The data analyzed in this study also enable the extraction of Mellin moments of the pion and kaon PDFs, providing an opportunity to compare with determinations based on local operators.
This work is embedded in a broader research program that calculates not only leading-twist PDFs, but also twist-3 PDFs and generalized parton distributions of the pion and kaon. Since some of these observables remain largely unexplored in lattice QCD, our initial effort has focused on establishing the methodology, producing high-statistics results, and extending the momentum boost to 2 GeV on a single pion-mass ensemble. As the program advances and the analysis of these datasets is completed, we plan to extend the study to additional ensembles at the same pion mass in order to quantify discretization effects, and subsequently incorporate ensembles at the physical point to enable controlled continuum and chiral extrapolations.

\begin{acknowledgements}
J. M. and M.~C. acknowledge financial support by the U.S. Department of Energy, Office of Nuclear Physics,  under Grant No.\ DE-SC0025218.
J.D received support from Argonne National Laboratory under the contract ``Pion and Kaon Form Factors using Lattice QCD''.
J. Torsiello acknowledges support by the U.S. Department of Energy, Office of Science, Office of Advanced Scientific Computing Research, Department of Energy Computational Science Graduate Fellowship under Award Number DE-SC0024386~\footnote{This manuscript was prepared as an account of work sponsored by an agency of the United States Government. Neither the United States Government nor any agency thereof, nor any of their employees, makes any warranty, express or implied, or assumes any legal liability or responsibility for the accuracy, completeness, or usefulness of any information, apparatus, product, or process disclosed, or represents that its use would not infringe privately owned rights. Reference herein to any specific commercial product, process, or service by trade name, trademark, manufacturer, or otherwise does not necessarily constitute or imply its endorsement, recommendation, or favoring by the United States Government or any agency thereof. The views and opinions of authors expressed herein do not necessarily state or reflect those of the United States Government or any agency thereof.}
I. A. acknowledges support by the REU program HPC Tools, Techniques, and Research across the Physical Sciences, funded by the National Science Foundation under Grant No. 2348782. 
K.~C.\ is supported by the National Science Centre (Poland) grant OPUS No.\ 2021/43/B/ST2/00497. 
S. L. received support for this project from the Research Scholars Program of the College of Science and Technology (CST) at Temple University, as well as the Frances Velay Fellowship through CST.
The authors acknowledge partial support from the U.S. Department of Energy, Office of Science, Office of Nuclear Physics, under the umbrella of the Quark-Gluon Tomography (QGT) Topical Collaboration, with Award DE-SC0023646.
Computations for this work were carried out in part on facilities of the USQCD Collaboration, which are funded by the Office of Science of the U.S. Department of Energy. 
This research includes calculations carried out on HPC resources supported in part
by the National Science Foundation through major research instrumentation grant number 1625061 and by the US Army Research Laboratory under contract number W911NF-16-2-0189. This research used resources of the Oak Ridge Leadership Computing Facility, which is a DOE Office of Science User Facility supported under Contract DEAC05-
00OR22725.
This research used resources of the National Energy Research
Scientific Computing Center, a DOE Office of Science User Facility
using NERSC award ALCC-ERCAP0030652.
The gauge configurations have been generated by the Extended Twisted Mass Collaboration on the KNL (A2) Partition of Marconi at CINECA, through the Prace project Pra13\_3304 ``SIMPHYS".
Inversions were performed using the DD-$\alpha$AMG solver~\cite{Frommer:2013fsa} with twisted mass support~\cite{Alexandrou:2016izb}.

\end{acknowledgements}

\bibliography{references.bib}

\end{document}